\documentclass[preprintnumbers,superscriptaddress,amsmath,amssymb,prd,nofootinbib,onecolumn]{revtex4-2}
\usepackage{mathrsfs}
\usepackage{amsthm}
\usepackage{braket}
\usepackage{bm}
\usepackage{graphicx}
\usepackage{amsfonts}
\usepackage[colorlinks,linkcolor=red,citecolor=green]{hyperref}
\usepackage{subfigure}
\usepackage{float}
\usepackage{enumerate}
\usepackage{multirow}

\usepackage{array}
\usepackage{upgreek}

\begin{document}
	
	\title{Exact dynamics of spinning particles and the spin wall in Schwarzschild spacetime}

	\author{Chao-Jun Feng}
	\affiliation{Department of Physics, Shanghai Normal University,\\
		100 Guilin Rd, Shanghai 200234, P.R.China}
	
	\author{Rui-Hui Lin}
	\thanks{Corresponding author}
	\email{linrh@shnu.edu.cn}
	\affiliation{Department of Physics, Shanghai Normal University,\\
		100 Guilin Rd, Shanghai 200234, P.R.China}

\begin{abstract}
The motion of a spinning test body is governed by the Mathisson--Papapetrou--Dixon equations and
deviates from geodesics already at first order in the spin. Most studies of this problem truncate
the dynamics at linear order; here we solve the equatorial, spin-aligned problem in Schwarzschild
spacetime at all orders in the spin under the Tulczyjew--Dixon condition. Eliminating the
four-velocity casts the system into a closed algebraic form and reduces the motion to a single
first-order radial equation. Keeping track of both kinematical masses, the norm of the four-momentum
and its projection on the velocity, we find that the representative worldline ceases to be timelike
on a surface in the space of radius, spin and angular momentum. We call this surface the spin wall.
It bounds the radii accessible to a spinning body and is present for every nonvanishing value of the
momentum's angular component; only for the purely radial family does it collapse, and that family
moves on a geodesic. Its position follows in closed form and may lie inside the horizon, enclose it,
or reach the exterior, the last possibility arising already at small spin. We also examine how far
the pole--dipole truncation remains under control where the wall is accessible. Within that regime
we determine the circular orbits and their stability, the spin dependence of the marginally stable
orbit, and the perihelion advance beyond first order in the spin.
		
	\end{abstract}
	
	\maketitle
	
	\tableofcontents
	
\section{Introduction}

The motion of a point test particle in a gravitational field is fully determined by the geodesic equation, 
and its observable consequences---light bending, perihelion precession, gravitational time delay
---constitute the classical tests of general relativity. Real astrophysical bodies, however, 
carry intrinsic angular momentum (spin), and a spinning body does not follow a geodesic: 
the coupling between the spin and the spacetime curvature backreacts on its worldline, 
displacing it from the geodesic by an amount that grows with the spin magnitude. 
With the advent of gravitational-wave astronomy, which has already delivered a large catalog of compact-binary 
coalescences \cite{gw1,LIGOScientific:2018mvr,LIGOScientific:2020ibl,KAGRA:2021vkt}, 
and with next-generation detectors on the ground \cite{Maggiore:2019uih,Reitze:2019iox,Kalogera:2021bya,ET:2025xjr} 
as well as the space-based Laser Interferometer Space Antenna (LISA) \cite{amaro2017,LISA:2022kgy}, 
the motion of spinning compact objects---neutron stars, stellar-mass and supermassive black holes, 
and hypothetical exotic compact objects---has become a problem of direct relevance for gravitational-wave modelling.

Among the most promising targets of LISA are extreme-mass-ratio inspirals (EMRIs), 
in which a stellar-mass compact object spirals into a supermassive black hole \cite{babak2017}. 
These systems provide a unique laboratory for testing whether the central object is truly a black hole described by Einstein gravity, 
or instead a horizonless exotic compact object \cite{Cardoso:2019rvt,Kesden:2004qx}; 
complementary information is being gathered from the direct imaging of rotating black holes, 
including spacetimes with a cosmological constant \cite{Wang:2023fge} 
and photon rings in exotic gravitational backgrounds \cite{GurtasDogan:2025vim}. 
EMRIs are best modeled within a mass-ratio expansion \cite{Barack:2018yvs,Pound:2021qin,poisson2011}: 
at leading adiabatic order the inspiral is driven by the gravitational-wave fluxes sourced by the secondary \cite{Skoupy:2023lih,Skoupy:2024jsi}, 
while post-1-adiabatic corrections---including second-order flux effects, the conservative self-force, 
and crucially the spin of the smaller companion---are essential for accurate waveform templates 
\cite{Huerta:2011kt,Warburton:2017sxk,Piovano:2021iwv,Mathews:2025nyb,Albertini:2024agg}. 
These corrections receive further contributions from the quadrupole and quadratic-in-spin sectors \cite{Rahman:2026qho}, 
from shifted-geodesic approximations to the spinning-body flux \cite{Drummond:2026haw}, 
and from environmental effects on inspirals with spinning secondaries \cite{Lui:2026uai}. 
At linear order in the spin of the secondary, these effects are fully captured by the motion of a spinning test particle 
in the spacetime of the massive primary, together with the outgoing gravitational-wave flux it sources 
\cite{Skoupy:2022adh,Drummond:2023wqc,Piovano:2024yks,Skoupy:2025nie}.

The equations of motion for a spinning test body were derived by Mathisson \cite{mathisson1937} and Papapetrou \cite{papapetrou1951} 
and completed within Dixon's multipole formalism \cite{dixon1964,dixon1970}; 
the resulting Mathisson--Papapetrou--Dixon (MPD) equations govern the evolution of the momentum and of the spin tensor along the body's worldline. 
Because an extended body does not define a unique representative worldline, 
the MPD system must be supplemented by a spin supplementary condition (SSC). 
The most widely adopted choice is the Tulczyjew--Dixon condition \cite{tulczyjew1959}, 
which fixes the center of mass to lie in the momentum rest frame; 
it renders the dynamical mass an exact constant of motion and yields a well-defined, 
causal evolution of the representative point \cite{ehlers1977,costa2015}; recently, 
the resulting dynamics has also been recast into a canonical symplectic form \cite{Ramond:2024sfp}.
The MPD equations are the pole--dipole truncation of the multipolar expansion and presuppose that the
multipole scale of the body, for a spinning body its M\o ller radius \cite{moller1949,semerak1999}, is
small compared with the local curvature scale; how far this condition can be met will have to be
watched once the spin is taken large.

Most studies of the MPD equations in black-hole spacetimes have truncated the dynamics at first order in the spin. 
This linearized treatment already produces interesting effects: 
a linear-in-spin shift of the innermost stable circular orbit (ISCO) of a spinning particle 
\cite{Suzuki:1997by,Jefremov:2015gza,tsupko2016parameters,Favata:2010ic,Bizyaev:2025mva}, 
spin corrections to the periastron advance and to the two-body dynamics \cite{LeTiec:2011bk,Bini:2019zjj}, 
and the emergence of chaos and resonances that have no geodesic counterpart \cite{suzuki1997,suzuki1999,Zelenka:2019nyp,Mukherjee:2019jhd}. 
The spin dynamics has by now been studied in many further directions: spin precession in the strong-deflection regime 
\cite{Geng:2026xcs,Pang:2024tco}, lensing of spinning massive particles in Gauss--Bonnet gravity \cite{Pantig:2026qcf}, 
motion around black holes immersed in dark-matter halos \cite{Tan:2024hzw} or on the brane \cite{Liu:2024lda}, 
in charged and magnetized backgrounds \cite{Ciou:2025ygb,Chen:2025ncm,Jumaniyozov:2025irx,Jumaniyozov:2026lbf}, 
and as a probe of scalar-hairy \cite{Chen:2024luw} or modified-gravity solutions \cite{Turakhonov:2026lia,Umarov:2025ihy}; 
its observational imprints on the S2 star orbit around Sgr~A$^{*}$ have been constrained \cite{Uktamov:2025bth,Uktamov:2026gtm}, 
and general orbital perturbation theory in Schwarzschild spacetime was recently revisited \cite{Yanchyshen:2026bmy}. 
Yet the linear truncation is not uniformly adequate: chaotic behaviour and resonant growth are
genuinely nonlinear, and the accuracy now demanded in the strong-field regime calls for the spin
dependence to be known beyond first order---while the pole--dipole truncation itself bounds how far
in spin any such treatment can be pushed. A reliable understanding of spinning-particle dynamics
therefore requires either a systematic higher-order treatment or an exact one, together with an
account of where the underlying truncation remains valid.

Recently, Witzany and Piovano \cite{Witzany:2023bmq} constructed analytic solutions of the MPD equations 
with the Tulczyjew--Dixon condition near spherically symmetric compact objects, 
expressing the linear-in-spin motion in terms of one-dimensional closed-form integrals and thereby proving its integrability in arbitrary static, 
spherically symmetric spacetimes. 
For equatorial, spin-aligned motion, closed-form expressions for the momentum components are known
\cite{Armaza:2015eha}, and a divergence radius that appears in them has been noted.
These results have since been extended to the full Kerr spacetime 
\cite{Skoupy:2024uan,Piovano:2025aro,Piovano:2026wpz} and to the computation of accurate bound orbits 
\cite{Drummond:2022xej,vandeMeent:2017bcc}, opening the door to a fully nonlinear description of the orbital dynamics. 
Analytic progress has likewise been made in plane gravitational-wave and pp-wave backgrounds, 
where the spin--curvature coupling drives a characteristic deviation of the spin vector and contributes to the gravitational-wave memory effect 
\cite{Wang:2023eqj,Wang:2024dmn,Chen:2025tok,Andrzejewski:2026wmm,Wang:2026grz}. 
In this paper we pursue a complementary algebraic method. Instead of expanding in the spin, 
we eliminate the four-velocity from the MPD equations altogether. 
The resulting system determines the momentum and the spin through closed-form, velocity-independent evolution equations, 
and the four-velocity is recovered a posteriori from the transport relation. For equatorial, 
spin-aligned motion in Schwarzschild spacetime this reformulation yields an exactly integrable dynamics: 
the two Killing charges $E$ and $J$, together with the conserved spin magnitude $s$ and the dynamical mass $\mathcal{M}$, 
fully determine the trajectory.

Here we exploit this reformulation to analyse the equatorial, spin-aligned problem at all
orders in the spin. The analysis keeps track of the distinction between the two kinematical masses,
and in doing so exhibits a feature that no truncated treatment contains: the representative
worldline ceases to be timelike on a surface in the space of radius, spin and angular momentum. We
call this surface the \emph{spin wall}. It bounds the radii accessible to a spinning body and is present
for all but one value of the momentum's angular component. Its position is determined in closed
form, and where it lies outside the horizon we examine how far the pole--dipole truncation remains
under control. In the regime in which that truncation is reliable we determine the circular orbits
and their stability, the displacement of the marginally stable orbit, and the perihelion advance
beyond first order in the spin.

The paper is organized as follows. In Sec.~\ref{sec:MPD} we reformulate the MPD system without the
four-velocity and introduce the two kinematical masses. Section~\ref{sec:planarSchw} constructs the
exact solution for equatorial, spin-aligned motion in Schwarzschild spacetime and analyses the
structure of the resulting radial equation. Section~\ref{sec:spinwall} is devoted to the spin wall:
its origin, its shape in the parameter space, the domain it bounds, and the orbits it filters.
Circular orbits and the weak-field advance are treated in Sec.~\ref{sec:co-prec}, and
Sec.~\ref{sec:dc} collects our conclusions. We use units with $G=c=1$ and the metric signature
$(-,+,+,+)$.

\section{Basic definitions and equations}
\label{sec:MPD}

\subsection{Reformulating the MPD equations}

For a spinning particle in a curved spacetime, the spin--curvature coupling already affects the motion at first order in the spin: the particle no longer follows a geodesic, and its internal rotation feeds back on the motion of the center of mass. Within the pole--dipole approximation this dynamics is described by the Mathisson--Papapetrou--Dixon (MPD) equations \cite{tulczyjew1959,dixon1970}, which read
\begin{align}
	\frac{DP^\mu\vphantom{^{\mu\nu}}}{d\tau} &= -\frac{1}{2}R^\mu{}_{\nu\kappa\lambda} \dot x^\nu S^{\kappa\lambda} \,, \label{eq:mpd-momentum} \\[8pt]
	\frac{DS^{\mu\nu}}{d\tau} &= P^{\mu}\dot{x}^{\nu}-P^{\nu}\dot{x}^{\mu},
	\label{eq:mpd-spin}
\end{align}
where $P^{\mu}$ is the four-momentum, $\dot{x}^{\mu}\equiv dx^{\mu}/d\tau$ is the four-velocity, $\tau$ is the proper time along the worldline, $S^{\mu\nu}=-S^{\nu\mu}$ is the antisymmetric spin tensor, and $R^{\mu}{}_{\nu\kappa\lambda}$ is the spacetime curvature tensor. 

The MPD system consists of 10 component equations for the 10 unknowns $P^\mu$ and $S^{\mu\nu}$. It does not, however, determine the worldline uniquely: the four-velocity is fixed only up to the normalization $\dot{x}^\mu\dot{x}_\mu=-1$, leaving three independent degrees of freedom, which must be removed by a spin supplementary condition (SSC). Throughout this paper we adopt the Tulczyjew--Dixon (TD) condition
\begin{equation}
	S^{\mu\nu} P_\nu = 0\,.
	\label{eq:td-ssc}
\end{equation}
This TD-SSC defines the reference worldline such that the mass dipole moment vanishes in the momentum rest frame, thereby placing the center of mass on the worldline and reducing the spin to a purely spatial three-vector orthogonal to the four-momentum. 

For later convenience we introduce two kinematical scalars,
\begin{equation}
	\mathcal{M}^2 \equiv -P^\mu P_\mu,\qquad
	m \equiv -P^\mu \dot x_\mu,
	\label{eq:def-Mm}
\end{equation}
where $\mathcal{M}$ is the dynamical mass (the norm of the four-momentum) and $m$ measures the projection of the four-momentum onto the four-velocity; they coincide only in the absence of spin, $m=\mathcal{M}$ when $P^\mu\propto\dot{x}^\mu$.

The TD-SSC immediately yields a relation between the four-velocity and the four-momentum. Covariantly differentiating $S^{\mu\nu}P_\nu=0$ along the worldline gives
\begin{equation}
	0 = \frac{D}{d\tau}(S^{\mu\nu}P_\nu)
	  = \frac{DS^{\mu\nu}}{d\tau}P_\nu + S^{\mu\nu}\frac{DP_\nu}{d\tau}.
	\label{eq:dSSC}
\end{equation}
Substituting the spin equation~\eqref{eq:mpd-spin} for $DS^{\mu\nu}/d\tau$,
\[
	(P^\mu\dot{x}^\nu - P^\nu\dot{x}^\mu)P_\nu + S^{\mu\nu}\frac{DP_\nu}{d\tau}
	= -m P^\mu + \mathcal{M}^2\dot{x}^\mu + S^{\mu\nu}\frac{DP_\nu}{d\tau} = 0,
\]
which rearranges to the standard transport relation
\begin{equation}
	\dot{x}^\mu = \frac{m}{\mathcal{M}^2} P^\mu - \frac{S^{\mu\nu}}{\mathcal{M}^2} \frac{DP_\nu}{d\tau}\,.
	\label{eq:transport}
\end{equation}
Due to the TD-SSC condition, $\mathcal{M}^2$ is constant along the motion (see Appendix~\ref{app:MPD-derivation}). Substituting the above relation into Eq.~\eqref{eq:mpd-momentum}, the momentum equation becomes
\begin{equation}
	A_\mu^{\ \eta} \frac{DP_\eta}{d\tau}
	= -\frac{m}{2\mathcal{M}^2} R_{\mu\nu\kappa\lambda} S^{\kappa\lambda} P^\nu,
	\label{eq:mpd-A}
\end{equation}
where we have introduced two matrices
\begin{equation}
	A_\mu^{\ \eta} \equiv \delta_\mu^{\ \eta} - B_\mu^{\ \eta},
	\qquad
	B_\mu^{\ \eta} \equiv \frac{1}{2\mathcal{M}^2} R_{\mu\nu\kappa\lambda} S^{\kappa\lambda} S^{\nu\eta}.
	\label{eq:AB}
\end{equation}
Provided $\det(A^{\ \eta}_\mu) \neq 0$, one can invert $A$ to obtain
\begin{equation}
	\frac{DP_\eta}{d\tau}
	= -\frac{m}{2\mathcal{M}^2} (A^{-1})_\eta{}^{\rho} V_{\rho\nu} P^\nu,
	\label{eq:mpd-momentum2}
\end{equation}
with the auxiliary tensor
\begin{equation}
	V_{\rho\nu} \equiv R_{\rho\nu\kappa\lambda} S^{\kappa\lambda}.
	\label{eq:Vdef}
\end{equation}
Since $S^{\mu\nu}$ is antisymmetric and satisfies the supplementary condition~\eqref{eq:td-ssc}, the matrix $B$ has rank at most~2 (see Appendix~\ref{app:Ainv-proof} for a detailed proof). Consequently,
\begin{align}
	\Delta_A &\equiv \det(A) = 1 - b_1 + \frac{1}{2}(b_1^2 - b_2), \label{eq:detA-gen} \\[4pt]
	b_1 &\equiv \operatorname{Tr}(B)
		= \frac{1}{2\mathcal{M}^2} R_{\mu\nu\kappa\lambda} S^{\kappa\lambda} S^{\nu\mu}
		= \frac{1}{2\mathcal{M}^2} V_{\mu\nu} S^{\nu\mu}, \label{eq:b1} \\[4pt]
	b_2 &\equiv \operatorname{Tr}(B^2)
		= \frac{1}{4\mathcal{M}^4} R_{\mu\nu\kappa\lambda} S^{\kappa\lambda} S^{\nu\eta}
		R_{\eta\rho\sigma\tau} S^{\sigma\tau} S^{\rho\mu}
		= \frac{1}{4\mathcal{M}^4} V_{\mu\nu} S^{\nu\eta} V_{\eta\rho} S^{\rho\mu}. \label{eq:b2}
\end{align}
The inverse of $A$ can then be written explicitly as (the algebraic derivation via the Cayley--Hamilton theorem is given in Appendix~\ref{app:Ainv-proof})
\begin{equation}
	(A^{-1})_\mu^{\ \rho}
	= \delta_\mu^{\ \rho}
	+ \frac{1-b_1}{\Delta_A} B_\mu^{\ \rho}
	+ \frac{1}{\Delta_A} (B^2)_\mu^{\ \rho}.
	\label{eq:Ainv-gen}
\end{equation}
Finally, substituting the momentum equation~\eqref{eq:mpd-momentum2} into the transport relation~\eqref{eq:transport} yields the general velocity formula
\begin{equation}
	\dot{x}^\mu
	= \frac{m}{\mathcal{M}^2} P^\mu
	+ \frac{m}{2\mathcal{M}^4} S^{\mu\nu} (A^{-1})_\nu{}^{\rho} V_{\rho\lambda} P^\lambda.
	\label{eq:xdot}
\end{equation}

Substituting the transport relation~\eqref{eq:transport} into the spin equation~\eqref{eq:mpd-spin} eliminates the four-velocity and yields the velocity-independent form
\begin{equation}
	\frac{DS^{\mu\nu}}{d\tau}
	= \frac{1}{\mathcal{M}^2}\bigl(P^{\nu} S^{\mu\lambda} - P^{\mu} S^{\nu\lambda}\bigr)\frac{DP_\lambda}{d\tau},
	\label{eq:spin-selfconsistent}
\end{equation}
which, together with the momentum equation~\eqref{eq:mpd-momentum2}, determines the evolution of $S^{\mu\nu}$ without reference to $\dot{x}^\mu$. 

By introducing a $4\times4$ matrix
\begin{equation}
	C^\mu_{\ \lambda} \equiv \delta^\mu_{\ \lambda} + \frac{2P^\mu P_\lambda}{\mathcal{M}^2},
	\label{eq:C-matrix}
\end{equation}
which acts as a \emph{reflector} along the four-momentum direction ($C^\mu_{\ \lambda}P^\lambda = -P^\mu$) and has trace $\operatorname{Tr}(C)=4-2=2$, Eq.~\eqref{eq:spin-selfconsistent} can be cast into the compact form
\begin{equation}
	C^\mu_{\ \lambda} \frac{D S^{\lambda\nu}}{d\tau}
	= C^\nu_{\ \lambda} \frac{D S^{\lambda\mu}}{d\tau}.
	\label{eq:C-identity}
\end{equation}
Geometrically, Eq.~\eqref{eq:C-identity} states that the action of $C$ on the first index of $DS^{\mu\nu}/d\tau$ yields a tensor symmetric in $(\mu,\nu)$; since $DS^{\mu\nu}/d\tau$ is itself antisymmetric, the $C$ operator precisely cancels the part of $DS$ projected along $P^\mu$.

In this subsection, we have performed a key reformulation of the original MPD equations~\eqref{eq:mpd-momentum}--\eqref{eq:mpd-spin}, in which $\dot{x}^\mu$ enters both the momentum equation (via the $R\,\dot{x}\,S$ coupling) and the spin equation (as $P^\mu\dot{x}^\nu-P^\nu\dot{x}^\mu$). In the reformulated system~\eqref{eq:mpd-momentum2} and~\eqref{eq:C-identity}, the four-velocity $\dot{x}^\mu$ does \emph{not} appear explicitly. This property is essential for the subsequent nonperturbative analysis: once $DP_\mu/d\tau$ and $DS^{\mu\nu}/d\tau$ are determined, the worldline velocity can be recovered \emph{a~posteriori} from the transport relation~\eqref{eq:transport}.

\subsection{Spin four-vector formulation}

Under the TD-SSC condition~\eqref{eq:td-ssc}, one can introduce a spin four-vector
\begin{equation}
	s^{\mu}\equiv \frac{1}{2\mathcal{M}}\,\varepsilon^{\mu\nu\rho\sigma}P_{\nu}S_{\rho\sigma},
	\label{eq:spin-vector-def}
\end{equation}
where $\varepsilon^{\mu\nu\rho\sigma}=-(-g)^{-1/2}\,\tilde\varepsilon^{\mu\nu\rho\sigma}$ (with $\tilde\varepsilon^{0123}=+1$) is the Levi-Civita tensor for the $(-,+,+,+)$ signature, and its covariant form is $\varepsilon_{\mu\nu\rho\sigma}=(-g)^{1/2}\,\tilde\varepsilon_{\mu\nu\rho\sigma}$ (so that $\varepsilon_{0123}=\sqrt{-g}$). The complete antisymmetry of $\varepsilon^{\mu\nu\rho\sigma}$ immediately implies $s^{\mu}P_{\mu}=0$. 

The inverse relations are given by
\begin{equation}
	S_{\alpha\beta}=-\frac{1}{\mathcal{M}}\,\varepsilon_{\alpha\beta\mu\lambda}P^{\mu}s^{\lambda},
	\quad 
	S^{\mu\nu}=-\frac{1}{\mathcal{M}}\,\varepsilon^{\mu\nu\rho\sigma}P_{\rho}s_{\sigma},
	\label{eq:spin-tensor-inverse}
\end{equation}
and the equation of motion for $s^\mu$ reads
\begin{equation}
	\frac{Ds^{\mu}}{d\tau}
	= \frac{s^{\nu}}{\mathcal{M}^{2}}\frac{DP_{\nu}}{d\tau}\,P^{\mu}.
	\label{eq:spin-ev}
\end{equation}
Thus $Ds^{\mu}/d\tau$ is parallel to $P^{\mu}$ and is entirely controlled by the scalar contraction $s^{\nu}DP_{\nu}/d\tau$.

The exact relation between $\mathcal{M}$ and $m$ reads
\begin{align}
	\mathcal{M}^{2} = m^{2} + \dot x_{\mu} S^{\mu\nu}\frac{DP_{\nu}}{d\tau}.
	\label{eq:Mm-relation}
\end{align}
Substituting the transport relation for $\dot{x}_{\mu}$ and eliminating $S^{\mu\nu}$ in favour of $s^{\mu}$ via the identity $S^{\mu\rho}S_{\rho}^{\ \nu}=s^{2}(g^{\mu\nu}+P^{\mu}P^{\nu}/\mathcal{M}^{2})-s^{\mu}s^{\nu}$ (derived in Appendix~\ref{app:MPD-derivation}), one obtains the compact four-vector form
\begin{equation}
	\mathcal{M}^{2}=m^{2}
	+ \frac{1}{\mathcal{M}^{2}}\bigl(s^{\rho}s^{\nu} - s^{2}g^{\rho\nu}\bigr)
	  \frac{DP_{\rho}}{d\tau}\frac{DP_{\nu}}{d\tau},
	\label{eq:Mm-s}
\end{equation}
where $s^{2}=s^{\mu}s_{\mu}$ and the $P^{\mu}P^{\nu}$ term has been eliminated by $P^{\nu}DP_{\nu}/d\tau=0$.  
In the spinless limit $s\to0$ the correction vanishes and $m=\mathcal{M}$; 
more generally $\mathcal{M}^{2}-m^{2}$ is quadratic in the covariant force $DP_{\nu}/d\tau$ and hence of order $s^{4}$.
Since Mathisson \cite{mathisson1937} it is known that a spinning body admits no unique center
of mass, so that under the TD-SSC the representative four-velocity need not be parallel to the
four-momentum. For later use we record the quantitative form of this
non-alignment. With $n^{\mu}\equiv P^{\mu}/\mathcal M$ and
$w^{\mu}\equiv-\mathcal M^{-2}S^{\mu\nu}DP_\nu/d\tau$, the transport
relation~\eqref{eq:transport} reads
\begin{equation}
	\dot x^{\mu}=\gamma\,n^{\mu}+w^{\mu},
	\qquad
	\gamma\equiv-n_{\mu}\dot x^{\mu}=\frac{m}{\mathcal M},
	\qquad
	n_{\mu}w^{\mu}=0,
	\label{eq:xdot-decomp}
\end{equation}
where the orthogonality follows from $P_{\mu}S^{\mu\nu}=0$. Thus $\gamma$ is the Lorentz
factor between the worldline frame and the momentum frame, while $w^{\mu}$ is the residual
velocity of the representative point in the momentum rest frame. Since
$\dot x^{\mu}\dot x_{\mu}=-1$, these quantities satisfy the exact identity
\begin{equation}
	\gamma^{2}=1+w^{2}\equiv 1+w^{\mu}w_{\mu}\,,
	\label{eq:gamma-norm}
\end{equation}
so that $\gamma\ge1$, with equality if and only if $w^{\mu}=0$. Equation~\eqref{eq:gamma-norm}
is simply the geometric form of Eq.~\eqref{eq:Mm-s}. Because $S^{\mu\nu}$ is linear in $s$ and,
by Eq.~\eqref{eq:mpd-momentum}, the covariant force $DP_{\nu}/d\tau$ is itself of order $s$, 
$w^{\mu}=\mathcal O(s^{2})$ and consequently $\gamma-1=\mathcal O(s^{4})$.
That is, the non-alignment is second order in the spin as a velocity but
fourth order as a boost, which is the counting used in Sec.~\ref{sec:boost}.

The exact momentum formula in spin-tensor form is
\begin{equation}
	P^{\mu} = m\dot{x}^{\mu} + \frac{1}{m}
	\bigl( \dot x^{\mu}\dot x_\rho S^{\rho\nu} + S^{\mu\nu} \bigr) \frac{DP_\nu}{d\tau},
	\label{eq:momentum-exact}
\end{equation}
from which it follows that $\dot x_{\mu}(\dot x^{\mu}\dot x_\rho S^{\rho\nu}+S^{\mu\nu}) = 0$.
The orthogonality $s^{\mu}P_{\mu}=0$ is preserved by the MPD evolution:
\begin{align}
	\frac{D}{d\tau}(s^{\mu}P_{\mu})
	&= P_{\mu}\frac{Ds^{\mu}}{d\tau} + s_{\mu}\frac{DP^{\mu}}{d\tau} \nonumber \\
	&= \frac{1}{P^{2}}\bigl( P_{\mu}s^{\mu}P^{\nu} - P_{\mu}P^{\mu}s^{\nu} \bigr) \frac{DP_{\nu}}{d\tau}
	   + s_{\mu}\frac{DP^{\mu}}{d\tau} = 0\,.
	\label{eq:SSC-preserved}
\end{align}
Differentiating $s^{2}\equiv s_{\mu}s^{\mu}$ along the worldline and using Eq.~\eqref{eq:spin-ev} yields
\begin{equation}
	\frac{d(s^{2})}{d\tau}
	= 2s_{\mu}\frac{Ds^{\mu}}{d\tau}
	= 2\,s_{\mu}P^{\mu}\,\frac{s^{\nu}}{\mathcal{M}^{2}}\frac{DP_{\nu}}{d\tau}
	= 0,
	\label{eq:sconst}
\end{equation}
where the last equality follows from $s^{\mu}P_{\mu}=0$. Hence $s^{2}$ is an exact constant of the motion for the MPD system under the Tulczyjew--Dixon supplementary condition.
For more detailed calculations, see Appendix~\ref{app:MPD-derivation}.

The pole--dipole truncation used above presupposes that the multipole scale of the body is small
compared with the local curvature scale. For a spinning body the relevant scale is the M\o ller
radius \cite{moller1949,semerak1999}
\begin{equation}
	\rho_M\equiv\frac{|s|}{\mathcal M},
	\label{eq:rhom}
\end{equation}
which does not exceed the radius of the body. All results obtained below are exact within the
pole--dipole model; whenever a statement requires $\rho_M$ to be small compared with the
curvature scale we say so explicitly, and in the next section we shall express this
condition in closed form for Schwarzschild spacetime.

\section{The planar motion trajectory in Schwarzschild spacetime}
\label{sec:planarSchw}

\subsection{Equatorial spin-aligned motion}
\begin{figure}[htb]
\centering
\includegraphics[width=0.45\textwidth]{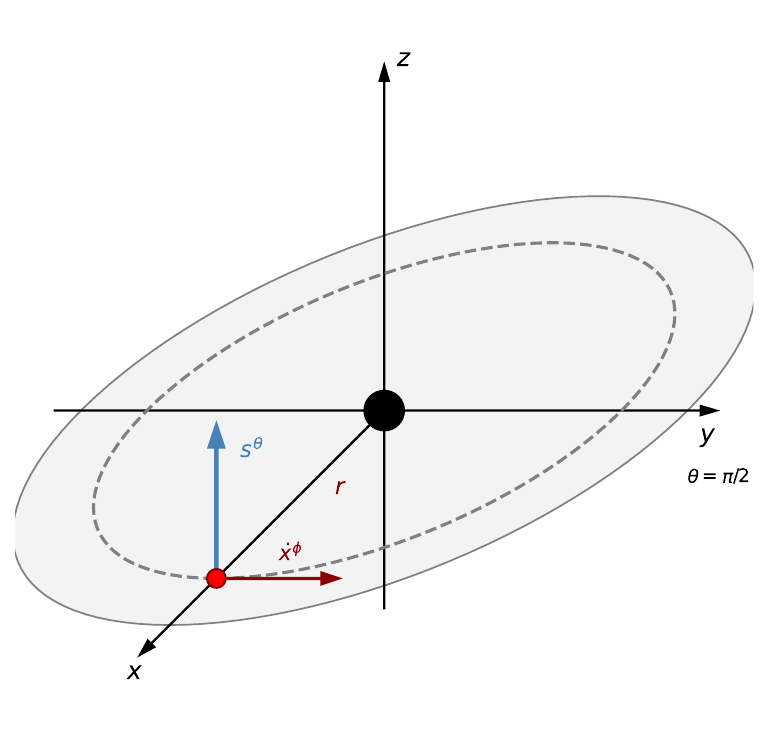}
\caption{The particle orbits in the equatorial plane $\theta=\pi/2$; the spin vector $s^{\theta}$ is orthogonal to the orbital plane.}
\label{fig:planar-setup}
\end{figure}

In what follows, we focus on the so-called equatorial spin-aligned motion, for which the orbit is confined to the equatorial plane $\theta=\pi/2$, with four-velocity $\dot{x}^{\mu}=(\dot{x}^{t},\dot{x}^{r},0,\dot{x}^{\phi})$ (i.e., $\dot{x}^{\theta}=0$), and the spin vector is orthogonal to the orbital plane, possessing only a nonvanishing polar component: 
\begin{equation}
	s^{\mu}=(0,0,s^{\theta},0),\qquad
	s_{\mu}=g_{\mu\nu}s^{\nu}=(0,0,s_{\theta},0),
\end{equation}
see Fig.~\ref{fig:planar-setup} for an illustration.
The TD-SSC~\eqref{eq:td-ssc} condition forces the polar momentum to vanish, $P^{\theta}=0$. Consequently, the independent momentum components are
\begin{equation}
	P_{\mu}=(P_{t},P_{r},0,P_{\phi}).
\end{equation}
Therefore, it is clear that only the $\sigma=\theta$ component survives in the contraction $\varepsilon^{\kappa\lambda\rho\sigma}P_{\rho}s_{\sigma}$. The remaining indices of $\varepsilon$ then lie entirely in the in-plane set $\{t,r,\phi\}$. Denoting these in-plane indices by $b,c,d,e$ and using Eq.~\eqref{eq:spin-tensor-inverse}, the momentum equation~\eqref{eq:mpd-momentum} reduces to
\begin{equation}
	\frac{DP^{\mu}}{d\tau}
	= \frac{s_{\theta}}{2\mathcal{M}}\,R^{\mu}{}_{bcd}\,\dot{x}^{b}\,\varepsilon^{cde\theta}P_{e}.
	\label{eq:mpd-equatorial}
\end{equation}
For the normal component $\mu=\theta$, one finds
\begin{equation}
	\frac{DP^{\theta}}{d\tau}=0,
	\label{eq:DPtheta-zero}
\end{equation}
since $R^{\theta}{}_{bcd}=0$ for $b,c,d\in\{t,r,\phi\}$. 
Hence, once the particle is initialized on the equatorial plane with $P^{\theta}=0$, no out-of-plane momentum develops during the subsequent evolution.

Combining Eqs.~\eqref{eq:spin-ev} and \eqref{eq:DPtheta-zero}, we obtain $Ds^{\mu}/d\tau=0$. Expanding the covariant derivative, we have
\begin{equation}
	\frac{Ds^{\mu}}{d\tau}
	= \frac{ds^{\mu}}{d\tau}
	+ \Gamma^{\mu}_{\nu\rho}\,\dot{x}^{\nu}s^{\rho}
	= \frac{ds^{\mu}}{d\tau}
	+ \Gamma^{\mu}_{\nu\theta}\,\dot{x}^{\nu}s^{\theta}.
\end{equation}
For $\mu\neq\theta$, the relevant connection coefficients vanish at $\theta=\pi/2$ (the only surviving term $\Gamma^{r}_{\theta\theta}$ is eliminated by $\dot{x}^{\theta}=0$), while $s^{t}=s^{r}=s^{\phi}=0$ identically; hence $Ds^{t}/d\tau=Ds^{r}/d\tau=Ds^{\phi}/d\tau=0$. For $\mu=\theta$,
\begin{equation}
	\frac{Ds^{\theta}}{d\tau}
	= \dot{s}^{\theta}
	+ \Gamma^{\theta}_{r\theta}\,\dot{x}^{r}s^{\theta}
	= \dot{s}^{\theta}
	+ \frac{1}{2g_{\theta\theta}}\,\partial_{r}g_{\theta\theta}\,\dot{x}^{r}s^{\theta},
	\label{eq:ds}
\end{equation}
where we used $\Gamma^{\theta}_{r\theta}=\tfrac{1}{2}g^{\theta\theta}\partial_{r}g_{\theta\theta}$, valid for any axisymmetric equatorial metric. The scalar magnitude \(s^{2}=s_{\mu}s^{\mu}\) is conserved, as shown in Sec.~\ref{sec:MPD}. On the equatorial plane, \(s^{2}=g_{\theta\theta}(s^{\theta})^{2}\), and therefore conservation of \(s^{2}\) implies
\begin{equation}
	\frac{1}{s^{\theta}}\frac{ds^{\theta}}{d\tau}
	= -\frac{\dot{x}^{r}}{2g_{\theta\theta}}\,\partial_{r}g_{\theta\theta},
	\label{eq:ds-conserved}
\end{equation}
which is equivalent to Eq.~\eqref{eq:ds} when \(Ds^{\theta}/d\tau=0\) is imposed. The above equation possesses a simple first integral. In Schwarzschild spacetime, $g_{\theta\theta}=r^{2}$ and $\partial_{r}g_{\theta\theta}=2r$, which gives
\begin{equation}
	\frac{d}{d\tau}\ln s^{\theta}
	= -\frac{\dot{x}^{r}}{r}
	= -\frac{d}{d\tau}\ln r,
\end{equation}
and hence integrates to $s^{\theta}=C/r$. The invariant spin magnitude $s^{2}=g_{\theta\theta}(s^{\theta})^{2}=r^{2}(s^{\theta})^{2}=C^{2}$ then fixes the integration constant to $C=\pm s$, yielding
\begin{equation}
	s^{\theta}=\pm\frac{s}{r}.
	\label{eq:stheta-integrated}
\end{equation}
Thus the coordinate component $s^{\theta}$ decreases as $1/r$ along the worldline, exactly compensating the growth of the metric factor $g_{\theta\theta}=r^{2}$ so that the invariant spin magnitude remains constant. This behavior is generic for any axisymmetric equatorial metric and not specific to Schwarzschild; the equatorial aligned-spin configuration defines an invariant submanifold of the full MPD dynamics.

\subsection{Equations of motion of the four-momentum in Schwarzschild spacetime}
We now specialize to Schwarzschild spacetime and obtain the exact four-momentum equations \eqref{eq:mpd-momentum2} for the equatorial spin-aligned motion, keeping all orders in the spin. 
In Schwarzschild coordinates the metric reads
$\mathrm ds^2=-f\,\mathrm dt^2+\mathrm dr^2/f+r^2\mathrm d\Omega^2$, with
$f\equiv1-2M/r$ and $r_s\equiv2M$ the Schwarzschild radius.
The detailed derivations are collected in Appendix~\ref{app:Schwarzschild}.
In this case, only three components  in Eq.~\eqref{eq:spin-tensor-inverse}  survive:
\begin{equation}
	S^{rt}=\frac{s}{\mathcal{M} r}P_{\phi},
	\qquad
	S^{r\phi}=-\frac{s}{\mathcal{M} r}P_{t},
	\qquad
	S^{t\phi}=\frac{s}{\mathcal{M} r}P_{r}\,.
	\label{eq:spin-tensor-equatorial}
\end{equation}
Hereafter $s\equiv\pm\sqrt{s^{2}}$ denotes the signed spin magnitude.

Upon substituting the Riemann tensor and the spin tensor~\eqref{eq:spin-tensor-equatorial} into Eq.~\eqref{eq:AB}, the only nonvanishing components are confined to the $(t,r,\phi)$ block. The $\theta$ row and column vanish identically, yielding $A^{\ \theta}_{\theta}=1$ and reducing the full determinant to that of the $3\times3$ block. Factoring out the common overall scale  
\begin{equation}
	\beta_r\equiv  \frac{ M s^{2} }{ \mathcal{M}^{4}r^{5} }\,,
\end{equation}
 the mixed matrix reads (with the lower index $\mu$ denoting the row and the upper index $\eta$ the column, in the ordering $t,r,\phi$):
\begin{equation}
	B^{\ \eta}_\mu=\beta_r\,
	\begin{pmatrix}
		2P_{\phi}^{2}-fr^{2}P_{r}^{2} & fr^{2}P_{r}P_{t} & -2P_{\phi}P_{t} \\[8pt]
		-\dfrac{P_{r}P_{t}\,r^{2}}{f} & 2P_{\phi}^{2}+\dfrac{P_{t}^{2}r^{2}}{f} & -2P_{\phi}P_{r} \\[8pt]
		-\dfrac{P_{\phi}P_{t}\,r^{2}}{f} & fr^{2}P_{\phi}P_{r} & \dfrac{P_{t}^{2}r^{2}}{f}-fr^{2}P_{r}^{2}
	\end{pmatrix},
	\label{eq:B-schw}
\end{equation}
and $B^{\ \theta}_\mu=B^{\ \mu}_\theta=0$.
Crucially, on this block $B^{\ \eta}_\mu$ has rank two and a degenerate (double) nonzero eigenvalue $\alpha_r$,
\begin{equation}
	\alpha_r=\beta_r\left(2P_{\phi}^{2}-fr^{2}P_{r}^{2}+\frac{P_{t}^{2}r^{2}}{f}\right)\,,
	\label{eq:alpha-schw}
\end{equation}
 so that
\begin{equation}
	B^{2}=\alpha_r\,B,\quad 
	b_{1}\equiv\operatorname{Tr}(B)=2\alpha_r,\quad 
  b_{2}\equiv\operatorname{Tr}(B^{2})=2\alpha_r^{2}.
\end{equation}
Consequently the general determinant \eqref{eq:detA-gen} collapses to a perfect square,
\begin{equation}
	\Delta_{A}=\det(A)=(1-\alpha_r)^{2}\,,
	\label{eq:deltaA-schw}
\end{equation}
and the general inverse \eqref{eq:Ainv-gen} simplifies dramatically to
\begin{equation}
	(A^{-1})^{\ \rho}_\mu=\delta^{\ \rho}_\mu+\frac{1}{1-\alpha_r}\,B^{\ \rho}_\mu\,.
	\label{eq:Ainv-schw}
\end{equation}
Substituting Eq.~\eqref{eq:Ainv-schw} into Eq.~\eqref{eq:mpd-momentum2}, we obtain the explicit form
\begin{equation}
	\frac{DP_\eta}{d\tau}
	=-\frac{m}{2\mathcal{M}^{2}}\Bigl(\delta_\eta^{\ \rho}+\frac{1}{1-\alpha_r}B^{\ \rho}_\eta\Bigr)V_{\rho\nu}P^\nu\,.
	\label{eq:mpd-momentum-schw}
\end{equation}
A direct evaluation of $V_{\rho\nu}=R_{\rho\nu\kappa\lambda}S^{\kappa\lambda}$ yields the nonvanishing contractions
\begin{eqnarray}
	V_t\equiv	V_{t\nu}P^{\nu}=\frac{6Ma fP_{r}P_{\phi}}{r^{3}},\quad
	V_r \equiv V_{r\nu}P^{\nu}=\frac{6Ma P_{t}P_{\phi}}{fr^{3}},\quad
	V_\phi \equiv V_{\phi\nu}P^{\nu}=0\,,
	\label{eq:V-components}
\end{eqnarray}
with the shorthand $a\equiv s/(\mathcal{M}r)$.
Inserting these together with the $B$ matrix~\eqref{eq:B-schw} into Eq.~\eqref{eq:mpd-momentum-schw}, we obtain the three nontrivial component equations:
\begin{subequations}\label{eq:mpd-components}
	\begin{align}
		\frac{DP_{t}}{d\tau}
		&=-\frac{m}{2\mathcal{M}^{2}}\Biggl[V_{t}+\frac{1}{1-\alpha_r}\bigl(B^{\ t}_t V_{t}+B^{\ r}_t V_{r}\bigr)\Biggr]
		=-\frac{m}{2\mathcal{M}^{2}(1-\alpha_r)}V_{t}\,,\label{eq:DPt-comp}\\[6pt]
		\frac{DP_{r}}{d\tau}
		&=-\frac{m}{2\mathcal{M}^{2}}\Biggl[V_{r}+\frac{1}{1-\alpha_r}\bigl(B^{\ t}_r V_{t}+B^{\ r}_r V_{r}\bigr)\Biggr]
		=-\frac{m}{2\mathcal{M}^{2}(1-\alpha_r)}V_{r}\,,\label{eq:DPr-comp}\\[6pt]
		\frac{DP_{\phi}}{d\tau}
		&=-\frac{m}{2\mathcal{M}^{2}}\Biggl[V_{\phi}+\frac{1}{1-\alpha_r}\bigl(B^{\ t}_\phi V_{t}+B^{\ r}_\phi V_{r}\bigr)\Biggr]
		=0\,.\label{eq:DPphi-comp}
	\end{align}
\end{subequations}
Substituting these expressions~\eqref{eq:DPt-comp}--\eqref{eq:DPphi-comp} into Eq.~\eqref{eq:xdot} and simplifying with the explicit forms of $V_{t},V_{r}$ from~\eqref{eq:V-components}, the nontrivial components of $\dot{x}^a$ reduce to the compact factorized forms
\begin{subequations}\label{eq:xdot-components}
	\begin{align}
		\dot{x}^t &= \frac{m}{\mathcal{M}^2}P^{t} + \frac{s}{\mathcal{M}^{3}r}\Bigl(P_{\phi}\frac{DP_{r}}{d\tau}-P_{r}\frac{DP_{\phi}}{d\tau}\Bigr)
		           = \frac{m}{\mathcal{M}^{2}}P^{t}\,\frac{1-\beta_{r}\mathcal{M}^{2}r^{2}}{1-\alpha_r}, \label{eq:xdot-t-components}\\[8pt]
		\dot{x}^r &= \frac{m}{\mathcal{M}^2}P^{r} + \frac{s}{\mathcal{M}^{3}r}\Bigl(P_{t}\frac{DP_{\phi}}{d\tau}-P_{\phi}\frac{DP_{t}}{d\tau}\Bigr)
		           = \frac{m}{\mathcal{M}^{2}}P^{r}\,\frac{1-\beta_{r}\mathcal{M}^{2}r^{2}}{1-\alpha_r}, \label{eq:xdot-r-components}\\[8pt]
		\dot{x}^\phi &= \frac{m}{\mathcal{M}^2}P^{\phi} + \frac{s}{\mathcal{M}^{3}r}\Bigl(P_{r}\frac{DP_{t}}{d\tau}-P_{t}\frac{DP_{r}}{d\tau}\Bigr)
		              = \frac{m}{\mathcal{M}^{2}}P^{\phi}\,\frac{1+2\beta_{r}\mathcal{M}^{2}r^{2}}{1-\alpha_r}.\label{eq:xdot-p-components}
	\end{align}
\end{subequations}
where we have used the mass-shell relation
\begin{equation}
	\mathcal{M}^{2} = -P^2 = -g^{\mu\nu} P_\mu P_\nu = \frac{P_{t}^{2}}{f} - fP_{r}^{2} - \frac{P_{\phi}^{2}}{r^{2}},
	\label{eq:mass-shell}
\end{equation}
together with $P_{t}=-fP^{t}$, $P^{r}=fP_{r}$, and $P^{\phi}=P_{\phi}/r^{2}$.
Taking the covariant derivative of the mass-shell relation~\eqref{eq:mass-shell} along the worldline and using metric compatibility $Dg^{\mu\nu}/d\tau=0$, we obtain
\begin{equation}
	\frac{D}{d\tau}(-P^{2})
	= -\frac{D}{d\tau}(g^{\mu\nu}P_{\mu}P_{\nu})
	= -2\,g^{\mu\nu}P_{\mu}\frac{DP_{\nu}}{d\tau}
	= -2\,P^{\nu}\frac{DP_{\nu}}{d\tau}=0,
	\label{eq:mass-shell-deriv}
\end{equation}
which yields the simple orthogonality condition
\begin{equation}
	P^{\nu}\frac{DP_{\nu}}{d\tau}=0,
	\qquad\text{or}\qquad
	P_{t}\frac{DP_{t}}{d\tau} =
	 f^2P_{r}\frac{DP_{r}}{d\tau} \,,
	\label{eq:P-DP-orth}
\end{equation}
where we have used Eq.~\eqref{eq:DPphi-comp}.
The relation~\eqref{eq:P-DP-orth} is satisfied identically once the explicit forms~\eqref{eq:V-components} are inserted, and we retain it below as a convenient bookkeeping identity.

Substituting $g^{tt}=-1/f$, $g^{rr}=f$, $g^{\phi\phi}=1/r^{2}$ and the covariant derivatives~\eqref{eq:DPt-comp}--\eqref{eq:DPphi-comp} into Eq.~\eqref{eq:Mm-s} gives
\begin{align}
	\mathcal{M}^{2}
	&= m^{2} + \frac{s^{2}}{\mathcal{M}^{2}}\Biggl[
	   \frac{1}{f}\Bigl(\frac{DP_{t}}{d\tau}\Bigr)^{2}
	   - f\Bigl(\frac{DP_{r}}{d\tau}\Bigr)^{2}
	   \Biggr] \nonumber \\
	&= m^{2}\Biggl[\,1 - \frac{9\beta_{r}^{2}\,P_{\phi}^{2}\,(P_{\phi}^{2}+\mathcal{M}^{2}r^{2})}{(1-\alpha_{r})^{2}}\,\Biggr] \nonumber \\
	&= m^{2}\,\frac{\big(1-\beta_r \mathcal{M}^2r^2\big)^{2} - 3\beta_{r}P_{\phi}^{2}(\beta_r\mathcal{M}^{2}r^{2}+2)}{(1-\alpha_{r})^{2}}\,,
	\label{eq:Mm-schw}
\end{align}
where we have used the mass-shell relation \eqref{eq:mass-shell}. 

\subsection{Conserved quantities from Killing symmetries}

Schwarzschild spacetime admits the time-translation Killing vector $\xi_{(t)}^{\mu}=(\partial_{t})^{\mu}$ and the axial Killing vector $\xi_{(\phi)}^{\mu}=(\partial_{\phi})^{\mu}$. For any Killing vector $\xi^{\mu}$, the MPD system possesses the conserved charge\cite{dixon1970}
\begin{equation}
	P_{\mu}\xi^{\mu}+\frac{1}{2}S^{\mu\nu}\nabla_{\mu}\xi_{\nu}.
	\label{eq:Killing-charge}
\end{equation}
For $\xi_{(t)}^{\mu}$ we have $\xi^{(t)}_{\nu}=g_{\nu t}=(-f,0,0,0)$, so $(r,t)$ is the only contributing index pair:
\begin{align}
	\nabla_{r}\xi^{(t)}_{t}
	= \partial_{r}g_{tt}-\Gamma^{t}_{rt}\xi^{(t)}_{t}
	= \partial_{r}(-f)-\Gamma^{t}_{rt}(-f)
	= -\frac{2M}{r^{2}}+\frac{M}{r^{2}}
	= -\frac{M}{r^{2}},
\end{align}
and $\nabla_{t}\xi^{(t)}_{r}=-\nabla_{r}\xi^{(t)}_{t}=M/r^{2}$ by the Killing equation. 
Using the equatorial spin-tensor components~\eqref{eq:spin-tensor-equatorial}, we therefore have the conserved charge
\begin{align}
	E
	&= -P_{t}-\frac{1}{2}\bigl(S^{rt}\nabla_{r}\xi^{(t)}_{t}+S^{tr}\nabla_{t}\xi^{(t)}_{r}\bigr)
	   = -P_{t}+\frac{M}{r^{2}}S^{rt}
	   = -P_{t}+\frac{Ms}{\mathcal{M}r^{3}}P_{\phi}.
	\label{eq:qt}
\end{align}

For $\xi_{(\phi)}^{\mu}$ we have $\xi^{(\phi)}_{\nu}=g_{\nu\phi}=(0,0,0,r^{2})$. Computing
\begin{align}
	\nabla_{r}\xi^{(\phi)}_{\phi}
	= \partial_{r}(r^{2})-\Gamma^{\phi}_{r\phi}r^{2}
	= 2r-r = r,
\end{align}
with $\nabla_{\phi}\xi^{(\phi)}_{r}=-r$. The conserved charge reads
\begin{align}
	J
	&= P_{\phi}+\frac{1}{2}\bigl(S^{r\phi}\nabla_{r}\xi^{(\phi)}_{\phi}+S^{\phi r}\nabla_{\phi}\xi^{(\phi)}_{r}\bigr)
	   = P_{\phi}+rS^{r\phi}
	   = P_{\phi}-\frac{s}{\mathcal{M}}P_{t}.
	\label{eq:qphi}
\end{align}

Together with $s^{2}$ and $\mathcal{M}^{2}$, these two Killing integrals render the equatorial spin-aligned motion completely integrable. 
Eqs.~\eqref{eq:qt} and \eqref{eq:qphi} can be written in matrix form
\begin{equation}
	\begin{pmatrix} E \\ J \end{pmatrix}
	=
	\begin{pmatrix}
		-1 & \dfrac{Ms}{\mathcal{M}r^{3}} \\[8pt]
		-\dfrac{s}{\mathcal{M}} & 1
	\end{pmatrix}
	\begin{pmatrix} P_{t} \\ P_{\phi} \end{pmatrix}.
\end{equation}
The coefficient matrix is invertible provided
\begin{equation}
	F(r)\equiv 1-\frac{Ms^{2}}{\mathcal{M}^{2}r^{3}}\neq0 ,
	\label{eq:invertibility}
\end{equation}
which vanishes at the radius
\begin{equation}
	r_{*}=\Bigl(\frac{Ms^{2}}{\mathcal{M}^{2}}\Bigr)^{1/3},
	\label{eq:rstar}
\end{equation}
where the coefficient matrix drops to unit rank and the map $(P_{t},P_{\phi})\mapsto(E,J)$ becomes singular. 
A radius of the same form has been identified in Ref.~\cite{Armaza:2015eha}, 
where the question of whether an orbit can reach such a radius when it lies outside the horizon was also discussed. 
For ease of reference we shall call $r_{*}$ the \emph{degeneracy radius}.
For later reference we record that $r_{*}^{3}=M\rho_M^{2}$, with $\rho_M$ the M\o ller
radius~\eqref{eq:rhom}. That is, the degeneracy radius is the radius at which $\rho_M$ equals the
curvature scale $\ell_c\equiv(r^{3}/M)^{1/2}$ of Schwarzschild spacetime, so that the
pole--dipole expansion reaches its limit there. It is convenient to introduce the ratio
\begin{equation}
	\eta\equiv\frac{\rho_M}{\ell_c}
	=\frac{|s|}{\mathcal M}\Bigl(\frac{M}{r^{3}}\Bigr)^{1/2},
	\label{eq:eta}
\end{equation}
which serves below to delimit the range of validity of the results.

Solving for $P_{\phi}$,
\begin{equation}
	J=P_{\phi}-\frac{s}{\mathcal{M}}\Bigl(-E+\frac{Ms}{\mathcal{M}r^{3}}P_{\phi}\Bigr)
	  =FP_{\phi}+\frac{s}{\mathcal{M}}E,
	\label{eq:JE}
\end{equation}
hence
\begin{equation}
	P_{\phi}=\frac{1}{F}\Bigl(J-\frac{s}{\mathcal{M}}E\Bigr),
	\label{eq:Pphi-inv}
\end{equation}
and substituting back,
\begin{equation}
	P_{t}=-\frac{1}{F}\Bigl(E-\frac{Ms}{\mathcal{M}r^{3}}J\Bigr).
	\label{eq:Pt-inv}
\end{equation}
The inversion is valid for $r\neq r_{*}$; at $r=r_{*}$ only the single relation $J=sE/\mathcal{M}$ survives (cf.\ Eq.~\eqref{eq:JE}). 
Substituting Eqs.~\eqref{eq:Pphi-inv} and \eqref{eq:Pt-inv} into the mass-shell condition, we obtain:
\begin{equation}
	f P_{r}^{2} = \frac{N(r)}{F^{2}} - \mathcal{M}^{2},
	\qquad
	N(r) \equiv \frac{1}{f}\Bigl(E-\frac{Ms}{\mathcal{M}r^{3}}J\Bigr)^{2}
	          - \frac{1}{r^{2}}\Bigl(J-\frac{s}{\mathcal{M}}E\Bigr)^{2}.
	\label{eq:shell-barrier}
\end{equation}

Applying $DP_{\mu}/d\tau = dP_{\mu}/d\tau - \Gamma^{\lambda}_{\mu\nu}P_{\lambda}\dot{x}^{\nu}$ to the inversions~\eqref{eq:Pphi-inv}--\eqref{eq:Pt-inv} with the Schwarzschild Christoffel symbols, the ordinary derivatives of $E$ and $J$ vanish by conservation.  For $DP_{t}$ the Christoffel terms cancel pairwise, recovering~\eqref{eq:DPt-comp} upon substituting the transport relation; for $DP_{\phi}$ they combine to zero, consistent with~\eqref{eq:DPphi-comp}.

\subsection{Motion on the planar orbit}
With the momenta $P_{a}$ now fully determined by the conserved quantities $\{E,J\}$, the transport relation~\eqref{eq:xdot-components} yields the three velocity components in closed algebraic form. Substituting $P^{t}=-P_{t}/f$, $P^{r}=fP_{r}$, $P^{\phi}=P_{\phi}/r^{2}$, together with the inversions~\eqref{eq:Pphi-inv} and \eqref{eq:Pt-inv}, gives the explicit result.
\begin{align}
	\dot{x}^{t}
	&= \frac{m}{\mathcal{M}f}
	\Bigl(\mathcal{E}-\frac{\mathcal{SJ}r_s^3}{2r^{3}}\Bigr)\,
	\frac{1}{1-\alpha_{r}}, \label{eq:xdot-t-schw} \\[6pt]
	\dot{x}^{r} 
	&= \frac{m}{\mathcal{M}}
	\left( \frac{fP_{r}F}{\mathcal{M}}\right) \frac{1}{1-\alpha_{r}}, \label{eq:xdot-r-schw} \\[6pt]
	\dot{x}^{\phi} 
	&= \frac{mr_sL}{\mathcal{M}r^{2}F}
	\left(1+\frac{\mathcal{S}^2r_s^3}{r^3}\right)
	\frac{1}{1-\alpha_{r}}\,,\label{eq:xdot-phi-schw}
\end{align}
where we have used the relation $1-F =\beta_{r}\mathcal{M}^{2}r^{2} =\mathcal{S}^2r_s^3/(2r^3)$,
and defined the following dimensionless quantities
\begin{align}
	\mathcal{E} \equiv \frac{E}{\mathcal{M}}\,,\quad 
	\mathcal{J} \equiv \frac{J}{\mathcal{M}r_s}\,,\quad
	\mathcal{S} \equiv \frac{s}{\mathcal{M}r_s}\,,\quad 
	L\equiv \mathcal{J}-\mathcal{SE} \,.
	\label{eq:dimless}
\end{align}
In these variables the ratio~\eqref{eq:eta} reads
$\eta=(|\mathcal S|/\sqrt2)(r_s/r)^{3/2}$, so that
$F=1-\eta^{2}$.

Eq.~\eqref{eq:dimless} gives
\begin{align}
	\frac{f^2 P_{r}^{2}}{\mathcal{M}^{2}} F^{2} &= \Bigl(\mathcal{E}-\frac{\mathcal{SJ}r_s^3}{2r^{3}}\Bigr)^{2}
- fL^2\frac{r_s^2}{r^{2}} - fF^{2}\,, \label{eq:N-schw}
\end{align}
from Eq.~\eqref{eq:shell-barrier}.
Eliminating $fP_{r}$ by means of Eq.~\eqref{eq:N-schw} and using
Eq.~\eqref{eq:Mm-schw} to replace $m^{2}/\mathcal M^{2}$ by $(1-\alpha_{r})^{2}F^{2}/Q$, the
factor $(1-\alpha_{r})^{2}$ cancels and one obtains
\begin{equation}
	\Bigl(\frac{\mathrm dr}{\mathrm d\tau}\Bigr)^{2}
	=\frac{m^{2}}{\mathcal M^{2}}\,
	\frac{\bigl(\mathcal E-\frac{\mathcal S\mathcal J}{2}u^{3}\bigr)^{2}
	-fL^{2}u^{2}-fF^{2}}{(1-\alpha_{r})^{2}}
	=\frac{U(u)}{Q(u)}\,,
	\label{eq:drdtau-schw}
\end{equation}
where $u\equiv r_s/r$ and the entire spin dependence has been collected in two functions of $u$,
\begin{align}
	U(u)&\equiv F^{2}(u)\biggl[\mathcal E^{2}-1+u-L^{2}u^{2}
	     +(L^{2}+\mathcal S^{2}-\mathcal S\mathcal J\mathcal E)u^{3}
	     -\mathcal S^{2}u^{4}\nonumber\\
	     &\qquad\qquad\qquad
	     +\frac{\mathcal S^{2}(\mathcal J^{2}-\mathcal S^{2})}{4}u^{6}
	     +\frac{\mathcal S^{4}}{4}u^{7}\biggr],
	\label{eq:U-poly}\\[4pt]
	Q(u)&\equiv F^{4}(u)-3\,(1-F)(3-F)L^{2}u^{2}\nonumber\\
	     &=1-2\mathcal S^{2}u^{3}-3\mathcal S^{2}L^{2}u^{5}
	     +\frac{3\mathcal S^{4}}{2}u^{6}
	     -\frac{3\mathcal S^{4}L^{2}}{4}u^{8}
	     -\frac{\mathcal S^{6}}{2}u^{9}
	     +\frac{\mathcal S^{8}}{16}u^{12}.
	\label{eq:Q-poly}
\end{align}
Here we have used
$1-\alpha_{r}=F-3(1-F)L^{2}u^{2}/F^{2}$, which follows from
$\alpha_{r}=\beta_{r}(3P_{\phi}^{2}+\mathcal M^{2}r^{2})$ and $\beta_{r}\mathcal M^{2}r^{2}=1-F$.
Note that $Q$ involves the conserved charges only through $L^{2}$, whereas $U$ involves
$\mathcal J$ and $\mathcal E$ separately.

The same function $U$ governs the angular form of the radial equation. Dividing $\dot x^{r}$
by $\dot x^{\phi}$, squaring, and eliminating $fP_{r}$ once more with Eq.~\eqref{eq:N-schw} gives
\begin{equation}
	\Bigl(\frac{\mathrm dr}{\mathrm d\phi}\Bigr)^{2}
	=\frac{r_s^{2}F^{2}}{u^{4}L^{2}
	\bigl(1+\mathcal S^{2}u^{3}\bigr)^{2}}
	\biggl[\Bigl(\mathcal E-\frac{\mathcal S\mathcal J}{2}u^{3}\Bigr)^{2}
	-fL^{2}u^{2}-fF^{2}\biggr],
\end{equation}
so that, for $L\neq0$,
\begin{equation}
	\Bigl(\frac{\mathrm du}{\mathrm d\phi}\Bigr)^{2}
	=\frac{U(u)}{L^{2}\bigl(1+\mathcal S^{2}u^{3}\bigr)^{2}}.
	\label{eq:dudphi-schw}
\end{equation}
The two forms~\eqref{eq:drdtau-schw} and~\eqref{eq:dudphi-schw} thus
share the same numerator and differ only in the denominators $Q(u)$ and
$L^{2}(1+\mathcal S^{2}u^{3})^{2}$; the angular form degenerates at $L=0$, the proper-time form does not.

The values of $U(u)$ at spatial infinity ($u\to 0$, $r\to\infty$) and at the horizon ($u=1$, i.e.\ $r=r_s$) are given by
\begin{align}
	U(0) = \mathcal{E}^{2}-1\,,\quad 
	U(1) = \left(\mathcal{E}-\frac{\mathcal{J}\mathcal{S} }{2} \right)^2
	\left( 1-\frac{\mathcal{S}^{2}}{2}\right)^2 \geq 0 \,.
\end{align}

The polynomial~\eqref{eq:U-poly} factorizes naturally into a
spin-dependent prefactor and a seven-parameter bracket $U(u)=F^2(u)\,U_{7}(u)$ with
\begin{align}
	U_{7}(u)&\equiv\mathcal{E}^{2}-1+u-L^{2}u^{2}
	          + (L^{2}+\mathcal{S}^{2}-\mathcal{S}\mathcal{J}\mathcal{E})u^{3}
	          -\mathcal{S}^{2}u^{4}
	          +\frac{\mathcal{S}^{2}(\mathcal{J}^{2}-\mathcal{S}^{2})}{4}u^{6}
	          +\frac{\mathcal{S}^{4}}{4}u^{7}.
	\label{eq:U-factor}
\end{align}
The factor $F(u)$ is independent of the dynamical parameters $\mathcal E,\mathcal J,L$, so that
$u_{*}$ is a double root of $U$ for every $(\mathcal E,\mathcal J)$; $F$ contributes six roots
in all, including this double real root $u_{*}=\sqrt[3]{2/\mathcal S^{2}}=r_s/r_{*}$. These six roots sum
to zero and hence do not affect the total root sum of $U(u)$, which considerably simplifies the
analysis that follows. The radius $r_{*}$ is the degeneracy radius of Eq.~\eqref{eq:rstar}, and
the role it plays in the exact dynamics is discussed in Sec.~\ref{sec:wall}.

The physically relevant roots are the seven zeros of
$U_{7}(u)=0$.  Writing this equation in monic form
$u^{7}+a_{6}u^{6}+\cdots+a_{0}=0$ (divide the bracket by
$\mathcal{S}^{4}/4$) yields the coefficients
\begin{equation}
	\begin{aligned}
	a_{6}&=\frac{\mathcal{J}^{2}}{\mathcal{S}^{2}}-1,\qquad
	a_{5}=0,\qquad
	a_{4}=-\frac{4}{\mathcal{S}^{2}},\\
	a_{3}&=\frac{4(L^{2}+\mathcal{S}^{2}-\mathcal{S}\mathcal{J}\mathcal{E})}{\mathcal{S}^{4}},\qquad
	a_{2}=-\frac{4L^{2}}{\mathcal{S}^{4}},\qquad
	a_{1}=\frac{4}{\mathcal{S}^{4}},\qquad
	a_{0}=\frac{4(\mathcal{E}^{2}-1)}{\mathcal{S}^{4}}.
	\end{aligned}
	\label{eq:monic-coeffs}
\end{equation}
A striking feature is the absence of the $u^{5}$ term
($a_{5}=0$).  The consequences are most clearly seen through Vieta's
formulas.  Denoting the seven roots by $\{v_{1},\dots,v_{7}\}$, one
finds
\begin{equation}
	\begin{aligned}
	\sum_{i=1}^{7} v_{i}&=1-\frac{\mathcal{J}^{2}}{\mathcal{S}^{2}},\qquad
	\sum_{i<j}v_{i}v_{j}=0,\\[4pt]
	\sum_{i<j<k}v_{i}v_{j}v_{k}&=\frac{4}{\mathcal{S}^{2}},\qquad
	\sum_{i<j<k<l}v_{i}v_{j}v_{k}v_{l}
	   =\frac{4(L^{2}+\mathcal{S}^{2}-\mathcal{S}\mathcal{J}\mathcal{E})}{\mathcal{S}^{4}},\\[4pt]
	\sum v_{i}\cdots v_{5}&=\frac{4L^{2}}{\mathcal{S}^{4}},\qquad
	\sum v_{i}\cdots v_{6}=\frac{4}{\mathcal{S}^{4}},\\[4pt]
	\prod_{i=1}^{7} v_{i}&=\frac{4(1-\mathcal{E}^{2})}{\mathcal{S}^{4}}.
	\end{aligned}
	\label{eq:vieta}
\end{equation}
These relations imply three nontrivial constraints on the root distribution. 
First, since $\sum_{i<j}v_{i}v_{j}=0$, the identity $\sum v_{i}^{2}=(\sum v_{i})^{2}$ holds exactly, 
which yields the sharp bound $|v_{i}|\le|\mathcal{S}^{2}-\mathcal{J}^{2}|/\mathcal{S}^{2}$ for every real root. 
Second, a vanishing sum of pairwise products is impossible if all seven roots are real and of the same sign; 
hence $U_{7}(u)$ must have at least one negative real root or at least one pair of complex-conjugate roots. 
Finally, $\prod v_{i}=4(1-\mathcal{E}^{2})/\mathcal{S}^{4}$ shows that 
the product is negative for $\mathcal{E}>1$ (corresponding to an odd number of negative real roots) 
and positive for $\mathcal{E}<1$ (corresponding to an even number).

Equations~\eqref{eq:dudphi-schw} and~\eqref{eq:drdtau-schw}, together with the closed-form
velocities~\eqref{eq:xdot-t-schw}--\eqref{eq:xdot-phi-schw} and the constants
$E,J,s^{2},\mathcal M^{2}$, reduce the exact planar dynamics to two equivalent first-order
forms; the root structure of $U$ recorded above then classifies the possible orbits at every
order in the spin.

\section{The spin wall}

	\label{sec:spinwall}
\subsection{The relative boost of the representative point}
\label{sec:boost}

We now evaluate the non-alignment~\eqref{eq:gamma-norm} for the equatorial motion of Sec.~\ref{sec:planarSchw}. 
From Eq.~\eqref{eq:Mm-schw}, defining
\begin{equation}
	\mathcal K\equiv\frac{9\beta_r^{2}P_{\phi}^{2}\bigl(P_{\phi}^{2}+\mathcal M^{2}r^{2}\bigr)}
	{(1-\alpha_r)^{2}},
	\label{eq:K-def}
\end{equation}
we have the exact relation between the two kinematical scalars
$\mathcal M^{2}=m^{2}(1-\mathcal K)$, i.e.
\begin{equation}
	\gamma=\frac{m}{\mathcal M}=\frac{1}{\sqrt{1-\mathcal K}}\,.
	\label{eq:gamma-K}
\end{equation}
Expressing $\mathcal K$ in the dimensionless variables~\eqref{eq:dimless} by means of
$P_{\phi}^{2}=\mathcal M^{2}r_s^{2}L^{2}/F^{2}$, the mass-shell identity
$\alpha_r=\beta_r(3P_{\phi}^{2}+\mathcal M^{2}r^{2})$ and $\beta_r\mathcal M^{2}r^{2}=1-F$,
Eq.~\eqref{eq:K-def} becomes
\begin{equation}
	\mathcal K
	=\frac{9\,(1-F)^{2}\,L^{2}u^{2}\bigl(L^{2}u^{2}+F^{2}\bigr)}
	{\bigl[F^{3}-3(1-F)L^{2}u^{2}\bigr]^{2}} ,
	\label{eq:K-schw}
\end{equation}
with $F$ the factor of Eq.~\eqref{eq:invertibility}. Two features of~\eqref{eq:K-schw} are worth recording.
First, $\mathcal K\ge0$ identically; second, the boost depends on the two
conserved charges $\mathcal J$ and $\mathcal E$ only through the single combination $L=\mathcal J-\mathcal S\mathcal E$.

Equation~\eqref{eq:K-schw} turns the geometric discussion of Sec.~\ref{sec:MPD} into a transparent statement. 
By Eq.~\eqref{eq:gamma-norm}, $|w|^{2}=\gamma^{2}-1=\mathcal K/(1-\mathcal K)$, 
so $\sqrt{\mathcal K}$ is the speed with which the representative point drifts through its own momentum rest frame. 
A real Lorentz factor requires $\mathcal K<1$, which by Eq.~\eqref{eq:Mm-schw} is equivalent to $Q(u)>0$; 
on the surface $\mathcal K=1$ the drift speed reaches that of light and the transport relation~\eqref{eq:transport} admits no timelike solution.
The same transition was identified numerically in Ref.~\cite{Armaza:2015eha} through the vanishing of $u_\mu u^\mu$ at a finite radius, 
and Eq.~\eqref{eq:gamma-K} gives its invariant form.
Since $\mathcal K\to0$ both as $u\to0$ (spatial infinity) and as $u\to\infty$, 
while $\mathcal K\ge1$ throughout the range where $Q<0$,
the superluminal range of radii is a bounded interval whose endpoints are zeros of $Q$.
On the outer admissible branch the drift speed therefore takes values between zero and unity and
tends to unity as the branch end is approached; 
the shape of the interval, and the fact that it contains $r_{*}$, are determined in Sec.~\ref{sec:boundary}.
We shall refer to the surface $\mathcal K=1$, equivalently $Q=0$, as the \emph{spin wall}. 
It is a kinematic boundary, the locus on which no timelike representative worldline exists.

Two limiting cases are instructive. For $P_{\phi}=0$, i.e.\ $L=0$, the numerator
of~\eqref{eq:K-schw} vanishes and $\mathcal K\equiv0$, so that $\gamma\equiv1$ and 
the four-velocity and the four-momentum remain aligned. 
In this case the conserved angular momentum is carried entirely by the spin, 
for $J=P_{\phi}-(s/\mathcal M)P_{t}$ reduces to $-(s/\mathcal M)P_{t}=E s/\mathcal M$, 
so that $\mathcal J=\mathcal S\mathcal E$ while the momentum itself is purely radial. 
Consistently, Eqs.~\eqref{eq:V-components} and~\eqref{eq:mpd-momentum-schw} show that the covariant force vanishes identically for $P_{\phi}=0$.
The momentum is parallel transported, the worldline obeys the geodesic equation, 
and Eq.~\eqref{eq:drdtau-schw} collapses to $(\mathrm dr/\mathrm d\tau)^{2}=\mathcal E^{2}-1+r_s/r$. 
For small spin, by contrast, taking $\mathcal S\to0$ at fixed $L$ so that $1-F=\mathcal S^{2}u^{3}/2$, one finds
\begin{equation}
	\mathcal K=\frac94\,\mathcal S^{4}L^{2}u^{8}\bigl(1+L^{2}u^{2}\bigr)+\mathcal O(\mathcal S^{6}),
	\qquad
	\gamma-1=\tfrac12\mathcal K+\mathcal O(\mathcal S^{6}),
	\label{eq:K-smallspin}
\end{equation}
in agreement with the order counting following Eq.~\eqref{eq:gamma-norm}:
the boost contains no term of order $s^{2}$ or $s^{3}$ and appears only quartically.

To make the magnitude explicit, for $(\mathcal S,L)=(1,1)$ (hence $\mathcal J=1+\mathcal E$)
Eq.~\eqref{eq:K-schw} gives $\sqrt{\mathcal K}=0.013$ at $u=0.3$, $0.35$ at $u=0.6$ and $0.91$
at $u=0.68$; the drift speed reaches unity at $u\simeq0.686$, i.e.\ at $r\simeq1.46\,r_s$,
well outside the horizon.

\subsection{The wall as a surface in parameter space}
\label{sec:boundary}

The wall is the zero set of the function introduced in Eq.~\eqref{eq:Q-poly},
\begin{equation}
	Q(u;\mathcal S,L)=F^{4}-3\,(1-F)(3-F)\,L^{2}u^{2}=0,
	\label{eq:Qcompact}
\end{equation}
with $F$ the factor of Eq.~\eqref{eq:invertibility}, which is of degree twelve in $u$ but only linear in $L^{2}$, and even in $L$. Hence
$\Sigma\equiv\{Q=0\}$ is a real-analytic surface in the $(u,\mathcal S,L)$ space, and a graph
over the $(u,\mathcal S)$ plane. Its orientation is fixed by
\begin{equation}
	\frac{\partial Q}{\partial (L^{2})}=-3\,(1-F)(3-F)\,u^{2}<0 ,
	\label{eq:dQdL2}
\end{equation}
the sign following from $1-F=(\mathcal S^{2}/2)u^{3}>0$ and $3-F=2+(\mathcal S^{2}/2)u^{3}>0$:
at a fixed radius and a fixed spin a larger $|L|$ lowers $Q$, so that the wall moves outward as
$|L|$ grows. The surface does not reach spatial infinity, since $Q(0)=1$ identically.

For fixed $(\mathcal S,L\ne0)$ the wall consists of exactly two points. Indeed $Q(0)=1$,
$Q(u_*)=-9L^{2}u_*^{2}<0$ and $Q>0$ as $u\to\infty$, so at least one zero exists on either side of the degeneracy radius $u_*$; 
and since Eq.~\eqref{eq:Qcompact} is equivalent to $L^{2}=\Lambda(u;\mathcal S)$ with
\begin{equation}
	\Lambda(u;\mathcal S)\equiv\frac{F^{4}}{3\,(1-F)(3-F)\,u^{2}} ,
	\label{eq:Lambda}
\end{equation}
which can be shown to be strictly decreasing on $(0,u_*)$ and strictly increasing on $(u_*,\infty)$,
there is exactly one zero on each side. Denoting by $u_Q$ the zero in $(0,u_*)$ and by $u_R$ the zero in $(u_*,\infty)$, and writing
$r_Q\equiv r_s/u_Q$, $r_R\equiv r_s/u_R$ for the corresponding radii, we have
\begin{equation}
	0<u_Q(\mathcal S,L)<u_*<u_R(\mathcal S,L),
	\qquad
	r_Q\equiv\frac{r_s}{u_Q}>r_*>\frac{r_s}{u_R}\equiv r_R ,
	\label{eq:uQuR}
\end{equation}
so that a given orbit encounters the wall at $r_Q$, while $r_R$ bounds the second admissible branch $r<r_R$. 
As functions of $(\mathcal S,L^{2})$ the two branches are analytic wherever $\partial_uQ\ne0$; 
the only exception is $L=0$, where the two sheets meet at $u_*$ in a
square-root fold, whose explicit form is recorded in Sec.~\ref{sec:wall}. 
For $L=0$ the wall degenerates altogether: $Q=F^{4}$ is positive for every $u\ne u_*$.
\begin{figure}[htb]
\centering
\includegraphics[width=0.7\textwidth]{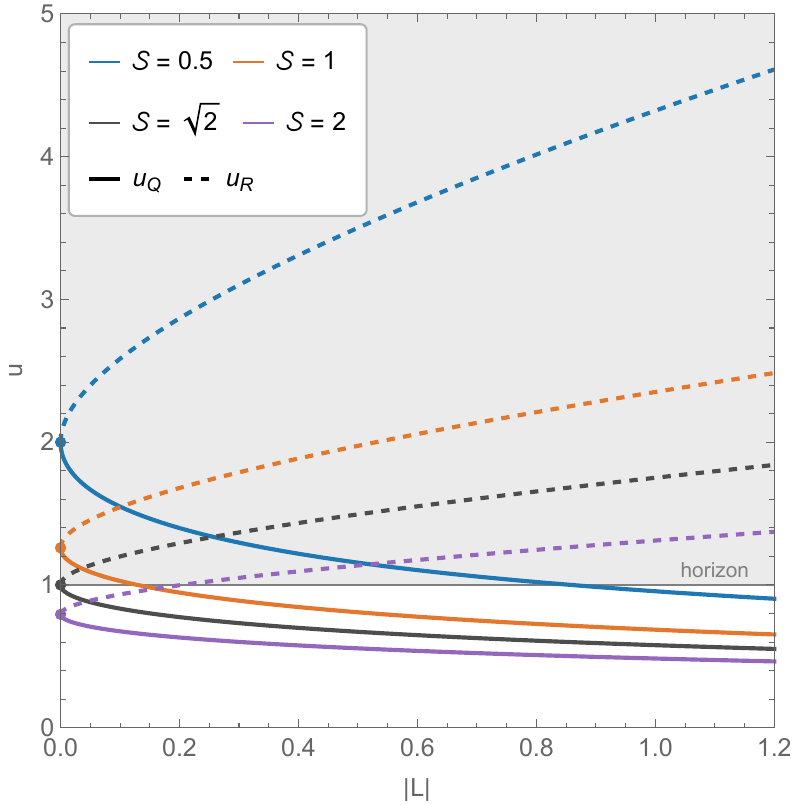}
\caption{Sections of the wall surface at fixed spin: the outer branch $u_Q$ (solid) and
	the inner branch $u_R$ (dashed) as functions of $|L|$, for $\mathcal S=0.5$ (blue), $1$
	(orange), $\sqrt2$ (black) and $2$ (purple). Dots mark the merging point $u_*$, the grey
	line is the horizon $u=1$, the shaded region is its interior, and $u$ increases inward.}
\label{fig:wall-branches}
\end{figure}
Figure~\ref{fig:wall-branches} displays the two branches for several spins. The outer branch
moves outward monotonically as $|L|$ grows, as required by Eq.~\eqref{eq:dQdL2}, while the inner
branch moves inward, so that the interval occupied by the wall widens with $|L|$.

The question of physical interest is where the wall lies relative to the horizon. Setting $u=1$
in Eq.~\eqref{eq:Qcompact} gives the explicit curve
\begin{equation}
	\Lambda(1;\mathcal S)
	=\frac{\left(1-\mathcal S^{2}/2\right)^{4}}
	{3\,(\mathcal S^{2}/2)\left(2+\mathcal S^{2}/2\right)} ,
	\label{eq:Lambda1}
\end{equation}
which, together with the line $\mathcal S=\sqrt2$, divides the parameter space into three regions. 
\begin{itemize}
	\item \textbf{(a)} For $\mathcal S<\sqrt2$ and $L^{2}<\Lambda(1;\mathcal S)$ one has $u_Q>1$: the wall lies entirely inside the horizon and the exterior motion is unaffected.
	\item \textbf{(b)} For $\mathcal S<\sqrt2$ and $L^{2}>\Lambda(1;\mathcal S)$ one has $u_Q<1<u_*<u_R$: the wall reaches the exterior, and since the horizon itself lies inside the interval $(u_Q,u_R)$ that separates the two admissible branches, no exterior orbit can reach it.
	\item \textbf{(c)} For $\mathcal S>\sqrt2$ one has $u_Q<u_*<1$ for every $L\ne0$, so the wall always extends into the exterior, and Eq.~\eqref{eq:Lambda1} instead separates the configurations whose horizon lies inside the forbidden interval from those whose horizon lies on the inner admissible branch.
\end{itemize}
Note that $\mathcal S>\sqrt2$ is also the condition for the degeneracy radius $r_{*}$ to lie outside the horizon.
It is sufficient but not necessary for the wall to affect the exterior; 
for $\mathcal S<\sqrt2$ a sufficiently large $|L|$ has the same effect. 
For $\mathcal S=1$, for
instance, the threshold is $|L|=\sqrt{\Lambda(1;1)}=0.129$, although $r_*>r_s$.
\begin{figure}[htb]
\centering
\includegraphics[width=0.7\textwidth]{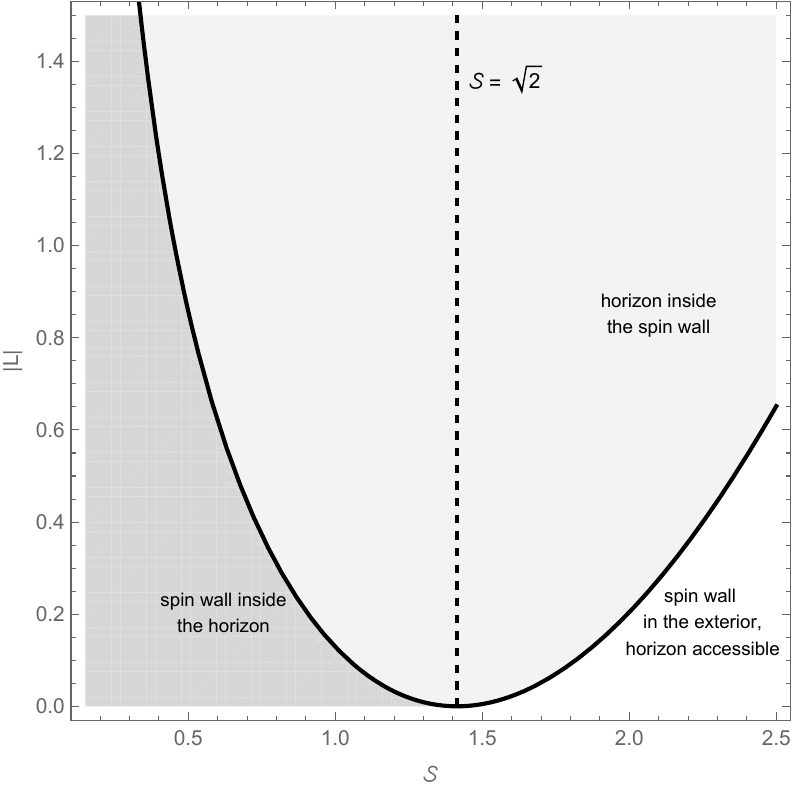}
\caption{The parameter plane $(\mathcal S,|L|)$. Thick curve: $|L|=\sqrt{\Lambda(1;\mathcal S)}$
	of Eq.~\eqref{eq:Lambda1}; dashed line: $\mathcal S=\sqrt2$. Dark grey, the wall lies inside
	the horizon; light grey, the horizon lies inside the wall; white, the wall lies in the
	exterior with the horizon on the inner admissible branch. The wall is absent at $|L|=0$.}
\label{fig:wall-regions}
\end{figure}
Figure~\ref{fig:wall-regions} shows this partition. At small spin the wall is hidden from the
exterior over most of the plane, because $\Lambda(1;\mathcal S)$ grows as $\mathcal S\to0$; the
hidden region terminates at $\mathcal S=\sqrt2$, beyond which every $|L|\ne0$ places the wall in
the exterior.

At the horizon $\eta=|\mathcal S|/\sqrt2$, so the condition $\mathcal S>\sqrt2$ is simply
$\eta(r_s)>1$. More generally $\eta(u_Q)=(u_Q/u_*)^{3/2}<1$: the wall always lies on the
controlled side of the surface $\eta=1$, which the wall itself touches only as $L\to0$. The value
of $\eta$ at the wall is not fixed by the spin alone; it can be made arbitrarily small by taking
a small spin with $|L|$ just above $\sqrt{\Lambda(1;\mathcal S)}$, and it tends to unity as
$|L|\to0$.

\subsection{The admissible domain and the fate of an orbit}
\label{sec:domain}

Equation~\eqref{eq:gamma-K} reduces the existence of a timelike representative worldline to the positivity of $Q$. 
For $L\ne0$, the admissible radii are therefore the two branches
\begin{equation}
	0<u<u_Q \qquad\text{or}\qquad u>u_R ,
	\label{eq:admissible}
\end{equation}
separated by the forbidden interval $(u_Q,u_R)$ bounded by the wall.
The inversion of $A$ remains regular throughout, since $1-\alpha_r=0$ requires $F^3=3(1-F)L^2u^2$ and hence $F>0$, 
while $Q=F^3(2F-3)<0$ on that locus; this singularity therefore lies inside the forbidden interval.
The two admissible components are strictly disconnected.
An orbit arriving from spatial infinity belongs to the outer branch, cannot cross the forbidden interval,
and therefore cannot reach the horizon whenever the horizon lies inside it; 
conversely, an orbit on the inner branch can never escape to large radii. 

On the outer branch $(\mathrm dr/\mathrm d\tau)^{2}=U/Q$ with $Q>0$, 
so the motion is possible only where $U\ge0$ and its turning points are the zeros of $U$. 
At the wall itself $Q$ vanishes while $U$ stays finite, 
so that generically $|\mathrm dr/\mathrm d\tau|$ diverges there instead of vanishing.
An orbit therefore ends in one of two ways. 
If $U$ vanishes somewhere in $(0,u_Q)$, the orbit turns around there and never reaches the wall; 
if $U>0$ throughout, it reaches the wall with a diverging radial velocity, 
and the proper time needed, $\int (Q/U)^{1/2}\,\mathrm dr$, converges. 
Only the first of these is a turning point; the wall itself never reflects an orbit, being not a
barrier of the effective potential but the end of the domain on which the description exists.

The degeneracy radius $u_{*}$ of \eqref{eq:rstar} is also a zero of $U$, but never a turning point.
Because $F^{2}$ vanishes there it is at least double; for $L\ne0$ it falls inside the forbidden interval, where no orbit exists, 
and for $L=0$ the simultaneous vanishing of $U$ and $Q$ leaves $(\mathrm dr/\mathrm d\tau)^{2}$ regular. 
Both cases are examined in Sec.~\ref{sec:wall}.

The angular form~\eqref{eq:dudphi-schw} behaves differently. 
Since $(\mathrm du/\mathrm d\phi)^{2}$ remains finite at $u_Q$, the curve $u(\phi)$ can be continued through the wall; 
but beyond $u_Q$ one has $Q<0$, no timelike four-velocity exists, 
and the continued curve is not a solution of the original system. 
This is also the sense in which the proper-time behaviour reported in Ref.~\cite{Armaza:2015eha} should be read.
For a trajectory that does reach the wall, the timelike part ends at $u_Q$ after a finite proper time, 
whereas the formal continuation towards the degeneracy radius would require a spacelike segment,
along which the proper-time integral diverges.

\begin{figure}[htb]
\centering
\includegraphics[width=0.7\textwidth]{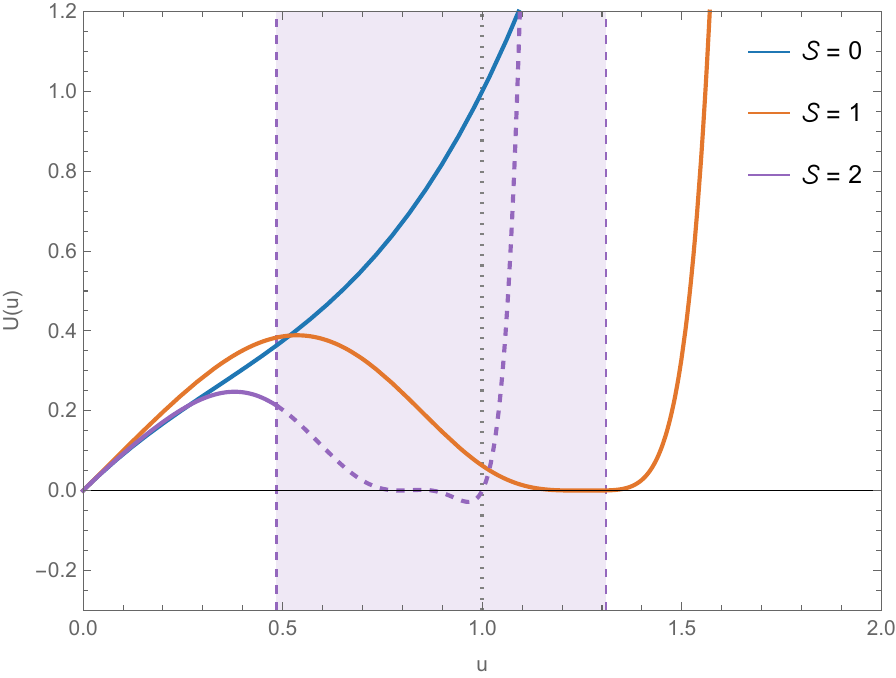}
\caption{The function $U(u)$ of Eq.~\eqref{eq:U-poly} for $\mathcal E=1$ and $\mathcal J=1$, with
$\mathcal S=0$ (blue), $1$ (orange) and $2$ (purple); the grey vertical dotted line is the horizon
$u=1$. The shaded strip marks the forbidden interval $(u_Q,u_R)$ of the $\mathcal S=2$ curve,
bounded by the dashed verticals, and the part of that curve lying inside the strip is drawn
dashed, since no motion exists there.}
\label{fig:effect-of-S}
\end{figure}

Figure~\ref{fig:effect-of-S} shows the three cases available at $(\mathcal E,\mathcal J)=(1,1)$,
for which $L=1-\mathcal S$. Without spin, $U=u(1-u+u^{2})$ has no zero beyond $u=0$, so the particle
falls through the horizon. For $\mathcal S=1$ one has $L=0$ and hence no wall, and
$U=u(1-u^{3}/2)^{4}$ only touches zero at the degeneracy radius, so that the motion passes
through it unaffected. For $\mathcal S=2$ the wall is present and $U$ stays positive all the way
to $u_Q$: the orbit therefore reaches the wall instead of turning at a zero of $U$.

\subsection{The radial family}
\label{sec:wall}

For $L\ne0$ the degeneracy radius lies inside the forbidden interval (see Sec.~\ref{sec:domain}), 
so no orbit reaches it; the exception is $L=0$, which is also the limit in which the two branches merge.
Inverting $L^{2}=\Lambda(u;\mathcal S)$ near $u_{*}$ gives
\begin{equation}
	u_{Q,R}=u_{*}\left[1\mp\frac{|L|^{1/2}}{\sqrt3\,(\mathcal S^{2}/2)^{1/6}}
	+\mathcal O(L)\right],
	\label{eq:ufold}
\end{equation}
so that the branches approach the degeneracy radius as a square root.
The width $u_R-u_Q$ of the forbidden interval scales as $|L|^{1/2}$ and vanishes only at $L=0$ exactly, 
and $u_Q$ admits no expansion in powers of $L$.

For $L=0$ one has $P_{\phi}=0$ by Eq.~\eqref{eq:Pphi-inv}, so the four-momentum is radial. 
By Eqs.~\eqref{eq:V-components} and~\eqref{eq:mpd-momentum-schw} the covariant force then vanishes identically, 
the momentum is parallel transported,
and, since $\gamma\equiv1$ by Eq.~\eqref{eq:K-schw} the four-velocity is parallel to it,
which means the worldline is a geodesic with $(\mathrm dr/\mathrm d\tau)^{2}=\mathcal E^{2}-1+r_s/r$. 
At $u_{*}$ the numerator of Eq.~\eqref{eq:Pphi-inv} vanishes together with the denominator, 
so neither the momenta nor the radial equation degenerates, and the orbit crosses the degeneracy radius $r_{*}$ without obstruction. 
The conserved angular momentum does not vanish, however, as $J=P_{\phi}-(s/\mathcal M)P_{t}$ reduces to
$-(s/\mathcal M)P_{t}=Es/\mathcal M$, i.e.\ $\mathcal J=\mathcal S\mathcal E$.
So it is carried entirely by the spin while the momentum itself has no angular component.

For these orbits characterized by $L=0$, the forbidden interval collapses onto the single point $u_{*}$, 
and the wall is transparent to the motion in the precise sense that no range of radii is excluded.
For $L\ne0$, by contrast, an interval of radii is excluded, and an orbit arriving from outside ends at $r_Q$ however small $|L|$ may be. 
We shall call the orbits with $L=0$ the \emph{radial family}.

\section{Circular orbits and weak-field comparison}
\label{sec:co-prec}
\subsection{Circular orbits}
\label{sec:circular-orbits}

Introducing the effective potential
\begin{equation}
	V_\text{eff}(r)\equiv-\Bigl(\frac{\mathrm dr}{\mathrm d\tau}\Bigr)^{2}=-\frac{U(u)}{Q(u)}\,,
	\label{eq:Vf-def}
\end{equation}
the conditions for a circular orbit at $u=u_c$ are $V_\text{eff}=0$ and $\mathrm dV_\text{eff}/\mathrm dr=0$, with
$\mathrm d^{2}V_\text{eff}/\mathrm dr^{2}>0$ for stability. Since $Q>0$ is required for the motion to exist
(Sec.~\ref{sec:domain}), circular orbits can only lie outside the wall. One finds
\begin{align}
	\frac{\mathrm dV_\text{eff}}{\mathrm dr}
	&=\frac{u^{2}}{r_s}\,\frac{U'Q-UQ'}{Q^{2}}\,,
	\label{eq:dVf-dr}\\[4pt]
	\frac{\mathrm d^{2}V_\text{eff}}{\mathrm dr^{2}}
	&=-\frac{u^{3}}{r_s^{2}}
	\biggl[\frac{2(U'Q-UQ')}{Q^{2}}
	+\frac{u\bigl[(U''Q-UQ'')Q-2Q'(U'Q-UQ')\bigr]}{Q^{3}}\biggr],
	\label{eq:d2Vf-dr2}
\end{align}
where a prime denotes $\mathrm d/\mathrm du$.

At $u=u_c$ the first two conditions reduce to $U(u_c)=U'(u_c)=0$ with $Q(u_c)>0$. Since
$U=F^{2}U_7$ and $F(u_c)\ne0$ outside the wall, this is equivalent to
\begin{equation}
	U_7(u_c)=0,\qquad U_7'(u_c)=0,\qquad Q(u_c)>0 ,
	\label{eq:circ-U7}
\end{equation}
and at such a point $U''=F^{2}U_7''$, so that Eq.~\eqref{eq:d2Vf-dr2} collapses to
$\mathrm d^{2}V_\text{eff}/\mathrm dr^{2}|_{u_c}=-(u_c^{4}/r_s^{2})U_7''(u_c)/Q(u_c)$. Stability is thus
equivalent to
\begin{equation}
	U_7''(u_c)<0\qquad(\text{stable}),\qquad
	U_7''(u_c)=0\qquad(\text{marginally stable}),
	\label{eq:stab-U7}
\end{equation}
exactly and at all orders in the spin. Whether the marginally stable orbit is also the
\emph{innermost} stable one depends on $U_7''$ changing sign only once along the branch; this is
what happens for the solutions discussed below.

The two conditions in Eq.~\eqref{eq:circ-U7} are linear in the three combinations
$X\equiv\mathcal E^{2}$, $Y\equiv\mathcal J^{2}$ and $Z\equiv\mathcal J\mathcal E$:
\begin{equation}
	U_7=\lambda_X X+\lambda_Y Y+\lambda_Z Z+\lambda_0 ,
	\label{eq:U7-linear}
\end{equation}
with
\begin{equation}
	\lambda_X=1-\mathcal S^{2}u^{2}(1-u),\quad
	\lambda_Y=-u^{2}(1-u)+\frac{\mathcal S^{2}}{4}u^{6},\quad
	\lambda_Z=\mathcal S u^{2}(2-3u),\quad
	\lambda_0=-(1-u)F^{2}.
	\label{eq:lambda}
\end{equation}
The same holds for $U_7'$, with the derivatives $\lambda_i'$. The solution set of the two
conditions is therefore a straight line in $(X,Y,Z)$, which the algebraic constraint $XY=Z^{2}$
cuts in at most two points. That is, even though the two conditions are quadratics whose
elimination would leave an equation of degree four, the system admits at most \emph{two} circular
orbits at any radius---the two branches, which we label by $\mathop{\rm sign}\mathcal J$.

The position of the marginally stable orbit follows from expanding Eq.~\eqref{eq:circ-U7} together
with $U_7''(u_c)=0$ order by order in $\mathcal S$ around the geodesic solution $u_c=\tfrac13$.
The expansion inherits a parity from Eq.~\eqref{eq:lambda}: $\lambda_X$, $\lambda_Y$ and
$\lambda_0$ are even in $\mathcal S$ while $\lambda_Z$ is odd, so that $(X,Y,Z)\to(X,Y,-Z)$ maps
solutions at $\mathcal S$ onto solutions at $-\mathcal S$. Since $\mathcal E>0$ and
$Z=\mathcal J\mathcal E$, this is precisely $\mathcal J\to-\mathcal J$, and hence
\begin{equation}
	u_{\rm ms}(\mathcal S;\ \mathcal J>0)=u_{\rm ms}(-\mathcal S;\ \mathcal J<0):
	\label{eq:isco-parity}
\end{equation}
even orders are common to the two branches, odd orders opposite. One finds
\begin{equation}
	u_{\rm ms}=\frac13\pm\frac{2\sqrt6}{27}\,\mathcal S+\frac{61}{324}\,\mathcal S^{2}
	+\mathcal O(\mathcal S^{3}),
	\qquad
	r_{\rm ms}=6M\biggl[1\mp\frac{2\sqrt6}{9}\,\mathcal S-\frac{29}{108}\,\mathcal S^{2}
	+\mathcal O(\mathcal S^{3})\biggr],
	\label{eq:isco-exp}
\end{equation}
the upper and lower signs referring to $\mathcal J>0$ and $\mathcal J<0$. The two effects are
distinct: the linear term separates the branches, moving one inward and the other outward, while
the quadratic term moves both towards smaller radii. The expansion is used here to exhibit this
structure; the quantitative values for $\mathcal S$ of order unity are obtained numerically, as in
the figures below. We note finally that different spin supplementary conditions differ already at
$\mathcal O(\mathcal S)$ by a redefinition of the representative worldline, so
Eq.~\eqref{eq:isco-exp} is specific to the Tulczyjew--Dixon condition and to our normalisation
$\mathcal S=s/(\mathcal M r_s)$.

\begin{figure}[htb]
	\centering
	\begin{minipage}{0.48\textwidth}
		\centering
		\includegraphics[width=\textwidth]{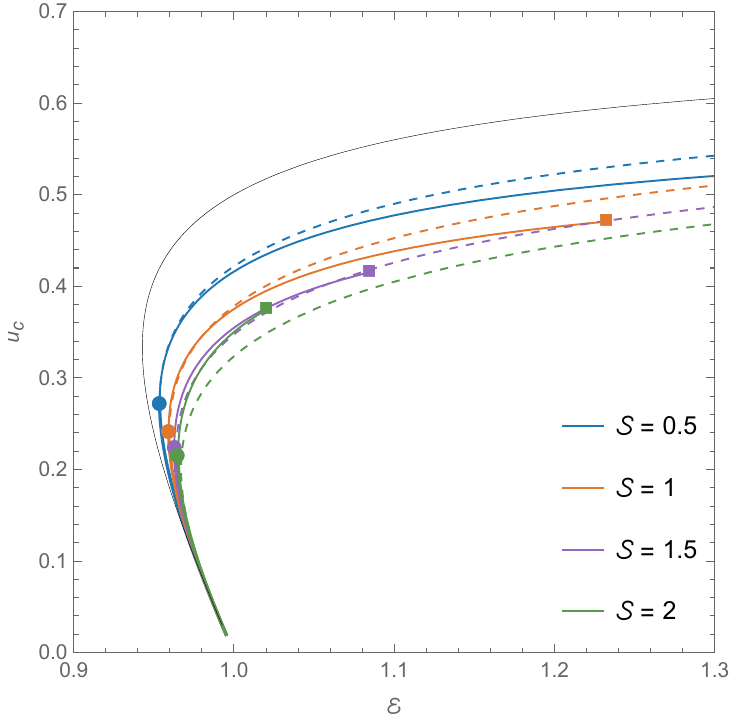}
	\end{minipage}
	\hfill
	\begin{minipage}{0.48\textwidth}
		\centering
		\includegraphics[width=\textwidth]{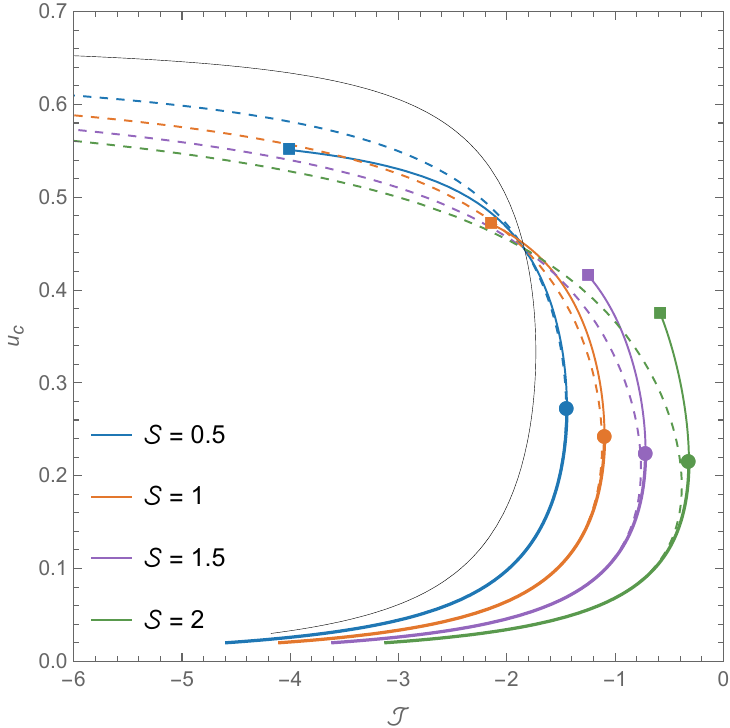}
	\end{minipage}
	\caption{Circular orbits with $\mathcal J<0$. Left: the energy $\mathcal E$ against
	$u_c=r_s/r_c$; right: the angular momentum $\mathcal J$ against $u_c$. Filled circles mark the
	marginally stable orbit, i.e.\ the ISCO; on either side of it the curves are thick (stable
	branch) and thin (unstable branch). Filled squares mark the points at which a branch meets the
	wall ($Q=0$). Thin dashed curves give the same quantities truncated at first order in
	$\mathcal S$, and the thin solid line is the geodesic ($\mathcal S=0$) branch, drawn up to the
	light-ring limit. For $\mathcal S=0.5,\,1,\,1.5,\,2$ the circles lie at
	$u_c=0.273,\,0.243,\,0.225,\,0.216$ and the squares at $u_c=0.551,\,0.471,\,0.415,\,0.375$.}
	\label{fig:circ-ret}
\end{figure}

\begin{figure}[htb]
	\centering
	\begin{minipage}{0.48\textwidth}
		\centering
		\includegraphics[width=\textwidth]{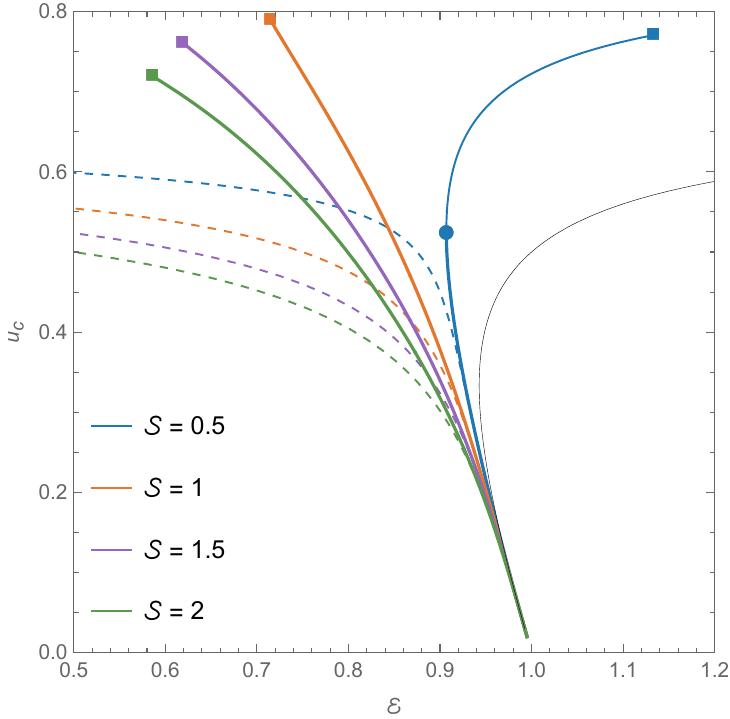}
	\end{minipage}
	\hfill
	\begin{minipage}{0.48\textwidth}
		\centering
		\includegraphics[width=\textwidth]{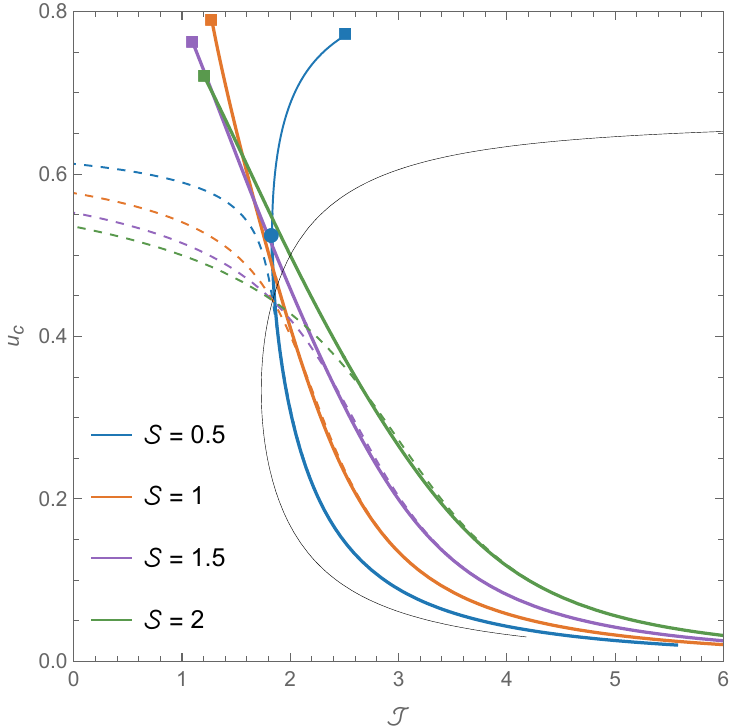}
	\end{minipage}
	\caption{As Fig.~\ref{fig:circ-ret}, for the branch with $\mathcal J>0$. The circle is drawn only
	for $\mathcal S=0.5$, at $u_c=0.525$, the sole case in which the marginally stable orbit lies
	outside the wall; the remaining curves are stable along the whole segment shown. The squares lie
	at $u_c=0.771,\,0.789,\,0.761,\,0.719$ for $\mathcal S=0.5,\,1,\,1.5,\,2$.}
	\label{fig:circ-prog}
\end{figure}

The two branches behave very differently. Along the branch with $\mathcal J<0$, as in Fig.~\ref{fig:circ-ret}, 
every curve has a turning point, at
which $\mathcal E$ and $\mathcal J$ stop growing inwards and turn back. Differentiating
Eq.~\eqref{eq:circ-U7} along the branch shows that $\mathcal E$ and $\mathcal J$ are stationary
precisely when $U_7''=0$, so this turning point is the marginally stable orbit; the thick and thin
segments lie on either side of it. As $\mathcal S$ grows the turning point moves outward, at
$u_c=0.273$, $0.243$, $0.225$ and $0.216$ for $\mathcal S=0.5$, $1$, $1.5$ and $2$, and it always
stays well outside the wall, which cuts the branch further in.

Along the branch with $\mathcal J>0$, as in Fig.~\ref{fig:circ-prog}, the turning point lies further in,
at $u_c=0.525$ for $\mathcal S=0.5$, and for the larger spins shown it no longer appears in the
plotted segment, which is stable along its entire length. In those cases $\mathcal E$ and
$\mathcal J$ decrease monotonically as the orbit shrinks, in contrast with the geodesic behaviour,
in which they turn upward before the light-ring limit. What the exact solution shows at these
spins is that all circular orbits outside the wall remain stable.

This is consistent with the analysis of Ref.~\cite{Jefremov:2015gza}, in which the
spin acts attractively on prograde orbits, displacing the ISCO inward, and repulsively on
retrograde ones, displacing it outward. The two branches obtained here bear this out. On the
branch with $\mathcal J<0$ (retrograde for $\mathcal S>0$), $u_c<\tfrac13$ for every spin shown, so the marginally stable orbit lies farther out than the
geodesic one; on the branch with $\mathcal J>0$ (prograde for $\mathcal S>0$), the sole case in which it remains outside the wall is
$\mathcal S=0.5$, with $u_c=0.525$, i.e.\ inside the geodesic ISCO radius. For the larger spins the
marginally stable orbit lies beyond the wall, at radii where no timelike orbit exists.

The radii at which the branches meet the wall also behave differently. For $\mathcal J<0$ they
decrease monotonically with $\mathcal S$, $u_c\simeq0.551,\,0.471,\,0.415,\,0.375$ for
$\mathcal S=0.5,\,1,\,1.5,\,2$, whereas for $\mathcal J>0$ they do not,
$u_c\simeq0.771,\,0.789,\,0.761,\,0.719$. Since the wall itself moves monotonically outward at
fixed $|L|$, Eq.~\eqref{eq:dQdL2}, the position of these endpoints cannot be inferred from the
monotonicity of the wall alone.

Moreover, the exact curves and their first-order truncations agree at large radii, where the spin
correction is of relative order $\mathcal S\sqrt{u_c}$, and that to the energy of order $\mathcal Su_c^{3/2}$,
and separate as the orbit shrinks, most
markedly near the inner end of the prograde branch, where the two descriptions differ
qualitatively.

Finally, every configuration displayed lies within the range of validity of the pole--dipole
truncation: at the points where the branches meet the wall, marked by squares in the figures,
the ratio~\eqref{eq:eta} never exceeds $0.862$, so that $\eta<1$ along each curve shown.

\subsection{Perihelion precession: spin corrections at $O(u^{3})$}

For weak-field orbits ($u\ll1$) the orbital equation~\eqref{eq:dudphi-schw}
can be expanded systematically.  Expanding $U(u)$ and the denominator to
$\mathcal{O}(u^{3})$ gives
\begin{equation}
	\Bigl(\frac{du}{d\phi}\Bigr)^{2}
	= \frac{\mathcal{E}^{2}-1}{L^{2}}
	  + \frac{u}{L^{2}}
	  - u^{2}
	  + \Bigl(1 + \frac{4\mathcal{S}^{2}}{L^{2}}
	           - \frac{3\mathcal{E}^{2}\mathcal{S}^{2}}{L^{2}}
	           - \frac{\mathcal{E}\mathcal{J}\mathcal{S}}{L^{2}}\Bigr)u^{3}
	  + O(u^{4}).
	\label{eq:dudphi-expand}
\end{equation}
Differentiating it with respect to $\phi$ yields the Binet-type equation
at this order,
\begin{equation}
	\frac{d^{2}u}{d\phi^{2}} + u
	= \frac{1}{2L^{2}}
	  +\mathcal B u^{2}
	  + O(u^{3})\,,\quad 
	  \mathcal B\equiv \frac{3}{2}
	  + \frac{6\mathcal{S}^{2}}{L^{2}}
	  - \frac{9\mathcal{E}^{2}\mathcal{S}^{2}}{2L^{2}}
	  - \frac{3\mathcal{E}\mathcal{J}\mathcal{S}}{2L^{2}}\,.
	\label{eq:binet-expand}
\end{equation}
For $\mathcal S=0$ both reduce to the standard Schwarzschild results. Of the spin-dependent terms,
the one linear in $\mathcal S$ is the spin--orbit contribution and has the form familiar from the
post-Newtonian treatment of spinning bodies~\cite{LeTiec:2011bk,Bini:2019zjj}, whereas those
quadratic in $\mathcal S$ are of second order in the spin and have no counterpart in the linear
treatment. In the limit of small eccentricity $e$ the perturbative solution is
\begin{equation}
	u(\phi) \simeq \frac{1}{2L^2}\biggl(1+e\cos\phi + \frac{e\mathcal B}{2L^2}\,\phi\sin\phi\biggr)
	        \simeq \frac{1}{2L^2}\ \biggl[1+e\cos\bigg(\phi(1-\frac{\mathcal B}{2L^2})\bigg)\biggr].
	\label{eq:secular}
\end{equation}
After one orbital period $\Delta\phi=2\pi$, the perihelion advances by
$\Delta\phi_{\rm total}=\pi \mathcal B/L^2$.  Substituting  $\mathcal B$ and
expressing $L^{2}\simeq a_s(1-e^{2})/(4GM)$ for a Keplerian orbit with semi-major axis $a_s$,
we obtain
\begin{equation}
		\Delta\phi = 	\Delta\phi_0  + 	\Delta\phi_s  
		=
	 \frac{6\pi GM}{a_s(1-e^{2})} 
		+  \frac{6\pi GM}{a_s(1-e^{2})} 
		\frac{4\mathcal{S}^{2}(1-\mathcal{E}^{2})
		-\mathcal{E}\mathcal{S}L}{L^2}\,,
\end{equation}
where the first term, $\Delta\phi_0$, is the geodesic advance, and the second, $\Delta\phi_s$, encodes the spin correction. 
The expansion presupposes that the orbit stays inside the admissible domain, $Q>0$.
For Mercury $\mathcal S\sim3\times10^{-6}$ and $L\sim3\times10^{3}$, so that
$|\Delta\phi_s/\Delta\phi_0|\sim\mathcal S/L\sim10^{-9}$. The wall is irrelevant at these
parameters in that $u_Q\simeq(3\mathcal S^{2}L^{2})^{-1/5}\simeq5$, that is $r_Q\simeq0.6$ km.
The degeneracy radius lies at $r_*\simeq0.5$ m. Both are far inside the Sun, where the exterior
Schwarzschild solution used here does not apply, and the estimate merely shows that no
observable effect is to be expected for planetary orbits.

\section{Discussions and conclusions}
\label{sec:dc}
In this work, we have presented a nonperturbative analysis of planar spinning-particle motion in Schwarzschild spacetime 
within the Mathisson--Papapetrou--Dixon framework, imposing the Tulczyjew--Dixon spin supplementary condition. 
The key step of our approach is the elimination of the four-velocity from the equations of motion, 
which recasts the MPD system into a closed algebraic form and reduces the orbital dynamics to an exactly integrable, 
one-dimensional problem. Our principal findings are summarized below.

\textbf{Exact reduction.} The Tulczyjew--Dixon constraint is used as an algebraic relation for
the four-velocity, so that both the momentum and the spin equations become velocity-free and the
velocity is recovered only afterwards from the transport relation. The momentum equation is
governed by a $4\times4$ matrix $A=1-B$, with $B$ the curvature contracted with two spin tensors;
since the supplementary condition restricts $B$ to rank two, $\det A$ and $A^{-1}$ are closed-form,
and this is what allows the spin to be kept to all orders. For equatorial motion in Schwarzschild,
$A^{-1}$ collapses to a single scalar factor and the radial equation becomes algebraic; its
numerator, the effective potential, factorizes into a factor depending on the spin alone and a
polynomial carrying the energy and the angular momentum, and this separation is what makes the
location of the wall independent of the orbit. The only approximation retained is the pole--dipole
truncation itself.

\textbf{The spin wall.} The dynamical
mass $\mathcal M^{2}=-P^\mu P_\mu$ and the projection $m=-P^{\mu}\dot x_{\mu}$ of the momentum onto the
velocity are kept distinct throughout. Their ratio
$\gamma=m/\mathcal M$ can be viewed as the Lorentz factor between the worldline and the momentum frames, and it
is fixed by the motion through $\gamma=(1-\mathcal K)^{-1/2}$, where $\mathcal K$ is the square of the speed with which
the representative point drifts through its own momentum rest frame.
The drift grows as the body moves inward, and generically $\mathcal K$ reaches unity
at a finite radius, beyond which the exact equations admit no timelike worldline at all. The locus
at which this happens bounds the region accessible to the body and is what we call the spin wall;
since $\gamma-1=\tfrac12\mathcal K+\mathcal O(\mathcal K^{2})$ is of fourth order in the spin, that
boundary is absent from any truncation below fourth order.
Its position can be obtained in closed
form from the radial equation and is fixed by two parameters, the spin $\mathcal S$ and $L=\mathcal J-\mathcal E\mathcal S$, the
latter being proportional to $P^\phi$. The parameter space of $\mathcal S$ and $L$ falls into three regions, according to
whether the wall lies entirely inside the horizon, encloses the horizon, or reaches the exterior
with the horizon resting on the innermost admissible branch. In particular, for small spin a
sufficiently large $|L|$ may still place the wall outside the horizon.

\textbf{Filtering of orbits.} For a body falling in from
spatial infinity the wall bounds the accessible radii, the four-velocity ceasing to be timelike
beyond it. The single exception is the family whose four-momentum has no angular component,
$P_{\phi}=0$ (equivalently $L=0$, or $\mathcal J=\mathcal S\mathcal E$). For that family the wall
degenerates onto the degeneracy radius $r_{*}=(Ms^{2}/\mathcal M^{2})^{1/3}$, the drift vanishes identically, and the motion is a
geodesic with $(\mathrm dr/\mathrm d\tau)^{2}=\mathcal E^{2}-1+r_s/r$ which passes the degeneracy
radius without obstruction, the conserved angular momentum being carried entirely by the spin. 
The wall therefore filters the orbits. It is a
surface present for every nonvanishing value of the momentum's angular component, and only the
radial family of orbits can reach arbitrarily small radii.

\textbf{Circular orbits.} The circular orbits were determined exactly, together with their stability,
and their energies and angular momenta were computed as functions of the radius over a range of
spins on both branches (Figs.~\ref{fig:circ-ret} and~\ref{fig:circ-prog}). A given radius supports
at most two circular orbits, distinguished by the sign of the angular momentum, and each branch
terminates where it meets the wall. The marginally stable orbit is displaced according to
\begin{equation}
	r_{\rm ms}=6M\Bigl[1\mp\frac{2\sqrt6}{9}\,\mathcal S-\frac{29}{108}\,\mathcal S^{2}
	+\mathcal O(\mathcal S^{3})\Bigr],
	\label{eq:concl-rms}
\end{equation}
the upper and lower signs referring to $\mathcal J>0$ and $\mathcal J<0$: the linear term separates
the two branches and the quadratic one moves both inward.

\textbf{Perihelion precession.} The weak-field expansion of the orbit gives the advance
\begin{equation}
	\Delta\phi=\Delta\phi_0+\Delta\phi_s,\qquad
	\frac{\Delta\phi_s}{\Delta\phi_0}
	=\frac{4\mathcal S^{2}(1-\mathcal E^{2})-\mathcal E\mathcal S L}{L^{2}},
\end{equation}
where $\Delta\phi_0$ is the geodesic advance, $\mathcal E$ the energy per unit dynamical mass and $L$
the quantity proportional to the angular component of the four-momentum. The term linear in the spin
has the same form as the spin--orbit contribution of the post-Newtonian treatment, while those
quadratic in the spin have no linear counterpart.

Several qualifications delimit these results. The wall is not everywhere in a regime where the
pole--dipole truncation is well controlled: when it lies close to the degeneracy radius---that is,
for a small angular component of the momentum---the M\o ller radius of the body is comparable to the
local curvature scale, so that the neglected quadrupole and higher multipoles could shift or remove
the boundary; and whether $Q<0$ signals the failure of the multipole expansion, of the supplementary
condition, or a genuine capture is not decided by the MPD equations alone. A body respecting the Kerr
bound, $s\le\mathcal M^{2}$, has $\mathcal S\le\mathcal M/(2M)\ll1$ in the test-body limit, so that
the wall lies behind the horizon; an exterior wall requires a spin of order unity, beyond what such
a body admits. Finally, the treatment is restricted to equatorial, spin-aligned motion and to the
Tulczyjew--Dixon condition with our normalization of the spin; a precessing spin, other supplementary
conditions, radiation reaction and the self-force all lie outside it.

The natural continuation is to include the quadrupole. Dixon's formalism supplies the leading
finite-size correction to the momentum equation explicitly, so that the displaced boundary can be
computed in the same algebraic manner and the leading correction to its position obtained in closed
form.

\acknowledgments
We acknowledge support from NSFC grant No.~12575060 and No.~12105179.

\appendix
\section{Detailed derivations}
\label{app:MPD-derivation}

\subsection{Derivation of the transport relation}

Differentiating the Tulczyjew--Dixon supplementary condition $S^{\mu\nu}P_\nu=0$ along the worldline yields
\begin{align}
	\frac{D}{d\tau}\bigl(S^{\mu\nu}P_{\nu}\bigr)
	&= \frac{DS^{\mu\nu}}{d\tau}P_{\nu} + S^{\mu\nu}\frac{DP_{\nu}}{d\tau} \nonumber \\
	&= \bigl(P^{\mu}\dot{x}^{\nu} - P^{\nu}\dot{x}^{\mu}\bigr)P_{\nu}
	   + S^{\mu\nu}\frac{DP_{\nu}}{d\tau} = 0,
	\label{eq:app-diff-ssc}
\end{align}
where we have substituted the spin equation~\eqref{eq:mpd-spin}. Recalling $m\equiv -P^{\nu}\dot{x}_{\nu}$ and $\mathcal{M}^{2}\equiv -P^{\mu}P_{\mu}$, this becomes
\begin{equation}
	P^{\mu}(-m) - \dot{x}^{\mu}(-\mathcal{M}^{2}) + S^{\mu\nu}\frac{DP_{\nu}}{d\tau} = 0,
\end{equation}
which rearranges to the transport relation~\eqref{eq:transport} in the main text.

\subsection{Kinematic relations}

Contracting the transport relation with $\dot{x}_{\mu}$ and using $\dot{x}^{\mu}\dot{x}_{\mu}=-1$ gives
\begin{equation}
	-1
	= \frac{m}{\mathcal{M}^{2}} P^{\mu}\dot{x}_{\mu}
	- \frac{1}{\mathcal{M}^{2}} \dot{x}_{\mu} S^{\mu\nu}\frac{DP_{\nu}}{d\tau}
	= \frac{m^{2}}{\mathcal{M}^{2}}
	- \frac{1}{\mathcal{M}^{2}} \dot{x}_{\mu} S^{\mu\nu}\frac{DP_{\nu}}{d\tau},
\end{equation}
which rearranges to $\mathcal{M}^{2}=m^{2}+\dot{x}_{\mu} S^{\mu\nu}DP_{\nu}/d\tau$, i.e.\ Eq.~\eqref{eq:Mm-relation}.  Substituting the transport relation $\dot{x}_{\mu}=(m/\mathcal{M}^{2})P_{\mu}-S_{\mu}^{\ \rho}DP_{\rho}/(\mathcal{M}^{2}d\tau)$ and noting $P_{\mu}S^{\mu\nu}=0$ by the TD-SSC eliminates the $P_{\mu}$ term, giving
\begin{equation}
	\dot{x}_{\mu}S^{\mu\nu}\frac{DP_{\nu}}{d\tau}
	= -\frac{1}{\mathcal{M}^{2}}\,g_{\mu\lambda}S^{\lambda\rho}S^{\mu\nu}\,
	  \frac{DP_{\rho}}{d\tau}\frac{DP_{\nu}}{d\tau}.
	\label{eq:app-xSDP}
\end{equation}
The product $g_{\mu\lambda}S^{\lambda\rho}S^{\mu\nu}$ is computed by inserting $S^{\mu\nu}=-\mathcal{M}^{-1}\varepsilon^{\mu\nu\kappa\lambda}P_{\kappa}s_{\lambda}$:
\begin{align}
	g_{\mu\lambda}S^{\lambda\rho}S^{\mu\nu}
	= \frac{1}{\mathcal{M}^{2}}\,
	   \varepsilon^{\mu\nu\kappa\lambda}\,
	   \varepsilon_{\mu}^{\ \rho\alpha\beta}\,
	   P_{\kappa}s_{\lambda}\,P_{\alpha}s_{\beta}.
	\label{eq:app-S2-epsilon}
\end{align}
Lowering the three raised indices of $\varepsilon_{\mu}^{\ \rho\alpha\beta}$ to $\varepsilon_{\mu\rho'\alpha'\beta'}$ via the metric, the single-index identity $\varepsilon^{\mu\nu\kappa\lambda}\varepsilon_{\mu\rho'\alpha'\beta'}= -\delta^{\nu\kappa\lambda}_{\rho'\alpha'\beta'}$ followed by raising the contracted indices back yields a $6$-term determinant.  Contracting with $P_{\kappa}s_{\lambda}P_{\alpha}s_{\beta}$, the three terms containing $P^{\alpha}s_{\alpha}=0$ (TD-SSC) vanish.  The remaining three reduce to
\begin{align}
	g_{\mu\lambda}S^{\lambda\rho}S^{\mu\nu}
	&= \frac{1}{\mathcal{M}^{2}}\Bigl[
	   \mathcal{M}^{2}s^{2}\,g^{\rho\nu}
	   + s^{2}\,P^{\rho}P^{\nu}
	   - \mathcal{M}^{2}\,s^{\rho}s^{\nu}
	   \Bigr] \nonumber \\[4pt]
	&= s^{2}\,g^{\rho\nu}
	   + \frac{s^{2}}{\mathcal{M}^{2}}\,P^{\rho}P^{\nu}
	   - s^{\rho}s^{\nu},
	\label{eq:app-SS-identity}
\end{align}
with the overall sign confirmed by the rest-frame derivation below.

\medskip
\noindent\textbf{Rest-frame verification.}
In the local rest frame $P^{\mu}=(\mathcal{M},\mathbf{0})$ with $g_{\mu\nu}=\eta_{\mu\nu}$, the TD-SSC forces $S^{0i}=0$, so $S^{ij}=-\varepsilon^{ijk}s_{k}$ ($\varepsilon^{123}=+1$).  The contraction $g_{\mu\lambda}S^{\lambda\rho}S^{\mu\nu}=S^{\mu\nu}S_{\mu}^{\ \rho}$ then becomes
\begin{equation}
	S^{ij}S_{i}^{\ k}
	= \varepsilon^{ij\ell}\,\varepsilon_{i}^{\ kn}\,s_{\ell}s_{n},
\end{equation}
where $\nu=j$, $\rho=k$ are spatial.  Lowering $\varepsilon_{i}^{\ kn}=g_{im}g^{kp}g^{nq}\varepsilon^{mpq}$ and using $\varepsilon^{ij\ell}\varepsilon_{ipq}= \delta^{j}_{p}\delta^{\ell}_{q}-\delta^{j}_{q}\delta^{\ell}_{p}$ gives $g^{kj}g^{n\ell}-g^{k\ell}g^{nj}$, which in Minkowski space reduces to $\delta^{kj}\delta^{n\ell}-\delta^{k\ell}\delta^{nj}$.  Hence
\begin{equation}
	S^{ij}S_{i}^{\ k}
	= (\delta^{kj}\delta^{n\ell}-\delta^{k\ell}\delta^{nj})\,s_{\ell}s_{n}
	= s^{2}\delta^{jk} - s^{j}s^{k}.
\end{equation}
The Kronecker symbol $\delta^{jk}$ acts as the spatial projector; its covariant extension to an arbitrary frame is the projector orthogonal to $P^{\mu}$, namely $g^{\mu\nu}+P^{\mu}P^{\nu}/\mathcal{M}^{2}$, which satisfies $P_{\mu}(g^{\mu\nu}+P^{\mu}P^{\nu}/\mathcal{M}^{2})=0$, has trace~$3$, and reduces to $\operatorname{diag}(0,1,1,1)$ in the rest frame.  Consequently,
\begin{equation}
	S^{\mu\nu}S_{\mu}^{\ \rho}
	= s^{2}\Bigl(g^{\nu\rho}+\frac{P^{\nu}P^{\rho}}{\mathcal{M}^{2}}\Bigr) - s^{\nu}s^{\rho}
	= s^{2}\,g^{\rho\nu} + \frac{s^{2}}{\mathcal{M}^{2}}\,P^{\rho}P^{\nu} - s^{\rho}s^{\nu},
\end{equation}
which coincides with~\eqref{eq:app-SS-identity}.

Contracting the transport relation with $DP_{\mu}/d\tau$ instead leads to
\begin{equation}
	\dot{x}_{\mu}\frac{DP^{\mu}}{d\tau}
	= \frac{m}{\mathcal{M}^{2}} P_{\mu}\frac{DP^{\mu}}{d\tau}
	- \frac{1}{\mathcal{M}^{2}} S^{\mu\nu}\frac{DP_{\mu}}{d\tau}\frac{DP_{\nu}}{d\tau}
	= 0,
	\label{eq:app-M2-const}
\end{equation}
where the first term vanishes because $P_{\mu}DP^{\mu}/d\tau = -\tfrac{1}{2}d\mathcal{M}^{2}/d\tau = 0$, and the second term vanishes by the antisymmetry of $S^{\mu\nu}$. Thus $\mathcal{M}^{2}$ is constant along the worldline, a fact used throughout the main text.

Combining the transport relation with Eq.~\eqref{eq:Mm-relation} yields the exact momentum formula~\eqref{eq:momentum-exact}.

Regarding the kinematical scalar $m\equiv -P^{\mu}\dot{x}_{\mu}$, its derivative reads
\begin{equation}
	\frac{dm}{d\tau} = -\frac{D}{d\tau}(P_{\mu}\dot{x}^{\mu})
	= -\dot{x}_{\mu}\frac{DP^{\mu}}{d\tau} - P_{\mu}\frac{D\dot{x}^{\mu}}{d\tau}
	= -P_{\mu}\frac{D\dot{x}^{\mu}}{d\tau},
\end{equation}
where we have used $\dot{x}_{\mu}DP^{\mu}/d\tau=0$ from Eq.~\eqref{eq:app-M2-const}. Hence $m$ is conserved only when the four-acceleration is orthogonal to the four-momentum; in general, the spin-curvature coupling misaligns $P^{\mu}$ and $\dot{x}^{\mu}$, so $m$ should be regarded as a kinematical quantity rather than a strict constant of motion.

\subsection{Spin tensor inverse relations}

Given the spin four-vector definition~\eqref{eq:spin-vector-def}, we recover the spin tensor by contracting with $\varepsilon_{\alpha\beta\mu\lambda}P^{\lambda}$:
\begin{equation}
	\varepsilon_{\alpha\beta\mu\lambda}P^{\lambda}s^{\mu}
	= \frac{1}{2\mathcal{M}}\,\varepsilon_{\alpha\beta\mu\lambda}
	  \varepsilon^{\mu\nu\rho\sigma}P^{\lambda}P_{\nu}S_{\rho\sigma}.
\end{equation}
Using the single-index Levi-Civita contraction identity
\begin{equation}
	\varepsilon_{\alpha\beta\mu\lambda}\varepsilon^{\mu\nu\rho\sigma}
	= -\delta_{\alpha\beta\lambda}^{\nu\rho\sigma},
\end{equation}
we obtain
\begin{equation}
	\varepsilon_{\alpha\beta\mu\lambda}P^{\lambda}s^{\mu}
	= -\frac{1}{2\mathcal{M}}\,\delta_{\alpha\beta\lambda}^{\nu\rho\sigma}
	  P^{\lambda}P_{\nu}S_{\rho\sigma}.
\end{equation}
Expanding the generalized Kronecker symbol,
\begin{align}
	\delta_{\alpha\beta\lambda}^{\nu\rho\sigma}P^{\lambda}P_{\nu}S_{\rho\sigma}
	&= P^{\lambda}P_{\alpha}S_{\beta\lambda}
	   + P^{\lambda}P_{\beta}S_{\lambda\alpha}
	   + P^{2}S_{\alpha\beta} \nonumber \\
	&\quad - P^{\lambda}P_{\alpha}S_{\lambda\beta}
	   - P^{\lambda}P_{\beta}S_{\alpha\lambda}
	   - P^{2}S_{\beta\alpha}.
\end{align}
Imposing the supplementary condition $S_{\mu\nu}P^{\nu}=0$ from Eq.~\eqref{eq:td-ssc} eliminates all terms containing $S_{\mu\nu}P^{\nu}$ or its transpose, leaving
\begin{equation}
	\delta_{\alpha\beta\lambda}^{\nu\rho\sigma}P^{\lambda}P_{\nu}S_{\rho\sigma}
	= P^{2}S_{\alpha\beta} - P^{2}S_{\beta\alpha}
	= 2P^{2}S_{\alpha\beta}.
\end{equation}
Therefore $\varepsilon_{\alpha\beta\mu\lambda}P^{\lambda}s^{\mu} = \mathcal{M}S_{\alpha\beta}$, which is precisely the inverse relation~\eqref{eq:spin-tensor-inverse} in the main text.

\subsection{Spin four-vector evolution}

Taking the covariant derivative of $s^{\mu}$ defined in Eq.~\eqref{eq:spin-vector-def} gives
\begin{align}
	\frac{Ds^{\mu}}{d\tau}
	&= \frac{1}{2\mathcal{M}}\,\varepsilon^{\mu\nu\rho\sigma}
	   \Bigl( \frac{DP_{\nu}}{d\tau}S_{\rho\sigma} + P_{\nu}\frac{DS_{\rho\sigma}}{d\tau} \Bigr) \nonumber \\
	&= \frac{1}{2\mathcal{M}}\,\varepsilon^{\mu\nu\rho\sigma}
	   \frac{DP_{\nu}}{d\tau}S_{\rho\sigma},
	\label{eq:app-spin-evol-intermediate}
\end{align}
where the $DS_{\rho\sigma}/d\tau$ term vanishes because
\begin{equation}
	\varepsilon^{\mu\nu\rho\sigma}P_{\nu}\frac{DS_{\rho\sigma}}{d\tau}
	= \varepsilon^{\mu\nu\rho\sigma}P_{\nu}
	  \bigl(P_{\rho}\dot{x}_{\sigma} - P_{\sigma}\dot{x}_{\rho}\bigr) = 0,
\end{equation}
owing to the antisymmetry of $\varepsilon^{\mu\nu\rho\sigma}$ against the symmetric products $P_{\nu}P_{\rho}$ and $P_{\nu}P_{\sigma}$. Substituting the inverse relation~\eqref{eq:spin-tensor-inverse} for $S_{\rho\sigma}$ and using the contraction identity
\begin{equation}
	\varepsilon^{\mu\nu\rho\sigma}\varepsilon_{\rho\sigma\gamma\delta}
	= -2\bigl(\delta^{\mu}_{\gamma}\delta^{\nu}_{\delta} - \delta^{\mu}_{\delta}\delta^{\nu}_{\gamma}\bigr),
\end{equation}
we find
\begin{align}
	\frac{Ds^{\mu}}{d\tau}
	&= -\frac{1}{2\mathcal{M}^{2}}\,\varepsilon^{\mu\nu\rho\sigma}
	   \varepsilon_{\rho\sigma\gamma\delta}P^{\gamma}s^{\delta}\frac{DP_{\nu}}{d\tau} \nonumber \\
	&= \frac{1}{\mathcal{M}^{2}}
	   \bigl(\delta^{\mu}_{\gamma}\delta^{\nu}_{\delta} - \delta^{\mu}_{\delta}\delta^{\nu}_{\gamma}\bigr)
	   P^{\gamma}s^{\delta}\frac{DP_{\nu}}{d\tau} \nonumber \\
	&= \frac{1}{\mathcal{M}^{2}}
	   \bigl(P^{\mu}s^{\nu} - s^{\mu}P^{\nu}\bigr)\frac{DP_{\nu}}{d\tau}.
\end{align}
Invoking $P^{\nu}DP_{\nu}/d\tau = 0$ (equivalently $d\mathcal{M}^{2}/d\tau=0$), the term proportional to $P^{\nu}$ drops out, yielding Eq.~\eqref{eq:spin-ev} in the main text.

\section{Determinant and inverse of $A=I-B$ for a rank-2 matrix}
\label{app:Ainv-proof}

In this appendix we provide the detailed derivation of the determinant formula~\eqref{eq:detA-gen} and the inverse formula~\eqref{eq:Ainv-gen}, which rely solely on the algebraic fact that the $4\times4$ matrix
\begin{equation}
	B_\mu^{\ \eta} \equiv \frac{1}{2\mathcal{M}^2} R_{\mu\nu\kappa\lambda} S^{\kappa\lambda} S^{\nu\eta}
\end{equation}
has rank at most~2.

\subsection{Rank of $B$}

We first establish $\operatorname{rank}(B)\leq 2$.
The antisymmetric spin tensor $S^{\mu\nu}$, together with the Tulczyjew--Dixon supplementary condition $S^{\mu\nu}P_{\nu}=0$, can have at most two independent nonvanishing components.
To see this, choose a local frame in which the momentum takes the rest-frame form $P^{\mu}=(\mathcal{M},0,0,0)$.
The condition $S^{\mu\nu}P_{\nu}=0$ then forces $S^{\mu0}=0$ for all $\mu$, so that only the spatial $3\times3$ block $S^{ij}$ ($i,j=1,2,3$) survives.
Since $S^{ij}=-S^{ji}$ is an antisymmetric $3\times3$ matrix, it can be mapped to a 3-dimensional dual vector:
\begin{equation}
	S^{ij} = \varepsilon^{ijk}S_{k},
\end{equation}
which immediately implies that $S^{ij}$ has rank at most~2 (generically rank~2 unless $S^{ij}=0$ identically).
Hence $\operatorname{rank}(S)\leq 2$ as a $4\times4$ matrix.

Now express $B$ as the matrix product
\begin{equation}
	B_\mu^{\ \eta} = T_{\mu\nu}\,S^{\nu\eta},\qquad
	T_{\mu\nu} \equiv \frac{1}{2\mathcal{M}^{2}} R_{\mu\nu\kappa\lambda} S^{\kappa\lambda}.
\end{equation}
Since the product of two matrices cannot exceed the rank of either factor, we obtain
\begin{equation}
	\operatorname{rank}(B) \leq \operatorname{rank}(S) \leq 2.
\end{equation}
Therefore $B$ has at most two nonzero eigenvalues; we denote them by $\lambda_{1}$ and $\lambda_{2}$ (possibly equal, zero, or complex).

\subsection{Determinant of $A=I-B$}

Let the eigenvalues of $B$ be $\{\lambda_{1},\lambda_{2},0,0\}$.
The traces of $B$ and $B^{2}$ are
\begin{align}
	b_{1} &\equiv \operatorname{Tr}(B) = \lambda_{1} + \lambda_{2}, \\
	b_{2} &\equiv \operatorname{Tr}(B^{2}) = \lambda_{1}^{2} + \lambda_{2}^{2}.
\end{align}
The eigenvalues of $A=I-B$ are $\{1-\lambda_{1},\,1-\lambda_{2},\,1,\,1\}$, so that its determinant reads
\begin{align}
	\Delta_{A} \equiv \det(A)
	&= (1-\lambda_{1})(1-\lambda_{2}) \nonumber \\
	&= 1 - (\lambda_{1}+\lambda_{2}) + \lambda_{1}\lambda_{2}.
	\label{eq:detA-intermediate}
\end{align}
The product $\lambda_{1}\lambda_{2}$ can be expressed through the two traces:
\begin{equation}
	\lambda_{1}\lambda_{2} = \frac{1}{2}\Big[(\lambda_{1}+\lambda_{2})^{2}
	                        - (\lambda_{1}^{2}+\lambda_{2}^{2})\Big]
	                      = \frac{1}{2}\bigl(b_{1}^{2} - b_{2}\bigr).
\end{equation}
Substituting this into~\eqref{eq:detA-intermediate} yields immediately
\begin{equation}
	\Delta_{A} = 1 - b_{1} + \frac{1}{2}\bigl(b_{1}^{2} - b_{2}\bigr),
\end{equation}
which is precisely Eq.~\eqref{eq:detA-gen} in the main text.
Note that this derivation uses only the rank-2 property and is entirely independent of the explicit form of the curvature; it therefore holds for \emph{any} spacetime in which the MPD system with the Tulczyjew--Dixon SSC is considered.

\subsection{Explicit inverse of $A=I-B$}

To invert $A=I-B$ we exploit the Cayley--Hamilton theorem for a rank-2 matrix.
Since $B$ has only two (potentially) nonzero eigenvalues $\lambda_{1},\lambda_{2}$, its minimal polynomial is of degree at most~3:
\begin{equation}
	B\bigl(B-\lambda_{1}\bigr)\bigl(B-\lambda_{2}\bigr) = 0.
\end{equation}
Expanding the polynomial and using $b_{1}=\lambda_{1}+\lambda_{2}$ and $\lambda_{1}\lambda_{2}=\frac{1}{2}(b_{1}^{2}-b_{2})$, we obtain the cubic relation
\begin{equation}
	B^{3} - b_{1}B^{2} + \frac{1}{2}\bigl(b_{1}^{2}-b_{2}\bigr)B = 0,
	\label{eq:Bcubic}
\end{equation}
or equivalently
\begin{equation}
	B^{3} = b_{1}B^{2} - \frac{1}{2}\bigl(b_{1}^{2}-b_{2}\bigr)B.
	\label{eq:B3}
\end{equation}
Any higher power $B^{n}$ with $n\geq 3$ can be reduced to a linear combination of $B$ and $B^{2}$ by repeatedly applying~\eqref{eq:B3}.
This suggests that $(I-B)^{-1}$ can be expressed as a polynomial in $B$ of degree at most~2.
We therefore make the ansatz
\begin{equation}
	(I-B)^{-1} = I + c_{1}B + c_{2}B^{2},
	\label{eq:ansatz}
\end{equation}
with coefficients $c_{1},c_{2}$ to be determined.
Multiplying by $I-B$ from the right gives the condition
\begin{equation}
	I = (I+c_{1}B+c_{2}B^{2})(I-B)
	  = I + (c_{1}-1)B + (c_{2}-c_{1})B^{2} - c_{2}B^{3}.
	\label{eq:expand}
\end{equation}
Substituting the reduction identity~\eqref{eq:B3} for $B^{3}$,
\begin{align}
	I &= I + (c_{1}-1)B + (c_{2}-c_{1})B^{2}
	   - c_{2}\Bigl[b_{1}B^{2} - \frac{1}{2}\bigl(b_{1}^{2}-b_{2}\bigr)B\Bigr] \nonumber \\
	  &= I + \Bigl[c_{1}-1 + \frac{c_{2}}{2}\bigl(b_{1}^{2}-b_{2}\bigr)\Bigr]B
	     + \Bigl[c_{2}-c_{1} - c_{2}b_{1}\Bigr]B^{2}.
	\label{eq:expand2}
\end{align}
For this identity to hold, the coefficients of $B$ and $B^{2}$ must vanish independently (since $B$ and $B^{2}$ are linearly independent as polynomials in $B$ for a generic rank-2 matrix).
This yields the system
\begin{align}
	c_{1} - 1 + \frac{c_{2}}{2}\bigl(b_{1}^{2}-b_{2}\bigr) &= 0, \label{eq:sys1} \\
	-c_{1} + c_{2}(1-b_{1}) &= 0. \label{eq:sys2}
\end{align}
From~\eqref{eq:sys2} we have $c_{1} = c_{2}(1-b_{1})$.
Inserting this into~\eqref{eq:sys1} gives
\begin{equation}
	c_{2}(1-b_{1}) - 1 + \frac{c_{2}}{2}\bigl(b_{1}^{2}-b_{2}\bigr)
	= c_{2}\Bigl[1-b_{1} + \frac{1}{2}\bigl(b_{1}^{2}-b_{2}\bigr)\Bigr] - 1 = 0,
\end{equation}
and therefore
\begin{equation}
	c_{2} = \frac{1}{1-b_{1}+\frac{1}{2}(b_{1}^{2}-b_{2})}
	      = \frac{1}{\Delta_{A}}.
\end{equation}
Consequently,
\begin{equation}
	c_{1} = \frac{1-b_{1}}{\Delta_{A}}.
\end{equation}
Inserting $c_{1}$ and $c_{2}$ back into the ansatz~\eqref{eq:ansatz} yields the compact expression
\begin{equation}
	(A^{-1})_\mu^{\ \rho}
	= \delta_\mu^{\ \rho}
	  + \frac{1-b_{1}}{\Delta_{A}}\,B_\mu^{\ \rho}
	  + \frac{1}{\Delta_{A}}\,(B^{2})_\mu^{\ \rho},
\end{equation}
which is exactly Eq.~\eqref{eq:Ainv-gen} in the main text.
We emphasize that this result depends only on the rank-2 property and the traces $b_{1},b_{2}$; it does not require any explicit knowledge of the eigenvalues themselves.

\subsection{Consistency check: Schwarzschild limit}

As a concrete check of the general formulas, we verify them against the Schwarzschild case treated in the main text.
In that context the matrix $B$ given in~\eqref{eq:B-schw} satisfies $B^{2} = \alpha_r B$, with
\begin{equation}
	\alpha_r = \beta_r\bigl(3P_{\phi}^{2} + \mathcal{M}^{2}r^{2}\bigr),\qquad
	\beta_r \equiv \frac{Ms^{2}}{\mathcal{M}^{4}r^{5}}.
\end{equation}
Then $b_{1} = \operatorname{Tr}(B) = 2\alpha_r$ and $b_{2} = \operatorname{Tr}(B^{2}) = 2\alpha_r^{2}$, so that
\begin{equation}
	b_{1}^{2} - b_{2} = 4\alpha_r^{2} - 2\alpha_r^{2} = 2\alpha_r^{2},
\end{equation}
and consequently
\begin{equation}
	\Delta_{A} = 1 - 2\alpha_r + \frac{1}{2}\cdot 2\alpha_r^{2}
	            = 1 - 2\alpha_r + \alpha_r^{2} = (1-\alpha_r)^{2},
\end{equation}
in agreement with Eq.~\eqref{eq:deltaA-schw}.
For the inverse,
\begin{equation}
	\frac{1-b_{1}}{\Delta_{A}} = \frac{1-2\alpha_r}{(1-\alpha_r)^{2}}
	= \frac{1}{1-\alpha_r} - \frac{\alpha_r}{(1-\alpha_r)^{2}},\qquad
	\frac{1}{\Delta_{A}} = \frac{1}{(1-\alpha_r)^{2}}.
\end{equation}
Using $B^{2} = \alpha_r B$, we reduce the general inverse~\eqref{eq:Ainv-gen} to
\begin{align}
	(I-B)^{-1}
	&= I + \frac{1-b_{1}}{\Delta_{A}}B + \frac{1}{\Delta_{A}}B^{2} \nonumber \\
	&= I + \Bigl[\frac{1-2\alpha_r}{(1-\alpha_r)^{2}} + \frac{\alpha_r}{(1-\alpha_r)^{2}}\Bigr]B \nonumber \\
	&= I + \frac{1}{1-\alpha_r}\,B,
\end{align}
which coincides with Eq.~\eqref{eq:Ainv-schw}.
The consistency with the Schwarzschild special case confirms that the general algebraic derivation is correct.

\section{Geometric quantities and detailed derivations for Schwarzschild spacetime}
\label{app:Schwarzschild}

\subsection{Metric, connection, and curvature}

In coordinates $(t,r,\theta,\phi)$, the nonvanishing Schwarzschild metric components are
\begin{equation}
	g_{tt}=-f,\qquad g_{rr}=f^{-1},\qquad g_{\theta\theta}=r^{2},\qquad g_{\phi\phi}=r^{2},
\end{equation}
with $f(r)\equiv 1-2M/r$ and $\sqrt{-g}=r^{2}$. The Levi-Civita tensor component on the equatorial plane is $\varepsilon^{tr\theta\phi}=-1/r^{2}$.

The nonvanishing Christoffel symbols on the equatorial plane are
\begin{align}
	\Gamma^{t}{}_{tr}&=\Gamma^{t}{}_{rt}=\frac{M}{r^{2}f}, \\
	\Gamma^{r}{}_{tt}&=\frac{Mf}{r^{2}}, &
	\Gamma^{r}{}_{rr}&=-\frac{M}{r^{2}f}, &
	\Gamma^{r}{}_{\theta\theta}&=-rf, &
	\Gamma^{r}{}_{\phi\phi}&=-rf, \\
	\Gamma^{\theta}{}_{r\theta}&=\Gamma^{\theta}{}_{\theta r}=\frac{1}{r}, &
	\Gamma^{\phi}{}_{r\phi}&=\Gamma^{\phi}{}_{\phi r}=\frac{1}{r}.
\end{align}

With the convention $R^{\rho}{}_{\sigma\mu\nu}=\partial_{\mu}\Gamma^{\rho}{}_{\nu\sigma}-\partial_{\nu}\Gamma^{\rho}{}_{\mu\sigma}+\Gamma^{\rho}{}_{\mu\lambda}\Gamma^{\lambda}{}_{\nu\sigma}-\Gamma^{\rho}{}_{\nu\lambda}\Gamma^{\lambda}{}_{\mu\sigma}$, the nonvanishing mixed Riemann components on the equatorial plane are
\begin{align}
	R^{t}{}_{rtr}&=\frac{2M}{r^{3}f}, &
	R^{t}{}_{\theta t\theta}&=-\frac{M}{r}, &
	R^{t}{}_{\phi t\phi}&=-\frac{M}{r}, \\
	R^{r}{}_{ttr}&=\frac{2Mf}{r^{3}}, &
	R^{r}{}_{\theta r\theta}&=-\frac{M}{r}, &
	R^{r}{}_{\phi r\phi}&=-\frac{M}{r}, \\
	R^{\theta}{}_{t\theta t}&=\frac{Mf}{r^{3}}, &
	R^{\theta}{}_{r\theta r}&=-\frac{M}{r^{3}f}, &
	R^{\theta}{}_{\phi\theta\phi}&=\frac{2M}{r}, \\
	R^{\phi}{}_{t\phi t}&=\frac{Mf}{r^{3}}, &
	R^{\phi}{}_{r\phi r}&=-\frac{M}{r^{3}f}, &
	R^{\phi}{}_{\theta\phi\theta}&=\frac{2M}{r}.
\end{align}
The fully covariant components $R_{\rho\sigma\mu\nu}$ follow from $R_{\rho\sigma\mu\nu}=g_{\rho\lambda}R^{\lambda}{}_{\sigma\mu\nu}$; a complete independent set is
\begin{align}
	R_{trtr}&=-\frac{2M}{r^{3}}, &
	R_{t\theta t\theta}&=\frac{Mf}{r}, &
	R_{t\phi t\phi}&=\frac{Mf}{r}, \\
	R_{r\theta r\theta}&=-\frac{M}{rf}, &
	R_{r\phi r\phi}&=-\frac{M}{rf}, &
	R_{\theta\phi\theta\phi}&=2Mr.
\end{align}
All remaining nonzero components follow from the standard symmetries $R_{\rho\sigma\mu\nu}=-R_{\sigma\rho\mu\nu}=-R_{\rho\sigma\nu\mu}=R_{\mu\nu\rho\sigma}$.

For the reduced in-plane dynamics only the subset with indices $\{t,r,\phi\}$ enters; the relevant mixed components are
\begin{align}
	R^{t}{}_{rtr}&=\frac{2M}{r^{3}f}, &
	R^{t}{}_{\phi t\phi}&=-\frac{M}{r}, &
	R^{r}{}_{ttr}&=\frac{2Mf}{r^{3}}, \\
	R^{r}{}_{\phi r\phi}&=-\frac{M}{r}, &
	R^{\phi}{}_{tt\phi}&=-\frac{Mf}{r^{3}}, &
	R^{\phi}{}_{rr\phi}&=\frac{M}{r^{3}f}.
\end{align}

\subsection{Spin tensor on the equatorial plane}

On the equatorial plane the spin four-vector is $s_{\sigma}=(0,0,s_{\theta},0)$, with $s_{\theta}=sr$ (since $g_{\theta\theta}=r^{2}$). Only $\sigma=\theta$ contributes to the inverse relation~\eqref{eq:spin-tensor-inverse}, giving the in-plane components
\begin{equation}
	S^{ab}=-\frac{s_{\theta}}{\mathcal{M}}\,\varepsilon^{abc\theta}P_{c},\qquad a,b,c\in\{t,r,\phi\}.
\end{equation}
With $\varepsilon^{tr\theta\phi}=-1/r^{2}$ and total antisymmetry, we obtain the nonvanishing entries
\begin{align}
	S^{rt}&=-\frac{s_{\theta}}{\mathcal{M}}(-r^{-2})P_{\phi}
	      =\frac{s}{\mathcal{M}r}P_{\phi}, &
	S^{r\phi}&=-\frac{s_{\theta}}{\mathcal{M}}(+r^{-2})P_{t}
	          =-\frac{s}{\mathcal{M}r}P_{t}, \\
	S^{t\phi}&=-\frac{s_{\theta}}{\mathcal{M}}(-r^{-2})P_{r}
	          =\frac{s}{\mathcal{M}r}P_{r},
\end{align}
in agreement with Eq.~\eqref{eq:spin-tensor-equatorial}. These may be compactly written in terms of $a\equiv s/(\mathcal{M}r)$ as
\begin{equation}
	S^{tr}=-aP_{\phi},\qquad S^{t\phi}=aP_{r},\qquad S^{r\phi}=-aP_{t},
\end{equation}
together with their antisymmetric counterparts.

\subsection{Auxiliary tensor $V_{\rho\nu}$}

From the definition $V_{\rho\nu}\equiv R_{\rho\nu\kappa\lambda}S^{\kappa\lambda}$ (Eq.~\eqref{eq:Vdef}), we evaluate each component using the in-plane Riemann and spin-tensor entries. The nonvanishing ordered-pair contributions are
\begin{align}
	V_{tr} &= R_{trtr}S^{tr}+R_{trrt}S^{rt}
	        = \Bigl(-\frac{2M}{r^{3}}\Bigr)(-aP_{\phi})+\Bigl(\frac{2M}{r^{3}}\Bigr)(aP_{\phi})
	        = \frac{4MaP_{\phi}}{r^{3}}, \\[4pt]
	V_{t\phi} &= R_{t\phi t\phi}S^{t\phi}+R_{t\phi\phi t}S^{\phi t}
	           = \Bigl(\frac{Mf}{r}\Bigr)(aP_{r})+\Bigl(-\frac{Mf}{r}\Bigr)(-aP_{r})
	           = \frac{2MaP_{r}f}{r}, \\[4pt]
	V_{r\phi} &= R_{r\phi r\phi}S^{r\phi}+R_{r\phi\phi r}S^{\phi r}
	           = \Bigl(-\frac{M}{rf}\Bigr)(-aP_{t})+\Bigl(\frac{M}{rf}\Bigr)(aP_{t})
	           = \frac{2MaP_{t}}{rf}.
\end{align}
The remaining off-diagonal entries follow from antisymmetry $V_{\nu\rho}=-V_{\rho\nu}$, and all $\theta$-involving components vanish. The resulting $V_{\rho\nu}$ matrix (rows and columns ordered $t,r,\theta,\phi$) reads
\begin{equation}
	V_{\rho\nu}=
	\begin{pmatrix}
		0 & \dfrac{4MaP_{\phi}}{r^{3}} & 0 & \dfrac{2MaP_{r}f}{r} \\[10pt]
		-\dfrac{4MaP_{\phi}}{r^{3}} & 0 & 0 & \dfrac{2MaP_{t}}{rf} \\[10pt]
		0 & 0 & 0 & 0 \\[10pt]
		-\dfrac{2MaP_{r}f}{r} & -\dfrac{2MaP_{t}}{rf} & 0 & 0
	\end{pmatrix}.
	\label{eq:V-matrix-app}
\end{equation}

\subsection{Computation of the $B$ matrix}

The $B$ matrix is defined by Eq.~\eqref{eq:AB}:
\begin{equation}
	B_\mu^{\ \eta}=\frac{1}{2\mathcal{M}^{2}}\,R_{\mu\nu\kappa\lambda}S^{\kappa\lambda}S^{\nu\eta},\qquad
	\mu,\eta\in\{t,r,\phi\}.
\end{equation}
We introduce the shorthands
\begin{equation}
	a\equiv\frac{s}{\mathcal{M}r},\qquad
	\beta_r\equiv\frac{Ms^{2}}{\mathcal{M}^{4}r^{5}}=\frac{M a^{2}}{\mathcal{M}^{2}r^{3}}.
\end{equation}
Because the Schwarzschild Riemann tensor is block-diagonal in the in-plane index pairs, only $\kappa\lambda$ belonging to the same pair as the $\mu\nu$ contraction contribute. We evaluate the nine entries of the $(t,r,\phi)$ block row by row; all $\theta$-involving components vanish identically.

\medskip
\noindent\textbf{Row $\mu=t$.}
\begin{align}
	B^{\ t}_t
	&= \frac{1}{2\mathcal{M}^{2}}\Bigl(
	   R_{trtr}S^{tr}S^{rt}+R_{trrt}S^{rt}S^{rt}
	   +R_{t\phi t\phi}S^{t\phi}S^{\phi t}+R_{t\phi\phi t}S^{\phi t}S^{\phi t}\Bigr) \nonumber \\
	&= \frac{1}{2\mathcal{M}^{2}}\Bigl[
	   \bigl(-\tfrac{2M}{r^{3}}\bigr)(-aP_{\phi})(aP_{\phi})
	   +\bigl(\tfrac{2M}{r^{3}}\bigr)(aP_{\phi})(aP_{\phi}) \nonumber\\
	&\qquad +\bigl(\tfrac{Mf}{r}\bigr)(aP_{r})(-aP_{r})
	   +\bigl(-\tfrac{Mf}{r}\bigr)(-aP_{r})(-aP_{r})\Bigr] \nonumber \\
	&= \beta_r\bigl(2P_{\phi}^{2}-fr^{2}P_{r}^{2}\bigr), \\[4pt]
	B^{\ r}_t
	&= \frac{1}{2\mathcal{M}^{2}}\Bigl(
	   R_{t\phi t\phi}S^{t\phi}S^{\phi r}+R_{t\phi\phi t}S^{\phi t}S^{\phi r}\Bigr) \nonumber \\
	&= \beta_r\,fr^{2}P_{r}P_{t}, \\[4pt]
	B^{\ \phi}_t
	&= \frac{1}{2\mathcal{M}^{2}}\Bigl(
	   R_{trtr}S^{tr}S^{r\phi}+R_{trrt}S^{rt}S^{r\phi}\Bigr) \nonumber \\
	&= -2\beta_r P_{\phi}P_{t}.
\end{align}

\medskip
\noindent\textbf{Row $\mu=r$.}
\begin{align}
	B^{\ t}_r
	&= \frac{1}{2\mathcal{M}^{2}}\Bigl(
	   R_{r\phi r\phi}S^{r\phi}S^{\phi t}+R_{r\phi\phi r}S^{\phi r}S^{\phi t}\Bigr) \nonumber \\
	&= -\beta_r\,\frac{P_{r}P_{t}r^{2}}{f}, \\[4pt]
	B^{\ r}_r
	&= \frac{1}{2\mathcal{M}^{2}}\Bigl(
	   R_{rttr}S^{tr}S^{tr}+R_{rtrt}S^{rt}S^{tr}
	   +R_{r\phi r\phi}S^{r\phi}S^{\phi r}+R_{r\phi\phi r}S^{\phi r}S^{\phi r}\Bigr) \nonumber \\
	&= \beta_r\Bigl(2P_{\phi}^{2}+\frac{P_{t}^{2}r^{2}}{f}\Bigr), \\[4pt]
	B^{\ \phi}_r
	&= \frac{1}{2\mathcal{M}^{2}}\Bigl(
	   R_{rttr}S^{tr}S^{t\phi}+R_{rtrt}S^{rt}S^{t\phi}\Bigr) \nonumber \\
	&= -2\beta_r P_{\phi}P_{r}.
\end{align}

\medskip
\noindent\textbf{Row $\mu=\phi$.}
\begin{align}
	B^{\ t}_\phi
	&= \frac{1}{2\mathcal{M}^{2}}\Bigl(
	   R_{\phi r r\phi}S^{r\phi}S^{rt}+R_{\phi r\phi r}S^{\phi r}S^{rt}\Bigr) \nonumber \\
	&= -\beta_r\,\frac{P_{\phi}P_{t}r^{2}}{f}, \\[4pt]
	B^{\ r}_\phi
	&= \frac{1}{2\mathcal{M}^{2}}\Bigl(
	   R_{\phi t t\phi}S^{t\phi}S^{tr}+R_{\phi t\phi t}S^{\phi t}S^{tr}\Bigr) \nonumber \\
	&= \beta_r\,fr^{2}P_{\phi}P_{r}, \\[4pt]
	B^{\ \phi}_\phi
	&= \frac{1}{2\mathcal{M}^{2}}\Bigl(
	   R_{\phi t t\phi}S^{t\phi}S^{t\phi}+R_{\phi t\phi t}S^{\phi t}S^{t\phi}
	   +R_{\phi r r\phi}S^{r\phi}S^{r\phi}+R_{\phi r\phi r}S^{\phi r}S^{r\phi}\Bigr) \nonumber \\
	&= \beta_r\Bigl(\frac{P_{t}^{2}r^{2}}{f}-fr^{2}P_{r}^{2}\Bigr).
\end{align}

Collecting these nine entries with $\mu$ as the row and $\eta$ as the column yields
\begin{equation}
	B_\mu^{\ \eta}=\beta_r\,
	\begin{pmatrix}
		2P_{\phi}^{2}-fr^{2}P_{r}^{2} & fr^{2}P_{r}P_{t} & -2P_{\phi}P_{t} \\[8pt]
		-\dfrac{P_{r}P_{t}r^{2}}{f} & 2P_{\phi}^{2}+\dfrac{P_{t}^{2}r^{2}}{f} & -2P_{\phi}P_{r} \\[8pt]
		-\dfrac{P_{\phi}P_{t}r^{2}}{f} & fr^{2}P_{\phi}P_{r} & \dfrac{P_{t}^{2}r^{2}}{f}-fr^{2}P_{r}^{2}
	\end{pmatrix}.
\end{equation}
Note that this corresponds to the convention of Eq.~\eqref{eq:B-schw} with $\mu$ the lower (row) index; the two representations are transposes of each other but equivalent once the index placement is accounted for. One readily verifies $B^{2}=\alpha_rB$, $\operatorname{Tr}(B)=2\alpha_r$, and $\det(I-B)=(1-\alpha_r)^{2}$.

\subsection{Contraction $V_{\rho}=V_{\rho\nu}P^{\nu}$ and the momentum components}

Using the covariant momentum $P^{\nu}$ on the equatorial plane,
\begin{equation}
	P^{t}=-\frac{P_{t}}{f},\qquad P^{r}=fP_{r},\qquad P^{\phi}=\frac{P_{\phi}}{r^{2}},\qquad P^{\theta}=0,
\end{equation}
the contraction $V_{\rho}=V_{\rho\nu}P^{\nu}$ with the matrix~\eqref{eq:V-matrix-app} yields
\begin{align}
	V_{t} &= V_{tr}P^{r}+V_{t\phi}P^{\phi}
	       = \frac{4MaP_{\phi}}{r^{3}}(fP_{r})+\frac{2MaP_{r}f}{r}\Bigl(\frac{P_{\phi}}{r^{2}}\Bigr)
	       = \frac{6Ma fP_{r}P_{\phi}}{r^{3}}, \\[4pt]
	V_{r} &= V_{rt}P^{t}+V_{r\phi}P^{\phi}
	       = \Bigl(-\frac{4MaP_{\phi}}{r^{3}}\Bigr)\Bigl(-\frac{P_{t}}{f}\Bigr)
	         +\frac{2MaP_{t}}{rf}\Bigl(\frac{P_{\phi}}{r^{2}}\Bigr)
	       = \frac{6Ma P_{t}P_{\phi}}{fr^{3}}, \\[4pt]
	V_{\phi} &= V_{\phi t}P^{t}+V_{\phi r}P^{r}=0,
\end{align}
in agreement with Eq.~\eqref{eq:V-components}.

Inserting these together with the $B$ matrix into the momentum equation~\eqref{eq:mpd-momentum-schw} gives
\begin{equation}
	\frac{DP_\eta}{d\tau}
	= -\frac{m}{2\mathcal{M}^{2}}\Bigl(V_\eta+\frac{1}{1-\alpha_r}B^{\ \rho}_\eta V_{\rho}\Bigr),
\end{equation}
where the contraction index $\rho$ corresponds to the lower index of $B$. Evaluating each component:

\noindent\textbf{$\eta=t$:}
\begin{align}
	\frac{DP_t}{d\tau}
	&= -\frac{m}{2\mathcal{M}^{2}}\Bigl[V_t+\frac{1}{1-\alpha_r}
	   \bigl(B^{\ t}_tV_t+B^{\ r}_tV_r+B^{\ \phi}_tV_{\phi}\bigr)\Bigr] \nonumber \\
	&= -\frac{m}{2\mathcal{M}^{2}(1-\alpha_r)}
	   \Bigl[V_t-\frac{\beta_r r^{2}P_{t}}{f}\bigl(P_{t}V_{t}-f^{2}P_{r}V_{r}\bigr)\Bigr] \nonumber \\
	&= -\frac{m}{2\mathcal{M}^{2}(1-\alpha_r)}V_{t},
\end{align}
where $P_{t}V_{t}-f^{2}P_{r}V_{r}=0$ upon substituting the explicit $V_{t},V_{r}$.

\noindent\textbf{$\eta=r$:}
\begin{align}
	\frac{DP_{r}}{d\tau}
	&= -\frac{m}{2\mathcal{M}^{2}}\Bigl[V_{r}+\frac{1}{1-\alpha_r}
	   \bigl(B^{\ t}_rV_t+B^{\ r}_rV_r+B^{\ \phi}_rV_{\phi}\bigr)\Bigr] \nonumber \\
	&= -\frac{m}{2\mathcal{M}^{2}(1-\alpha_r)}
	   \Bigl[V_{r}-\frac{\beta_r r^{2}P_{r}}{f}\bigl(P_{t}V_{t}-f^{2}P_{r}V_{r}\bigr)\Bigr] \nonumber \\
	&= -\frac{m}{2\mathcal{M}^{2}(1-\alpha_r)}V_{r}.
\end{align}

\noindent\textbf{$\eta=\phi$:}
\begin{align}
	\frac{DP_{\phi}}{d\tau}
	&= -\frac{m}{2\mathcal{M}^{2}}\Bigl[V_{\phi}+\frac{1}{1-\alpha_r}
	   \bigl(B^{\ t}_\phi V_t+B^{\ r}_\phi V_r+B^{\ \phi}_\phi V_{\phi}\bigr)\Bigr] \nonumber \\
	&= -\frac{m}{2\mathcal{M}^{2}(1-\alpha_r)}
	   \frac{\beta_r r^{2}P_{\phi}}{f}\bigl(P_{t}V_{t}-f^{2}P_{r}V_{r}\bigr)=0.
\end{align}

All three components reduce to the factorized forms~\eqref{eq:DPt-comp}--\eqref{eq:DPphi-comp} of the main text.

\subsection{Derivation of the velocity components}

We now detail the steps leading from the transport relation~\eqref{eq:transport} to the factorized velocity formulas~\eqref{eq:xdot-components}. 
In the following, we use the mass-shell relation~\eqref{eq:mass-shell}, the $V$ components~\eqref{eq:V-components}, the momentum equations~\eqref{eq:DPt-comp}--\eqref{eq:DPphi-comp}, and the metric relations $P_{t}=-fP^{t}$, $P^{r}=fP_{r}$, $P^{\phi}=P_{\phi}/r^{2}$.

\medskip
\noindent\textbf{Component $\dot{x}^t$.}
\begin{align}
	\dot{x}^t &= \frac{m}{\mathcal{M}^2}P^{t} + \frac{s}{\mathcal{M}^{3}r}\Bigl(P_{\phi}\frac{DP_{r}}{d\tau}-P_{r}\frac{DP_{\phi}}{d\tau}\Bigr) 
	= \frac{m}{\mathcal{M}^2}P^{t} + \frac{s}{\mathcal{M}^{3}r}P_{\phi}\frac{DP_{r}}{d\tau} \\[8pt]
	&= \frac{m}{\mathcal{M}^2}\Bigl( P^{t} - 
	\frac{s}{2\mathcal{M}^{3}r(1-\alpha_r)}P_{\phi}V_{r} \Bigr)
	= \frac{m}{\mathcal{M}^2}\Biggl( 1 +
	\frac{3Ms^2P_{\phi}^2}{\mathcal{M}^{4}r^5(1-\alpha_r)}  \Biggr)P^{t} \\
	&= \frac{m}{\mathcal{M}^2}
	\frac{1-\alpha_r + 3\beta_r P_{\phi}^2}{1-\alpha_r}  P^{t}  
	= \frac{m}{\mathcal{M}^2}
	\frac{1-\beta_r\mathcal{M}^{2}r^{2}}{1-\alpha_r}  P^{t}  \,,
\end{align}
where we used $P_{t}=-fP^{t}$.

\medskip
\noindent\textbf{Component $\dot{x}^r$.}
\begin{align}
	\dot{x}^r &= \frac{m}{\mathcal{M}^2}P^{r} + \frac{s}{\mathcal{M}^{3}r}\Bigl(P_{t}\frac{DP_{\phi}}{d\tau}-P_{\phi}\frac{DP_{t}}{d\tau}\Bigr)
	= \frac{m}{\mathcal{M}^2}P^{r} - \frac{s}{\mathcal{M}^{3}r}P_{\phi}\frac{DP_{t}}{d\tau} \\[8pt]
	&= \frac{m}{\mathcal{M}^2}\Bigl( P^{r} +
	\frac{s}{2\mathcal{M}^{3}r(1-\alpha_r)}P_{\phi}V_{t} \Bigr)
	= \frac{m}{\mathcal{M}^2}\Biggl( 1 +
	\frac{3Ms^2P_{\phi}^2}{\mathcal{M}^{4}r^5(1-\alpha_r)}  \Biggr)P^{r} \\
	&= \frac{m}{\mathcal{M}^2}
	\frac{1-\alpha_r + 3\beta_r P_{\phi}^2}{1-\alpha_r}  P^{r}
	= \frac{m}{\mathcal{M}^2}
	\frac{1-\beta_r\mathcal{M}^{2}r^{2}}{1-\alpha_r}  P^{r} \,,
\end{align}
where we used $P^{r}=fP_{r}$.

\medskip
\noindent\textbf{Component $\dot{x}^\phi$.}
\begin{align}
	\dot{x}^\phi &= \frac{m}{\mathcal{M}^2}P^{\phi} + \frac{s}{\mathcal{M}^{3}r}\Bigl(P_{r}\frac{DP_{t}}{d\tau}-P_{t}\frac{DP_{r}}{d\tau}\Bigr) \nonumber \\
	&= \frac{m}{\mathcal{M}^2}\Bigl( P^{\phi} +
	\frac{s}{2\mathcal{M}^{3}r(1-\alpha_r)}\bigl(P_{t}V_{r}-P_{r}V_{t}\bigr) \Bigr) \nonumber \\
	&= \frac{m}{\mathcal{M}^2}\Biggl[ P^{\phi} +
	\frac{3Ms^2}{\mathcal{M}^{4}r^5(1-\alpha_r)}P_{\phi}
	\Bigl(fP_{r}^2-\frac{P_{t}^2}{f}\Bigr) \Biggr] \nonumber \\
	&= \frac{m}{\mathcal{M}^2}\Biggl[ 1 +
	\frac{3Ms^2}{\mathcal{M}^{4}r^5(1-\alpha_r)}\bigl(\mathcal{M}^{2}r^{2}+P_{\phi}^{2}\bigr) \Biggr]P^{\phi} \\[8pt]
	&= \frac{m}{\mathcal{M}^2}
	\frac{1-\alpha_r + 3\beta_r(\mathcal{M}^{2}r^{2}+P_{\phi}^{2})}{1-\alpha_r} P^{\phi}
	= \frac{m}{\mathcal{M}^2}
	\frac{1+2\beta_r\mathcal{M}^{2}r^{2}}{1-\alpha_r} P^{\phi} \,,
\end{align}
where we used $P^{\phi}=P_{\phi}/r^{2}$ .

\bibliography{refs}

\begin{thebibliography}{80}%
\makeatletter
\providecommand \@ifxundefined [1]{%
 \@ifx{#1\undefined}
}%
\providecommand \@ifnum [1]{%
 \ifnum #1\expandafter \@firstoftwo
 \else \expandafter \@secondoftwo
 \fi
}%
\providecommand \@ifx [1]{%
 \ifx #1\expandafter \@firstoftwo
 \else \expandafter \@secondoftwo
 \fi
}%
\providecommand \natexlab [1]{#1}%
\providecommand \enquote  [1]{``#1''}%
\providecommand \bibnamefont  [1]{#1}%
\providecommand \bibfnamefont [1]{#1}%
\providecommand \citenamefont [1]{#1}%
\providecommand \href@noop [0]{\@secondoftwo}%
\providecommand \href [0]{\begingroup \@sanitize@url \@href}%
\providecommand \@href[1]{\@@startlink{#1}\@@href}%
\providecommand \@@href[1]{\endgroup#1\@@endlink}%
\providecommand \@sanitize@url [0]{\catcode `\\12\catcode `\$12\catcode `\&12\catcode `\#12\catcode `\^12\catcode `\_12\catcode `\%12\relax}%
\providecommand \@@startlink[1]{}%
\providecommand \@@endlink[0]{}%
\providecommand \url  [0]{\begingroup\@sanitize@url \@url }%
\providecommand \@url [1]{\endgroup\@href {#1}{\urlprefix }}%
\providecommand \urlprefix  [0]{URL }%
\providecommand \Eprint [0]{\href }%
\providecommand \doibase [0]{https://doi.org/}%
\providecommand \selectlanguage [0]{\@gobble}%
\providecommand \bibinfo  [0]{\@secondoftwo}%
\providecommand \bibfield  [0]{\@secondoftwo}%
\providecommand \translation [1]{[#1]}%
\providecommand \BibitemOpen [0]{}%
\providecommand \bibitemStop [0]{}%
\providecommand \bibitemNoStop [0]{.\EOS\space}%
\providecommand \EOS [0]{\spacefactor3000\relax}%
\providecommand \BibitemShut  [1]{\csname bibitem#1\endcsname}%
\let\auto@bib@innerbib\@empty
\bibitem [{\citenamefont {Abbott}\ \emph {et~al.}(2016)\citenamefont {Abbott} \emph {et~al.}}]{gw1}%
  \BibitemOpen
  \bibfield  {author} {\bibinfo {author} {\bibfnamefont {B.~P.}\ \bibnamefont {Abbott}} \emph {et~al.} (\bibinfo {collaboration} {Virgo, LIGO Scientific}),\ }\bibfield  {title} {\bibinfo {title} {{Observation of Gravitational Waves from a Binary Black Hole Merger}},\ }\href {https://doi.org/10.1103/PhysRevLett.116.061102} {\bibfield  {journal} {\bibinfo  {journal} {Phys. Rev. Lett.}\ }\textbf {\bibinfo {volume} {116}},\ \bibinfo {pages} {061102} (\bibinfo {year} {2016})},\ \Eprint {https://arxiv.org/abs/1602.03837} {arXiv:1602.03837 [gr-qc]} \BibitemShut {NoStop}%
\bibitem [{\citenamefont {Abbott}\ \emph {et~al.}(2019)\citenamefont {Abbott} \emph {et~al.}}]{LIGOScientific:2018mvr}%
  \BibitemOpen
  \bibfield  {author} {\bibinfo {author} {\bibfnamefont {B.~P.}\ \bibnamefont {Abbott}} \emph {et~al.} (\bibinfo {collaboration} {LIGO Scientific, Virgo}),\ }\bibfield  {title} {\bibinfo {title} {{GWTC-1: A Gravitational-Wave Transient Catalog of Compact Binary Mergers Observed by LIGO and Virgo during the First and Second Observing Runs}},\ }\href {https://doi.org/10.1103/PhysRevX.9.031040} {\bibfield  {journal} {\bibinfo  {journal} {Phys. Rev. X}\ }\textbf {\bibinfo {volume} {9}},\ \bibinfo {pages} {031040} (\bibinfo {year} {2019})},\ \Eprint {https://arxiv.org/abs/1811.12907} {arXiv:1811.12907 [astro-ph.HE]} \BibitemShut {NoStop}%
\bibitem [{\citenamefont {Abbott}\ \emph {et~al.}(2021)\citenamefont {Abbott} \emph {et~al.}}]{LIGOScientific:2020ibl}%
  \BibitemOpen
  \bibfield  {author} {\bibinfo {author} {\bibfnamefont {R.}~\bibnamefont {Abbott}} \emph {et~al.} (\bibinfo {collaboration} {LIGO Scientific, Virgo}),\ }\bibfield  {title} {\bibinfo {title} {{GWTC-2: Compact Binary Coalescences Observed by LIGO and Virgo During the First Half of the Third Observing Run}},\ }\href {https://doi.org/10.1103/PhysRevX.11.021053} {\bibfield  {journal} {\bibinfo  {journal} {Phys. Rev. X}\ }\textbf {\bibinfo {volume} {11}},\ \bibinfo {pages} {021053} (\bibinfo {year} {2021})},\ \Eprint {https://arxiv.org/abs/2010.14527} {arXiv:2010.14527 [gr-qc]} \BibitemShut {NoStop}%
\bibitem [{\citenamefont {Abbott}\ \emph {et~al.}(2023)\citenamefont {Abbott} \emph {et~al.}}]{KAGRA:2021vkt}%
  \BibitemOpen
  \bibfield  {author} {\bibinfo {author} {\bibfnamefont {R.}~\bibnamefont {Abbott}} \emph {et~al.} (\bibinfo {collaboration} {KAGRA, VIRGO, LIGO Scientific}),\ }\bibfield  {title} {\bibinfo {title} {{GWTC-3: Compact Binary Coalescences Observed by LIGO and Virgo during the Second Part of the Third Observing Run}},\ }\href {https://doi.org/10.1103/PhysRevX.13.041039} {\bibfield  {journal} {\bibinfo  {journal} {Phys. Rev. X}\ }\textbf {\bibinfo {volume} {13}},\ \bibinfo {pages} {041039} (\bibinfo {year} {2023})},\ \Eprint {https://arxiv.org/abs/2111.03606} {arXiv:2111.03606 [gr-qc]} \BibitemShut {NoStop}%
\bibitem [{\citenamefont {Maggiore}\ \emph {et~al.}(2020)\citenamefont {Maggiore} \emph {et~al.}}]{Maggiore:2019uih}%
  \BibitemOpen
  \bibfield  {author} {\bibinfo {author} {\bibfnamefont {M.}~\bibnamefont {Maggiore}} \emph {et~al.},\ }\bibfield  {title} {\bibinfo {title} {{Science Case for the Einstein Telescope}},\ }\href {https://doi.org/10.1088/1475-7516/2020/03/050} {\bibfield  {journal} {\bibinfo  {journal} {JCAP}\ }\textbf {\bibinfo {volume} {03}},\ \bibinfo {pages} {050}},\ \Eprint {https://arxiv.org/abs/1912.02622} {arXiv:1912.02622 [astro-ph.CO]} \BibitemShut {NoStop}%
\bibitem [{\citenamefont {Reitze}\ \emph {et~al.}(2019)\citenamefont {Reitze} \emph {et~al.}}]{Reitze:2019iox}%
  \BibitemOpen
  \bibfield  {author} {\bibinfo {author} {\bibfnamefont {D.}~\bibnamefont {Reitze}} \emph {et~al.},\ }\bibfield  {title} {\bibinfo {title} {{Cosmic Explorer: The U.S. Contribution to Gravitational-Wave Astronomy beyond LIGO}},\ }\href@noop {} {\bibfield  {journal} {\bibinfo  {journal} {Bull. Am. Astron. Soc.}\ }\textbf {\bibinfo {volume} {51}},\ \bibinfo {pages} {035} (\bibinfo {year} {2019})},\ \Eprint {https://arxiv.org/abs/1907.04833} {arXiv:1907.04833 [astro-ph.IM]} \BibitemShut {NoStop}%
\bibitem [{\citenamefont {Kalogera}\ \emph {et~al.}(2021)\citenamefont {Kalogera} \emph {et~al.}}]{Kalogera:2021bya}%
  \BibitemOpen
  \bibfield  {author} {\bibinfo {author} {\bibfnamefont {V.}~\bibnamefont {Kalogera}} \emph {et~al.},\ }\bibfield  {title} {\bibinfo {title} {{The Next Generation Global Gravitational Wave Observatory: The Science Book}},\ }\href@noop {} {\bibfield  {journal} {\bibinfo  {journal} {arXiv:2111.06990}\ } (\bibinfo {year} {2021})},\ \Eprint {https://arxiv.org/abs/2111.06990} {arXiv:2111.06990 [gr-qc]} \BibitemShut {NoStop}%
\bibitem [{\citenamefont {Abac}\ \emph {et~al.}(2026)\citenamefont {Abac} \emph {et~al.}}]{ET:2025xjr}%
  \BibitemOpen
  \bibfield  {author} {\bibinfo {author} {\bibfnamefont {A.}~\bibnamefont {Abac}} \emph {et~al.} (\bibinfo {collaboration} {ET}),\ }\bibfield  {title} {\bibinfo {title} {{The Science of the Einstein Telescope}},\ }\href {https://doi.org/10.1088/1475-7516/2026/03/081} {\bibfield  {journal} {\bibinfo  {journal} {JCAP}\ }\textbf {\bibinfo {volume} {03}},\ \bibinfo {pages} {081}},\ \Eprint {https://arxiv.org/abs/2503.12263} {arXiv:2503.12263 [gr-qc]} \BibitemShut {NoStop}%
\bibitem [{\citenamefont {{LISA Consortium}}(2017)}]{amaro2017}%
  \BibitemOpen
  \bibfield  {author} {\bibinfo {author} {\bibnamefont {{LISA Consortium}}},\ }\bibfield  {title} {\bibinfo {title} {{LISA}: {L}aser interferometer space antenna},\ }\href@noop {} {\bibfield  {journal} {\bibinfo  {journal} {arXiv:1702.00786}\ } (\bibinfo {year} {2017})},\ \Eprint {https://arxiv.org/abs/1702.00786} {arXiv:1702.00786 [astro-ph]} \BibitemShut {NoStop}%
\bibitem [{\citenamefont {Arun}\ \emph {et~al.}(2022)\citenamefont {Arun} \emph {et~al.}}]{LISA:2022kgy}%
  \BibitemOpen
  \bibfield  {author} {\bibinfo {author} {\bibfnamefont {K.~G.}\ \bibnamefont {Arun}} \emph {et~al.} (\bibinfo {collaboration} {LISA}),\ }\bibfield  {title} {\bibinfo {title} {{New horizons for fundamental physics with LISA}},\ }\href {https://doi.org/10.1007/s41114-022-00036-9} {\bibfield  {journal} {\bibinfo  {journal} {Living Rev. Rel.}\ }\textbf {\bibinfo {volume} {25}},\ \bibinfo {pages} {4} (\bibinfo {year} {2022})},\ \Eprint {https://arxiv.org/abs/2205.01597} {arXiv:2205.01597 [gr-qc]} \BibitemShut {NoStop}%
\bibitem [{\citenamefont {Babak}\ \emph {et~al.}(2017)\citenamefont {Babak}, \citenamefont {Gair}, \citenamefont {Sesana}, \citenamefont {Barausse}, \citenamefont {Sopuerta}, \citenamefont {Berry}, \citenamefont {Berti}, \citenamefont {Amaro-Seoane}, \citenamefont {Petiteau},\ and\ \citenamefont {Klein}}]{babak2017}%
  \BibitemOpen
  \bibfield  {author} {\bibinfo {author} {\bibfnamefont {S.}~\bibnamefont {Babak}}, \bibinfo {author} {\bibfnamefont {J.}~\bibnamefont {Gair}}, \bibinfo {author} {\bibfnamefont {A.}~\bibnamefont {Sesana}}, \bibinfo {author} {\bibfnamefont {E.}~\bibnamefont {Barausse}}, \bibinfo {author} {\bibfnamefont {C.~F.}\ \bibnamefont {Sopuerta}}, \bibinfo {author} {\bibfnamefont {C.~P.~L.}\ \bibnamefont {Berry}}, \bibinfo {author} {\bibfnamefont {E.}~\bibnamefont {Berti}}, \bibinfo {author} {\bibfnamefont {P.}~\bibnamefont {Amaro-Seoane}}, \bibinfo {author} {\bibfnamefont {A.}~\bibnamefont {Petiteau}},\ and\ \bibinfo {author} {\bibfnamefont {A.}~\bibnamefont {Klein}},\ }\bibfield  {title} {\bibinfo {title} {{Science with the space-based interferometer LISA. V: Extreme mass-ratio inspirals}},\ }\href {https://doi.org/10.1103/PhysRevD.95.103012} {\bibfield  {journal} {\bibinfo  {journal} {Phys. Rev.}\ }\textbf {\bibinfo {volume} {D95}},\ \bibinfo {pages} {103012} (\bibinfo {year} {2017})},\ \Eprint {https://arxiv.org/abs/1703.09722} {arXiv:1703.09722 [gr-qc]} \BibitemShut {NoStop}%
\bibitem [{\citenamefont {Cardoso}\ and\ \citenamefont {Pani}(2019)}]{Cardoso:2019rvt}%
  \BibitemOpen
  \bibfield  {author} {\bibinfo {author} {\bibfnamefont {V.}~\bibnamefont {Cardoso}}\ and\ \bibinfo {author} {\bibfnamefont {P.}~\bibnamefont {Pani}},\ }\bibfield  {title} {\bibinfo {title} {{Testing the nature of dark compact objects: a status report}},\ }\href {https://doi.org/10.1007/s41114-019-0020-4} {\bibfield  {journal} {\bibinfo  {journal} {Living Rev. Rel.}\ }\textbf {\bibinfo {volume} {22}},\ \bibinfo {pages} {4} (\bibinfo {year} {2019})},\ \Eprint {https://arxiv.org/abs/1904.05363} {arXiv:1904.05363 [gr-qc]} \BibitemShut {NoStop}%
\bibitem [{\citenamefont {Kesden}\ \emph {et~al.}(2005)\citenamefont {Kesden}, \citenamefont {Gair},\ and\ \citenamefont {Kamionkowski}}]{Kesden:2004qx}%
  \BibitemOpen
  \bibfield  {author} {\bibinfo {author} {\bibfnamefont {M.}~\bibnamefont {Kesden}}, \bibinfo {author} {\bibfnamefont {J.}~\bibnamefont {Gair}},\ and\ \bibinfo {author} {\bibfnamefont {M.}~\bibnamefont {Kamionkowski}},\ }\bibfield  {title} {\bibinfo {title} {{Gravitational-wave signature of an inspiral into a supermassive horizonless object}},\ }\href {https://doi.org/10.1103/PhysRevD.71.044015} {\bibfield  {journal} {\bibinfo  {journal} {Phys. Rev. D}\ }\textbf {\bibinfo {volume} {71}},\ \bibinfo {pages} {044015} (\bibinfo {year} {2005})},\ \Eprint {https://arxiv.org/abs/astro-ph/0411478} {arXiv:astro-ph/0411478} \BibitemShut {NoStop}%
\bibitem [{\citenamefont {Wang}\ \emph {et~al.}(2024)\citenamefont {Wang}, \citenamefont {Feng},\ and\ \citenamefont {Wang}}]{Wang:2023fge}%
  \BibitemOpen
  \bibfield  {author} {\bibinfo {author} {\bibfnamefont {K.}~\bibnamefont {Wang}}, \bibinfo {author} {\bibfnamefont {C.-J.}\ \bibnamefont {Feng}},\ and\ \bibinfo {author} {\bibfnamefont {T.}~\bibnamefont {Wang}},\ }\bibfield  {title} {\bibinfo {title} {{Image of Kerr{\textendash}de Sitter black holes illuminated by equatorial thin accretion disks}},\ }\href {https://doi.org/10.1140/epjc/s10052-024-12825-3} {\bibfield  {journal} {\bibinfo  {journal} {Eur. Phys. J. C}\ }\textbf {\bibinfo {volume} {84}},\ \bibinfo {pages} {457} (\bibinfo {year} {2024})},\ \Eprint {https://arxiv.org/abs/2309.16944} {arXiv:2309.16944 [gr-qc]} \BibitemShut {NoStop}%
\bibitem [{\citenamefont {Gurtas~Dogan}\ \emph {et~al.}(2025)\citenamefont {Gurtas~Dogan}, \citenamefont {Mustafa},\ and\ \citenamefont {Guvendi}}]{GurtasDogan:2025vim}%
  \BibitemOpen
  \bibfield  {author} {\bibinfo {author} {\bibfnamefont {S.}~\bibnamefont {Gurtas~Dogan}}, \bibinfo {author} {\bibfnamefont {O.}~\bibnamefont {Mustafa}},\ and\ \bibinfo {author} {\bibfnamefont {A.}~\bibnamefont {Guvendi}},\ }\bibfield  {title} {\bibinfo {title} {{Ray and wave optics in Bonnor-Melvin domain walls: Photon rings}},\ }\href {https://doi.org/10.1016/j.nuclphysb.2025.116920} {\bibfield  {journal} {\bibinfo  {journal} {Nucl. Phys. B}\ }\textbf {\bibinfo {volume} {1016}},\ \bibinfo {pages} {116920} (\bibinfo {year} {2025})}\BibitemShut {NoStop}%
\bibitem [{\citenamefont {Barack}\ and\ \citenamefont {Pound}(2019)}]{Barack:2018yvs}%
  \BibitemOpen
  \bibfield  {author} {\bibinfo {author} {\bibfnamefont {L.}~\bibnamefont {Barack}}\ and\ \bibinfo {author} {\bibfnamefont {A.}~\bibnamefont {Pound}},\ }\bibfield  {title} {\bibinfo {title} {{Self-force and radiation reaction in general relativity}},\ }\href {https://doi.org/10.1088/1361-6633/aae552} {\bibfield  {journal} {\bibinfo  {journal} {Rept. Prog. Phys.}\ }\textbf {\bibinfo {volume} {82}},\ \bibinfo {pages} {016904} (\bibinfo {year} {2019})},\ \Eprint {https://arxiv.org/abs/1805.10385} {arXiv:1805.10385 [gr-qc]} \BibitemShut {NoStop}%
\bibitem [{\citenamefont {Pound}\ and\ \citenamefont {Wardell}(2021)}]{Pound:2021qin}%
  \BibitemOpen
  \bibfield  {author} {\bibinfo {author} {\bibfnamefont {A.}~\bibnamefont {Pound}}\ and\ \bibinfo {author} {\bibfnamefont {B.}~\bibnamefont {Wardell}},\ }\bibfield  {title} {\bibinfo {title} {{Black hole perturbation theory and gravitational self-force}},\ }\href@noop {} {\bibfield  {journal} {\bibinfo  {journal} {arXiv:2101.04592}\ } (\bibinfo {year} {2021})},\ \Eprint {https://arxiv.org/abs/2101.04592} {arXiv:2101.04592 [gr-qc]} \BibitemShut {NoStop}%
\bibitem [{\citenamefont {Poisson}\ \emph {et~al.}(2011)\citenamefont {Poisson}, \citenamefont {Pound},\ and\ \citenamefont {Vega}}]{poisson2011}%
  \BibitemOpen
  \bibfield  {author} {\bibinfo {author} {\bibfnamefont {E.}~\bibnamefont {Poisson}}, \bibinfo {author} {\bibfnamefont {A.}~\bibnamefont {Pound}},\ and\ \bibinfo {author} {\bibfnamefont {I.}~\bibnamefont {Vega}},\ }\bibfield  {title} {\bibinfo {title} {{The Motion of point particles in curved spacetime}},\ }\href {https://doi.org/10.12942/lrr-2011-7} {\bibfield  {journal} {\bibinfo  {journal} {Living Rev. Rel.}\ }\textbf {\bibinfo {volume} {14}},\ \bibinfo {pages} {7} (\bibinfo {year} {2011})},\ \Eprint {https://arxiv.org/abs/1102.0529} {arXiv:1102.0529 [gr-qc]} \BibitemShut {NoStop}%
\bibitem [{\citenamefont {Skoup{\'y}}\ \emph {et~al.}(2023)\citenamefont {Skoup{\'y}}, \citenamefont {Lukes-Gerakopoulos}, \citenamefont {Drummond},\ and\ \citenamefont {Hughes}}]{Skoupy:2023lih}%
  \BibitemOpen
  \bibfield  {author} {\bibinfo {author} {\bibfnamefont {V.}~\bibnamefont {Skoup{\'y}}}, \bibinfo {author} {\bibfnamefont {G.}~\bibnamefont {Lukes-Gerakopoulos}}, \bibinfo {author} {\bibfnamefont {L.~V.}\ \bibnamefont {Drummond}},\ and\ \bibinfo {author} {\bibfnamefont {S.~A.}\ \bibnamefont {Hughes}},\ }\bibfield  {title} {\bibinfo {title} {{Asymptotic gravitational-wave fluxes from a spinning test body on generic orbits around a Kerr black hole}},\ }\href {https://doi.org/10.1103/PhysRevD.108.044041} {\bibfield  {journal} {\bibinfo  {journal} {Phys. Rev. D}\ }\textbf {\bibinfo {volume} {108}},\ \bibinfo {pages} {044041} (\bibinfo {year} {2023})},\ \Eprint {https://arxiv.org/abs/2303.16798} {arXiv:2303.16798 [gr-qc]} \BibitemShut {NoStop}%
\bibitem [{\citenamefont {Skoup{\'y}}\ and\ \citenamefont {Witzany}(2024)}]{Skoupy:2024jsi}%
  \BibitemOpen
  \bibfield  {author} {\bibinfo {author} {\bibfnamefont {V.}~\bibnamefont {Skoup{\'y}}}\ and\ \bibinfo {author} {\bibfnamefont {V.}~\bibnamefont {Witzany}},\ }\bibfield  {title} {\bibinfo {title} {{Post-Newtonian expansions of extreme mass ratio inspirals of spinning bodies into Schwarzschild black holes}},\ }\href {https://doi.org/10.1103/PhysRevD.110.084061} {\bibfield  {journal} {\bibinfo  {journal} {Phys. Rev. D}\ }\textbf {\bibinfo {volume} {110}},\ \bibinfo {pages} {084061} (\bibinfo {year} {2024})},\ \Eprint {https://arxiv.org/abs/2406.14291} {arXiv:2406.14291 [gr-qc]} \BibitemShut {NoStop}%
\bibitem [{\citenamefont {Huerta}\ and\ \citenamefont {Gair}(2011)}]{Huerta:2011kt}%
  \BibitemOpen
  \bibfield  {author} {\bibinfo {author} {\bibfnamefont {E.~A.}\ \bibnamefont {Huerta}}\ and\ \bibinfo {author} {\bibfnamefont {J.~R.}\ \bibnamefont {Gair}},\ }\bibfield  {title} {\bibinfo {title} {{Importance of including small body spin effects in the modelling of extreme and intermediate mass-ratio inspirals}},\ }\href {https://doi.org/10.1103/PhysRevD.84.064023} {\bibfield  {journal} {\bibinfo  {journal} {Phys. Rev. D}\ }\textbf {\bibinfo {volume} {84}},\ \bibinfo {pages} {064023} (\bibinfo {year} {2011})},\ \Eprint {https://arxiv.org/abs/1105.3567} {arXiv:1105.3567 [gr-qc]} \BibitemShut {NoStop}%
\bibitem [{\citenamefont {Warburton}\ \emph {et~al.}(2017)\citenamefont {Warburton}, \citenamefont {Osburn},\ and\ \citenamefont {Evans}}]{Warburton:2017sxk}%
  \BibitemOpen
  \bibfield  {author} {\bibinfo {author} {\bibfnamefont {N.}~\bibnamefont {Warburton}}, \bibinfo {author} {\bibfnamefont {T.}~\bibnamefont {Osburn}},\ and\ \bibinfo {author} {\bibfnamefont {C.~R.}\ \bibnamefont {Evans}},\ }\bibfield  {title} {\bibinfo {title} {{Evolution of small-mass-ratio binaries with a spinning secondary}},\ }\href {https://doi.org/10.1103/PhysRevD.96.084057} {\bibfield  {journal} {\bibinfo  {journal} {Phys. Rev. D}\ }\textbf {\bibinfo {volume} {96}},\ \bibinfo {pages} {084057} (\bibinfo {year} {2017})},\ \Eprint {https://arxiv.org/abs/1708.03720} {arXiv:1708.03720 [gr-qc]} \BibitemShut {NoStop}%
\bibitem [{\citenamefont {Piovano}\ \emph {et~al.}(2021)\citenamefont {Piovano}, \citenamefont {Brito}, \citenamefont {Maselli},\ and\ \citenamefont {Pani}}]{Piovano:2021iwv}%
  \BibitemOpen
  \bibfield  {author} {\bibinfo {author} {\bibfnamefont {G.~A.}\ \bibnamefont {Piovano}}, \bibinfo {author} {\bibfnamefont {R.}~\bibnamefont {Brito}}, \bibinfo {author} {\bibfnamefont {A.}~\bibnamefont {Maselli}},\ and\ \bibinfo {author} {\bibfnamefont {P.}~\bibnamefont {Pani}},\ }\bibfield  {title} {\bibinfo {title} {{Assessing the detectability of the secondary spin in extreme mass-ratio inspirals with fully relativistic numerical waveforms}},\ }\href {https://doi.org/10.1103/PhysRevD.104.124019} {\bibfield  {journal} {\bibinfo  {journal} {Phys. Rev. D}\ }\textbf {\bibinfo {volume} {104}},\ \bibinfo {pages} {124019} (\bibinfo {year} {2021})},\ \Eprint {https://arxiv.org/abs/2105.07083} {arXiv:2105.07083 [gr-qc]} \BibitemShut {NoStop}%
\bibitem [{\citenamefont {Mathews}\ and\ \citenamefont {Pound}(2025)}]{Mathews:2025nyb}%
  \BibitemOpen
  \bibfield  {author} {\bibinfo {author} {\bibfnamefont {J.}~\bibnamefont {Mathews}}\ and\ \bibinfo {author} {\bibfnamefont {A.}~\bibnamefont {Pound}},\ }\bibfield  {title} {\bibinfo {title} {{Postadiabatic waveform-generation framework for asymmetric precessing binaries}},\ }\href {https://doi.org/10.1103/rbkb-qnxv} {\bibfield  {journal} {\bibinfo  {journal} {Phys. Rev. D}\ }\textbf {\bibinfo {volume} {112}},\ \bibinfo {pages} {104078} (\bibinfo {year} {2025})},\ \Eprint {https://arxiv.org/abs/2501.01413} {arXiv:2501.01413 [gr-qc]} \BibitemShut {NoStop}%
\bibitem [{\citenamefont {Albertini}\ \emph {et~al.}(2025)\citenamefont {Albertini}, \citenamefont {Skoup{\'y}}, \citenamefont {Lukes-Gerakopoulos},\ and\ \citenamefont {Nagar}}]{Albertini:2024agg}%
  \BibitemOpen
  \bibfield  {author} {\bibinfo {author} {\bibfnamefont {A.}~\bibnamefont {Albertini}}, \bibinfo {author} {\bibfnamefont {V.}~\bibnamefont {Skoup{\'y}}}, \bibinfo {author} {\bibfnamefont {G.}~\bibnamefont {Lukes-Gerakopoulos}},\ and\ \bibinfo {author} {\bibfnamefont {A.}~\bibnamefont {Nagar}},\ }\bibfield  {title} {\bibinfo {title} {{Comparing effective-one-body and Mathisson-Papapetrou-Dixon results for a spinning test particle on circular equatorial orbits around a Kerr black hole}},\ }\href {https://doi.org/10.1103/PhysRevD.111.064086} {\bibfield  {journal} {\bibinfo  {journal} {Phys. Rev. D}\ }\textbf {\bibinfo {volume} {111}},\ \bibinfo {pages} {064086} (\bibinfo {year} {2025})},\ \Eprint {https://arxiv.org/abs/2412.16077} {arXiv:2412.16077 [gr-qc]} \BibitemShut {NoStop}%
\bibitem [{\citenamefont {Rahman}\ \emph {et~al.}(2026)\citenamefont {Rahman}, \citenamefont {Shahzadi}, \citenamefont {Pound},\ and\ \citenamefont {Mathews}}]{Rahman:2026qho}%
  \BibitemOpen
  \bibfield  {author} {\bibinfo {author} {\bibfnamefont {M.}~\bibnamefont {Rahman}}, \bibinfo {author} {\bibfnamefont {M.}~\bibnamefont {Shahzadi}}, \bibinfo {author} {\bibfnamefont {A.}~\bibnamefont {Pound}},\ and\ \bibinfo {author} {\bibfnamefont {J.}~\bibnamefont {Mathews}},\ }\bibfield  {title} {\bibinfo {title} {{Quadrupole and quadratic-in-spin effects in quasicircular, spinning, asymmetric binaries}},\ }\href {https://doi.org/10.1103/wmsn-pl9n} {\bibfield  {journal} {\bibinfo  {journal} {Phys. Rev. D}\ }\textbf {\bibinfo {volume} {114}},\ \bibinfo {pages} {064034} (\bibinfo {year} {2026})},\ \Eprint {https://arxiv.org/abs/2606.28937} {arXiv:2606.28937 [gr-qc]} \BibitemShut {NoStop}%
\bibitem [{\citenamefont {Drummond}\ \emph {et~al.}(2026)\citenamefont {Drummond}, \citenamefont {Hughes}, \citenamefont {Skoup{\'y}}, \citenamefont {Lynch},\ and\ \citenamefont {Piovano}}]{Drummond:2026haw}%
  \BibitemOpen
  \bibfield  {author} {\bibinfo {author} {\bibfnamefont {L.~V.}\ \bibnamefont {Drummond}}, \bibinfo {author} {\bibfnamefont {S.~A.}\ \bibnamefont {Hughes}}, \bibinfo {author} {\bibfnamefont {V.}~\bibnamefont {Skoup{\'y}}}, \bibinfo {author} {\bibfnamefont {P.}~\bibnamefont {Lynch}},\ and\ \bibinfo {author} {\bibfnamefont {G.~A.}\ \bibnamefont {Piovano}},\ }\bibfield  {title} {\bibinfo {title} {{Shifted-geodesic approximation for spinning-body gravitational wave fluxes}},\ }\href@noop {} {\bibfield  {journal} {\bibinfo  {journal} {arXiv:2603.12189}\ } (\bibinfo {year} {2026})},\ \Eprint {https://arxiv.org/abs/2603.12189} {arXiv:2603.12189 [gr-qc]} \BibitemShut {NoStop}%
\bibitem [{\citenamefont {Lui}\ \emph {et~al.}(2026)\citenamefont {Lui}, \citenamefont {Drummond},\ and\ \citenamefont {Torres-Orjuela}}]{Lui:2026uai}%
  \BibitemOpen
  \bibfield  {author} {\bibinfo {author} {\bibfnamefont {L.}~\bibnamefont {Lui}}, \bibinfo {author} {\bibfnamefont {L.~V.}\ \bibnamefont {Drummond}},\ and\ \bibinfo {author} {\bibfnamefont {A.}~\bibnamefont {Torres-Orjuela}},\ }\bibfield  {title} {\bibinfo {title} {{Pitching Cosmic Curveballs: Environmental Effects on Extreme-Mass-Ratio Inspirals with Spinning Secondaries}},\ }\href {https://doi.org/10.1103/dy3h-65ty} {\bibfield  {journal} {\bibinfo  {journal} {Phys. Rev. Lett.}\ }\textbf {\bibinfo {volume} {137}},\ \bibinfo {pages} {121403} (\bibinfo {year} {2026})},\ \Eprint {https://arxiv.org/abs/2606.01569} {arXiv:2606.01569 [gr-qc]} \BibitemShut {NoStop}%
\bibitem [{\citenamefont {Skoup\'y}\ and\ \citenamefont {Lukes-Gerakopoulos}(2022)}]{Skoupy:2022adh}%
  \BibitemOpen
  \bibfield  {author} {\bibinfo {author} {\bibfnamefont {V.}~\bibnamefont {Skoup\'y}}\ and\ \bibinfo {author} {\bibfnamefont {G.}~\bibnamefont {Lukes-Gerakopoulos}},\ }\bibfield  {title} {\bibinfo {title} {{Adiabatic equatorial inspirals of a spinning body into a Kerr black hole}},\ }\href {https://doi.org/10.1103/PhysRevD.105.084033} {\bibfield  {journal} {\bibinfo  {journal} {Phys. Rev. D}\ }\textbf {\bibinfo {volume} {105}},\ \bibinfo {pages} {084033} (\bibinfo {year} {2022})},\ \Eprint {https://arxiv.org/abs/2201.07044} {arXiv:2201.07044 [gr-qc]} \BibitemShut {NoStop}%
\bibitem [{\citenamefont {Drummond}\ \emph {et~al.}(2024)\citenamefont {Drummond}, \citenamefont {Lynch}, \citenamefont {Hanselman}, \citenamefont {Becker},\ and\ \citenamefont {Hughes}}]{Drummond:2023wqc}%
  \BibitemOpen
  \bibfield  {author} {\bibinfo {author} {\bibfnamefont {L.~V.}\ \bibnamefont {Drummond}}, \bibinfo {author} {\bibfnamefont {P.}~\bibnamefont {Lynch}}, \bibinfo {author} {\bibfnamefont {A.~G.}\ \bibnamefont {Hanselman}}, \bibinfo {author} {\bibfnamefont {D.~R.}\ \bibnamefont {Becker}},\ and\ \bibinfo {author} {\bibfnamefont {S.~A.}\ \bibnamefont {Hughes}},\ }\bibfield  {title} {\bibinfo {title} {{Extreme mass-ratio inspiral and waveforms for a spinning body into a Kerr black hole via osculating geodesics and near-identity transformations}},\ }\href {https://doi.org/10.1103/PhysRevD.109.064030} {\bibfield  {journal} {\bibinfo  {journal} {Phys. Rev. D}\ }\textbf {\bibinfo {volume} {109}},\ \bibinfo {pages} {064030} (\bibinfo {year} {2024})},\ \Eprint {https://arxiv.org/abs/2310.08438} {arXiv:2310.08438 [gr-qc]} \BibitemShut {NoStop}%
\bibitem [{\citenamefont {Piovano}\ \emph {et~al.}(2025)\citenamefont {Piovano}, \citenamefont {Pantelidou}, \citenamefont {Mac~Uilliam},\ and\ \citenamefont {Witzany}}]{Piovano:2024yks}%
  \BibitemOpen
  \bibfield  {author} {\bibinfo {author} {\bibfnamefont {G.~A.}\ \bibnamefont {Piovano}}, \bibinfo {author} {\bibfnamefont {C.}~\bibnamefont {Pantelidou}}, \bibinfo {author} {\bibfnamefont {J.}~\bibnamefont {Mac~Uilliam}},\ and\ \bibinfo {author} {\bibfnamefont {V.}~\bibnamefont {Witzany}},\ }\bibfield  {title} {\bibinfo {title} {{Spinning particles near Kerr black holes: Orbits and gravitational-wave fluxes through the Hamilton-Jacobi formalism}},\ }\href {https://doi.org/10.1103/PhysRevD.111.044009} {\bibfield  {journal} {\bibinfo  {journal} {Phys. Rev. D}\ }\textbf {\bibinfo {volume} {111}},\ \bibinfo {pages} {044009} (\bibinfo {year} {2025})},\ \Eprint {https://arxiv.org/abs/2410.05769} {arXiv:2410.05769 [gr-qc]} \BibitemShut {NoStop}%
\bibitem [{\citenamefont {Skoup{\'y}}\ \emph {et~al.}(2025)\citenamefont {Skoup{\'y}}, \citenamefont {Piovano},\ and\ \citenamefont {Witzany}}]{Skoupy:2025nie}%
  \BibitemOpen
  \bibfield  {author} {\bibinfo {author} {\bibfnamefont {V.}~\bibnamefont {Skoup{\'y}}}, \bibinfo {author} {\bibfnamefont {G.~A.}\ \bibnamefont {Piovano}},\ and\ \bibinfo {author} {\bibfnamefont {V.}~\bibnamefont {Witzany}},\ }\bibfield  {title} {\bibinfo {title} {{Spherical inspirals of spinning bodies into Kerr black holes}},\ }\href {https://doi.org/10.1103/x9yy-c2jq} {\bibfield  {journal} {\bibinfo  {journal} {Phys. Rev. D}\ }\textbf {\bibinfo {volume} {112}},\ \bibinfo {pages} {124054} (\bibinfo {year} {2025})},\ \Eprint {https://arxiv.org/abs/2506.20726} {arXiv:2506.20726 [gr-qc]} \BibitemShut {NoStop}%
\bibitem [{\citenamefont {Mathisson}(1937)}]{mathisson1937}%
  \BibitemOpen
  \bibfield  {author} {\bibinfo {author} {\bibfnamefont {M.}~\bibnamefont {Mathisson}},\ }\bibfield  {title} {\bibinfo {title} {{Neue mechanik materieller systemes}},\ }\href@noop {} {\bibfield  {journal} {\bibinfo  {journal} {Acta Phys. Polon.}\ }\textbf {\bibinfo {volume} {6}},\ \bibinfo {pages} {163} (\bibinfo {year} {1937})}\BibitemShut {NoStop}%
\bibitem [{\citenamefont {Papapetrou}(1951)}]{papapetrou1951}%
  \BibitemOpen
  \bibfield  {author} {\bibinfo {author} {\bibfnamefont {A.}~\bibnamefont {Papapetrou}},\ }\bibfield  {title} {\bibinfo {title} {Spinning test-particles in general relativity. i},\ }\href {https://doi.org/10.1098/rspa.1951.0200} {\bibfield  {journal} {\bibinfo  {journal} {Proc. Roy. Soc. Lond.}\ }\textbf {\bibinfo {volume} {A209}},\ \bibinfo {pages} {248} (\bibinfo {year} {1951})}\BibitemShut {NoStop}%
\bibitem [{\citenamefont {Dixon}(1964)}]{dixon1964}%
  \BibitemOpen
  \bibfield  {author} {\bibinfo {author} {\bibfnamefont {W.}~\bibnamefont {Dixon}},\ }\bibfield  {title} {\bibinfo {title} {A covariant multipole formalism for extended test bodies in general relativity},\ }\href {https://doi.org/10.1007/BF02734579} {\bibfield  {journal} {\bibinfo  {journal} {Il Nuovo Cimento (1955-1965)}\ }\textbf {\bibinfo {volume} {34}},\ \bibinfo {pages} {317} (\bibinfo {year} {1964})}\BibitemShut {NoStop}%
\bibitem [{\citenamefont {Dixon}(1970)}]{dixon1970}%
  \BibitemOpen
  \bibfield  {author} {\bibinfo {author} {\bibfnamefont {W.~G.}\ \bibnamefont {Dixon}},\ }\bibfield  {title} {\bibinfo {title} {{Dynamics of extended bodies in general relativity. I. Momentum and angular momentum}},\ }\href {https://doi.org/10.1098/rspa.1970.0020} {\bibfield  {journal} {\bibinfo  {journal} {Proc. Roy. Soc. Lond.}\ }\textbf {\bibinfo {volume} {A314}},\ \bibinfo {pages} {499} (\bibinfo {year} {1970})}\BibitemShut {NoStop}%
\bibitem [{\citenamefont {Tulczyjew}(1959)}]{tulczyjew1959}%
  \BibitemOpen
  \bibfield  {author} {\bibinfo {author} {\bibfnamefont {W.}~\bibnamefont {Tulczyjew}},\ }\bibfield  {title} {\bibinfo {title} {Motion of multipole particles in general relativity theory},\ }\href@noop {} {\bibfield  {journal} {\bibinfo  {journal} {Acta Phys. Pol.}\ }\textbf {\bibinfo {volume} {18}},\ \bibinfo {pages} {393} (\bibinfo {year} {1959})}\BibitemShut {NoStop}%
\bibitem [{\citenamefont {Ehlers}\ and\ \citenamefont {Rudolph}(1977)}]{ehlers1977}%
  \BibitemOpen
  \bibfield  {author} {\bibinfo {author} {\bibfnamefont {J.}~\bibnamefont {Ehlers}}\ and\ \bibinfo {author} {\bibfnamefont {E.}~\bibnamefont {Rudolph}},\ }\bibfield  {title} {\bibinfo {title} {Dynamics of extended bodies in general relativity center-of-mass description and quasirigidity},\ }\href {https://doi.org/10.1007/BF00763547} {\bibfield  {journal} {\bibinfo  {journal} {Gen. Relativ. Gravitation}\ }\textbf {\bibinfo {volume} {8}},\ \bibinfo {pages} {197} (\bibinfo {year} {1977})}\BibitemShut {NoStop}%
\bibitem [{\citenamefont {Costa}\ and\ \citenamefont {Nat{\'a}rio}(2015)}]{costa2015}%
  \BibitemOpen
  \bibfield  {author} {\bibinfo {author} {\bibfnamefont {L.~F.~O.}\ \bibnamefont {Costa}}\ and\ \bibinfo {author} {\bibfnamefont {J.}~\bibnamefont {Nat{\'a}rio}},\ }\bibfield  {title} {\bibinfo {title} {Center of mass, spin supplementary conditions, and the momentum of spinning particles},\ }in\ \href {https://doi.org/10.1007/978-3-319-18335-0_6} {\emph {\bibinfo {booktitle} {Equations of Motion in Relativistic Gravity}}}\ (\bibinfo  {publisher} {Springer},\ \bibinfo {year} {2015})\ pp.\ \bibinfo {pages} {215--258},\ \Eprint {https://arxiv.org/abs/1410.6443} {arXiv:1410.6443 [gr-qc]} \BibitemShut {NoStop}%
\bibitem [{\citenamefont {Ramond}\ and\ \citenamefont {Isoyama}(2025)}]{Ramond:2024sfp}%
  \BibitemOpen
  \bibfield  {author} {\bibinfo {author} {\bibfnamefont {P.}~\bibnamefont {Ramond}}\ and\ \bibinfo {author} {\bibfnamefont {S.}~\bibnamefont {Isoyama}},\ }\bibfield  {title} {\bibinfo {title} {{Symplectic mechanics of relativistic spinning compact bodies: Canonical formalism in the Schwarzschild spacetime}},\ }\href {https://doi.org/10.1103/PhysRevD.111.064027} {\bibfield  {journal} {\bibinfo  {journal} {Phys. Rev. D}\ }\textbf {\bibinfo {volume} {111}},\ \bibinfo {pages} {064027} (\bibinfo {year} {2025})},\ \Eprint {https://arxiv.org/abs/2402.05049} {arXiv:2402.05049 [gr-qc]} \BibitemShut {NoStop}%
\bibitem [{\citenamefont {M{\o}ller}(1949)}]{moller1949}%
  \BibitemOpen
  \bibfield  {author} {\bibinfo {author} {\bibfnamefont {C.}~\bibnamefont {M{\o}ller}},\ }\bibfield  {title} {\bibinfo {title} {On the definition of the centre of gravity of an arbitrary closed system in the theory of relativity},\ }\href@noop {} {\bibfield  {journal} {\bibinfo  {journal} {Commun. Dublin Inst. Adv. Stud., A}\ }\textbf {\bibinfo {volume} {5}},\ \bibinfo {pages} {42} (\bibinfo {year} {1949})}\BibitemShut {NoStop}%
\bibitem [{\citenamefont {Semer{\'a}k}(1999)}]{semerak1999}%
  \BibitemOpen
  \bibfield  {author} {\bibinfo {author} {\bibfnamefont {O.}~\bibnamefont {Semer{\'a}k}},\ }\bibfield  {title} {\bibinfo {title} {Spinning test particles in a kerr field---i},\ }\href {https://doi.org/10.1046/j.1365-8711.1999.02754.x} {\bibfield  {journal} {\bibinfo  {journal} {Mon. Not. R. Astron. Soc.}\ }\textbf {\bibinfo {volume} {308}},\ \bibinfo {pages} {863} (\bibinfo {year} {1999})}\BibitemShut {NoStop}%
\bibitem [{\citenamefont {Suzuki}\ and\ \citenamefont {Maeda}(1998)}]{Suzuki:1997by}%
  \BibitemOpen
  \bibfield  {author} {\bibinfo {author} {\bibfnamefont {S.}~\bibnamefont {Suzuki}}\ and\ \bibinfo {author} {\bibfnamefont {K.-i.}\ \bibnamefont {Maeda}},\ }\bibfield  {title} {\bibinfo {title} {{Innermost stable circular orbit of a spinning particle in Kerr space-time}},\ }\href {https://doi.org/10.1103/PhysRevD.58.023005} {\bibfield  {journal} {\bibinfo  {journal} {Phys. Rev. D}\ }\textbf {\bibinfo {volume} {58}},\ \bibinfo {pages} {023005} (\bibinfo {year} {1998})},\ \Eprint {https://arxiv.org/abs/gr-qc/9712095} {arXiv:gr-qc/9712095} \BibitemShut {NoStop}%
\bibitem [{\citenamefont {Jefremov}\ \emph {et~al.}(2015)\citenamefont {Jefremov}, \citenamefont {Tsupko},\ and\ \citenamefont {Bisnovatyi-Kogan}}]{Jefremov:2015gza}%
  \BibitemOpen
  \bibfield  {author} {\bibinfo {author} {\bibfnamefont {P.~I.}\ \bibnamefont {Jefremov}}, \bibinfo {author} {\bibfnamefont {O.~Y.}\ \bibnamefont {Tsupko}},\ and\ \bibinfo {author} {\bibfnamefont {G.~S.}\ \bibnamefont {Bisnovatyi-Kogan}},\ }\bibfield  {title} {\bibinfo {title} {{Innermost stable circular orbits of spinning test particles in Schwarzschild and Kerr space-times}},\ }\href {https://doi.org/10.1103/PhysRevD.91.124030} {\bibfield  {journal} {\bibinfo  {journal} {Phys. Rev. D}\ }\textbf {\bibinfo {volume} {91}},\ \bibinfo {pages} {124030} (\bibinfo {year} {2015})},\ \Eprint {https://arxiv.org/abs/1503.07060} {arXiv:1503.07060 [gr-qc]} \BibitemShut {NoStop}%
\bibitem [{\citenamefont {Tsupko}\ \emph {et~al.}(2016)\citenamefont {Tsupko}, \citenamefont {Bisnovatyi-Kogan},\ and\ \citenamefont {Jefremov}}]{tsupko2016parameters}%
  \BibitemOpen
  \bibfield  {author} {\bibinfo {author} {\bibfnamefont {O.~Y.}\ \bibnamefont {Tsupko}}, \bibinfo {author} {\bibfnamefont {G.}~\bibnamefont {Bisnovatyi-Kogan}},\ and\ \bibinfo {author} {\bibfnamefont {P.}~\bibnamefont {Jefremov}},\ }\bibfield  {title} {\bibinfo {title} {Parameters of innermost stable circular orbits of spinning test particles: Numerical and analytical calculations},\ }\href@noop {} {\bibfield  {journal} {\bibinfo  {journal} {Gravitation and Cosmology}\ }\textbf {\bibinfo {volume} {22}},\ \bibinfo {pages} {138} (\bibinfo {year} {2016})}\BibitemShut {NoStop}%
\bibitem [{\citenamefont {Favata}(2011)}]{Favata:2010ic}%
  \BibitemOpen
  \bibfield  {author} {\bibinfo {author} {\bibfnamefont {M.}~\bibnamefont {Favata}},\ }\bibfield  {title} {\bibinfo {title} {{Conservative corrections to the innermost stable circular orbit (ISCO) of a Kerr black hole: A New gauge-invariant post-Newtonian ISCO condition, and the ISCO shift due to test-particle spin and the gravitational self-force}},\ }\href {https://doi.org/10.1103/PhysRevD.83.024028} {\bibfield  {journal} {\bibinfo  {journal} {Phys. Rev. D}\ }\textbf {\bibinfo {volume} {83}},\ \bibinfo {pages} {024028} (\bibinfo {year} {2011})},\ \Eprint {https://arxiv.org/abs/1010.2553} {arXiv:1010.2553 [gr-qc]} \BibitemShut {NoStop}%
\bibitem [{\citenamefont {Bizyaev}(2026)}]{Bizyaev:2025mva}%
  \BibitemOpen
  \bibfield  {author} {\bibinfo {author} {\bibfnamefont {I.}~\bibnamefont {Bizyaev}},\ }\bibfield  {title} {\bibinfo {title} {{Dynamics of spinning test bodies in the Schwarzschild space-time: reduction and circular orbits}},\ }\href {https://doi.org/10.1007/s10714-026-03593-4} {\bibfield  {journal} {\bibinfo  {journal} {Gen. Rel. Grav.}\ }\textbf {\bibinfo {volume} {58}},\ \bibinfo {pages} {81} (\bibinfo {year} {2026})},\ \Eprint {https://arxiv.org/abs/2511.23154} {arXiv:2511.23154 [math.DS]} \BibitemShut {NoStop}%
\bibitem [{\citenamefont {Le~Tiec}\ \emph {et~al.}(2011)\citenamefont {Le~Tiec}, \citenamefont {Mroue}, \citenamefont {Barack}, \citenamefont {Buonanno}, \citenamefont {Pfeiffer}, \citenamefont {Sago},\ and\ \citenamefont {Taracchini}}]{LeTiec:2011bk}%
  \BibitemOpen
  \bibfield  {author} {\bibinfo {author} {\bibfnamefont {A.}~\bibnamefont {Le~Tiec}}, \bibinfo {author} {\bibfnamefont {A.~H.}\ \bibnamefont {Mroue}}, \bibinfo {author} {\bibfnamefont {L.}~\bibnamefont {Barack}}, \bibinfo {author} {\bibfnamefont {A.}~\bibnamefont {Buonanno}}, \bibinfo {author} {\bibfnamefont {H.~P.}\ \bibnamefont {Pfeiffer}}, \bibinfo {author} {\bibfnamefont {N.}~\bibnamefont {Sago}},\ and\ \bibinfo {author} {\bibfnamefont {A.}~\bibnamefont {Taracchini}},\ }\bibfield  {title} {\bibinfo {title} {{Periastron Advance in Black Hole Binaries}},\ }\href {https://doi.org/10.1103/PhysRevLett.107.141101} {\bibfield  {journal} {\bibinfo  {journal} {Phys. Rev. Lett.}\ }\textbf {\bibinfo {volume} {107}},\ \bibinfo {pages} {141101} (\bibinfo {year} {2011})},\ \Eprint {https://arxiv.org/abs/1106.3278} {arXiv:1106.3278 [gr-qc]} \BibitemShut {NoStop}%
\bibitem [{\citenamefont {Bini}\ and\ \citenamefont {Geralico}(2019)}]{Bini:2019zjj}%
  \BibitemOpen
  \bibfield  {author} {\bibinfo {author} {\bibfnamefont {D.}~\bibnamefont {Bini}}\ and\ \bibinfo {author} {\bibfnamefont {A.}~\bibnamefont {Geralico}},\ }\bibfield  {title} {\bibinfo {title} {{Analytical determination of the periastron advance in spinning binaries from self-force computations}},\ }\href {https://doi.org/10.1103/PhysRevD.100.121502} {\bibfield  {journal} {\bibinfo  {journal} {Phys. Rev. D}\ }\textbf {\bibinfo {volume} {100}},\ \bibinfo {pages} {121502} (\bibinfo {year} {2019})},\ \Eprint {https://arxiv.org/abs/1907.11083} {arXiv:1907.11083 [gr-qc]} \BibitemShut {NoStop}%
\bibitem [{\citenamefont {Suzuki}\ and\ \citenamefont {Maeda}(1997)}]{suzuki1997}%
  \BibitemOpen
  \bibfield  {author} {\bibinfo {author} {\bibfnamefont {S.}~\bibnamefont {Suzuki}}\ and\ \bibinfo {author} {\bibfnamefont {K.-i.}\ \bibnamefont {Maeda}},\ }\bibfield  {title} {\bibinfo {title} {{Chaos in Schwarzschild space-time: The motion of a spinning particle}},\ }\href {https://doi.org/10.1103/PhysRevD.55.4848} {\bibfield  {journal} {\bibinfo  {journal} {Phys. Rev.}\ }\textbf {\bibinfo {volume} {D55}},\ \bibinfo {pages} {4848} (\bibinfo {year} {1997})},\ \Eprint {https://arxiv.org/abs/gr-qc/9604020} {arXiv:gr-qc/9604020 [gr-qc]} \BibitemShut {NoStop}%
\bibitem [{\citenamefont {Suzuki}\ and\ \citenamefont {Maeda}(1999)}]{suzuki1999}%
  \BibitemOpen
  \bibfield  {author} {\bibinfo {author} {\bibfnamefont {S.}~\bibnamefont {Suzuki}}\ and\ \bibinfo {author} {\bibfnamefont {K.-i.}\ \bibnamefont {Maeda}},\ }\bibfield  {title} {\bibinfo {title} {Signature of chaos in gravitational waves from a spinning particle},\ }\href@noop {} {\bibfield  {journal} {\bibinfo  {journal} {Phys. Rev. D}\ }\textbf {\bibinfo {volume} {61}},\ \bibinfo {pages} {024005} (\bibinfo {year} {1999})}\BibitemShut {NoStop}%
\bibitem [{\citenamefont {Zelenka}\ \emph {et~al.}(2020)\citenamefont {Zelenka}, \citenamefont {Lukes-Gerakopoulos}, \citenamefont {Witzany},\ and\ \citenamefont {Kop\'a\v{c}ek}}]{Zelenka:2019nyp}%
  \BibitemOpen
  \bibfield  {author} {\bibinfo {author} {\bibfnamefont {O.}~\bibnamefont {Zelenka}}, \bibinfo {author} {\bibfnamefont {G.}~\bibnamefont {Lukes-Gerakopoulos}}, \bibinfo {author} {\bibfnamefont {V.}~\bibnamefont {Witzany}},\ and\ \bibinfo {author} {\bibfnamefont {O.}~\bibnamefont {Kop\'a\v{c}ek}},\ }\bibfield  {title} {\bibinfo {title} {{Growth of resonances and chaos for a spinning test particle in the Schwarzschild background}},\ }\href {https://doi.org/10.1103/PhysRevD.101.024037} {\bibfield  {journal} {\bibinfo  {journal} {Phys. Rev. D}\ }\textbf {\bibinfo {volume} {101}},\ \bibinfo {pages} {024037} (\bibinfo {year} {2020})},\ \Eprint {https://arxiv.org/abs/1911.00414} {arXiv:1911.00414 [gr-qc]} \BibitemShut {NoStop}%
\bibitem [{\citenamefont {Mukherjee}\ and\ \citenamefont {Tripathy}(2020)}]{Mukherjee:2019jhd}%
  \BibitemOpen
  \bibfield  {author} {\bibinfo {author} {\bibfnamefont {S.}~\bibnamefont {Mukherjee}}\ and\ \bibinfo {author} {\bibfnamefont {S.}~\bibnamefont {Tripathy}},\ }\bibfield  {title} {\bibinfo {title} {{Resonant orbits for a spinning particle in Kerr spacetime}},\ }\href {https://doi.org/10.1103/PhysRevD.101.124047} {\bibfield  {journal} {\bibinfo  {journal} {Phys. Rev. D}\ }\textbf {\bibinfo {volume} {101}},\ \bibinfo {pages} {124047} (\bibinfo {year} {2020})},\ \Eprint {https://arxiv.org/abs/1905.04061} {arXiv:1905.04061 [gr-qc]} \BibitemShut {NoStop}%
\bibitem [{\citenamefont {Geng}\ \emph {et~al.}(2026)\citenamefont {Geng}, \citenamefont {Xiang}, \citenamefont {Jiang},\ and\ \citenamefont {Pang}}]{Geng:2026xcs}%
  \BibitemOpen
  \bibfield  {author} {\bibinfo {author} {\bibfnamefont {J.}~\bibnamefont {Geng}}, \bibinfo {author} {\bibfnamefont {Y.}~\bibnamefont {Xiang}}, \bibinfo {author} {\bibfnamefont {Q.}~\bibnamefont {Jiang}},\ and\ \bibinfo {author} {\bibfnamefont {X.}~\bibnamefont {Pang}},\ }\bibfield  {title} {\bibinfo {title} {{Spin precession in the strong deflection limit}},\ }\href@noop {} {\bibfield  {journal} {\bibinfo  {journal} {arXiv:2607.25617}\ } (\bibinfo {year} {2026})},\ \Eprint {https://arxiv.org/abs/2607.25617} {arXiv:2607.25617 [gr-qc]} \BibitemShut {NoStop}%
\bibitem [{\citenamefont {Pang}\ \emph {et~al.}(2025)\citenamefont {Pang}, \citenamefont {Jiang}, \citenamefont {Xiang},\ and\ \citenamefont {Deng}}]{Pang:2024tco}%
  \BibitemOpen
  \bibfield  {author} {\bibinfo {author} {\bibfnamefont {X.}~\bibnamefont {Pang}}, \bibinfo {author} {\bibfnamefont {Q.}~\bibnamefont {Jiang}}, \bibinfo {author} {\bibfnamefont {Y.}~\bibnamefont {Xiang}},\ and\ \bibinfo {author} {\bibfnamefont {G.-M.}\ \bibnamefont {Deng}},\ }\bibfield  {title} {\bibinfo {title} {{The precession of particle spin in spherical symmetric spacetimes}},\ }\href {https://doi.org/10.1140/epjc/s10052-025-13894-8} {\bibfield  {journal} {\bibinfo  {journal} {Eur. Phys. J. C}\ }\textbf {\bibinfo {volume} {85}},\ \bibinfo {pages} {193} (\bibinfo {year} {2025})},\ \Eprint {https://arxiv.org/abs/2410.04323} {arXiv:2410.04323 [gr-qc]} \BibitemShut {NoStop}%
\bibitem [{\citenamefont {Pantig}\ and\ \citenamefont {{\"O}vg{\"u}n}(2026)}]{Pantig:2026qcf}%
  \BibitemOpen
  \bibfield  {author} {\bibinfo {author} {\bibfnamefont {R.~C.}\ \bibnamefont {Pantig}}\ and\ \bibinfo {author} {\bibfnamefont {A.}~\bibnamefont {{\"O}vg{\"u}n}},\ }\bibfield  {title} {\bibinfo {title} {{Gauss{\textendash}Bonnet lensing of spinning massive particles in static spherically symmetric spacetimes}},\ }\href {https://doi.org/10.1016/j.aop.2026.170421} {\bibfield  {journal} {\bibinfo  {journal} {Annals Phys.}\ }\textbf {\bibinfo {volume} {488}},\ \bibinfo {pages} {170421} (\bibinfo {year} {2026})},\ \Eprint {https://arxiv.org/abs/2603.00650} {arXiv:2603.00650 [gr-qc]} \BibitemShut {NoStop}%
\bibitem [{\citenamefont {Tan}\ \emph {et~al.}(2025)\citenamefont {Tan}, \citenamefont {Deng}, \citenamefont {Long},\ and\ \citenamefont {Jing}}]{Tan:2024hzw}%
  \BibitemOpen
  \bibfield  {author} {\bibinfo {author} {\bibfnamefont {Q.}~\bibnamefont {Tan}}, \bibinfo {author} {\bibfnamefont {W.}~\bibnamefont {Deng}}, \bibinfo {author} {\bibfnamefont {S.}~\bibnamefont {Long}},\ and\ \bibinfo {author} {\bibfnamefont {J.}~\bibnamefont {Jing}},\ }\bibfield  {title} {\bibinfo {title} {{Motion of spinning particles around black hole in a dark matter halo}},\ }\href {https://doi.org/10.1088/1475-7516/2025/05/044} {\bibfield  {journal} {\bibinfo  {journal} {JCAP}\ }\textbf {\bibinfo {volume} {05}},\ \bibinfo {pages} {044}},\ \Eprint {https://arxiv.org/abs/2409.17760} {arXiv:2409.17760 [gr-qc]} \BibitemShut {NoStop}%
\bibitem [{\citenamefont {Liu}\ and\ \citenamefont {Zhang}(2025)}]{Liu:2024lda}%
  \BibitemOpen
  \bibfield  {author} {\bibinfo {author} {\bibfnamefont {Y.}~\bibnamefont {Liu}}\ and\ \bibinfo {author} {\bibfnamefont {X.}~\bibnamefont {Zhang}},\ }\bibfield  {title} {\bibinfo {title} {{Analytic solutions for the motion of spinning particles near braneworld black hole}},\ }\href {https://doi.org/10.1103/PhysRevD.111.044056} {\bibfield  {journal} {\bibinfo  {journal} {Phys. Rev. D}\ }\textbf {\bibinfo {volume} {111}},\ \bibinfo {pages} {044056} (\bibinfo {year} {2025})},\ \Eprint {https://arxiv.org/abs/2408.06852} {arXiv:2408.06852 [gr-qc]} \BibitemShut {NoStop}%
\bibitem [{\citenamefont {Ciou}\ \emph {et~al.}(2025)\citenamefont {Ciou}, \citenamefont {Hsieh},\ and\ \citenamefont {Lee}}]{Ciou:2025ygb}%
  \BibitemOpen
  \bibfield  {author} {\bibinfo {author} {\bibfnamefont {S.-Y.}\ \bibnamefont {Ciou}}, \bibinfo {author} {\bibfnamefont {T.}~\bibnamefont {Hsieh}},\ and\ \bibinfo {author} {\bibfnamefont {D.-S.}\ \bibnamefont {Lee}},\ }\bibfield  {title} {\bibinfo {title} {{Dynamics of spinning particles in Reissner-Nordstr{\"o}m black hole exterior}},\ }\href {https://doi.org/10.1088/1475-7516/2025/05/086} {\bibfield  {journal} {\bibinfo  {journal} {JCAP}\ }\textbf {\bibinfo {volume} {05}},\ \bibinfo {pages} {086}},\ \Eprint {https://arxiv.org/abs/2503.12911} {arXiv:2503.12911 [gr-qc]} \BibitemShut {NoStop}%
\bibitem [{\citenamefont {Chen}\ \emph {et~al.}(2025{\natexlab{a}})\citenamefont {Chen}, \citenamefont {Hsieh},\ and\ \citenamefont {Lee}}]{Chen:2025ncm}%
  \BibitemOpen
  \bibfield  {author} {\bibinfo {author} {\bibfnamefont {Y.-P.}\ \bibnamefont {Chen}}, \bibinfo {author} {\bibfnamefont {T.}~\bibnamefont {Hsieh}},\ and\ \bibinfo {author} {\bibfnamefont {D.-S.}\ \bibnamefont {Lee}},\ }\bibfield  {title} {\bibinfo {title} {{Motion of spinning particles in the Kerr-Newman black hole exterior}},\ }\href@noop {} {\bibfield  {journal} {\bibinfo  {journal} {arXiv:2510.05603}\ } (\bibinfo {year} {2025}{\natexlab{a}})},\ \Eprint {https://arxiv.org/abs/2510.05603} {arXiv:2510.05603 [gr-qc]} \BibitemShut {NoStop}%
\bibitem [{\citenamefont {Jumaniyozov}\ \emph {et~al.}(2025)\citenamefont {Jumaniyozov}, \citenamefont {Rayimbaev},\ and\ \citenamefont {Turaev}}]{Jumaniyozov:2025irx}%
  \BibitemOpen
  \bibfield  {author} {\bibinfo {author} {\bibfnamefont {S.}~\bibnamefont {Jumaniyozov}}, \bibinfo {author} {\bibfnamefont {J.}~\bibnamefont {Rayimbaev}},\ and\ \bibinfo {author} {\bibfnamefont {Y.}~\bibnamefont {Turaev}},\ }\bibfield  {title} {\bibinfo {title} {{Dynamics of spinning particles around a charged black-bounce spacetime}},\ }\href {https://doi.org/10.1140/epjc/s10052-025-14834-2} {\bibfield  {journal} {\bibinfo  {journal} {Eur. Phys. J. C}\ }\textbf {\bibinfo {volume} {85}},\ \bibinfo {pages} {1247} (\bibinfo {year} {2025})}\BibitemShut {NoStop}%
\bibitem [{\citenamefont {Jumaniyozov}\ \emph {et~al.}(2026)\citenamefont {Jumaniyozov}, \citenamefont {Rayimbaev}, \citenamefont {Turaev}, \citenamefont {Akhmedov}, \citenamefont {Seytov},\ and\ \citenamefont {Khasanov}}]{Jumaniyozov:2026lbf}%
  \BibitemOpen
  \bibfield  {author} {\bibinfo {author} {\bibfnamefont {S.}~\bibnamefont {Jumaniyozov}}, \bibinfo {author} {\bibfnamefont {J.}~\bibnamefont {Rayimbaev}}, \bibinfo {author} {\bibfnamefont {Y.}~\bibnamefont {Turaev}}, \bibinfo {author} {\bibfnamefont {M.}~\bibnamefont {Akhmedov}}, \bibinfo {author} {\bibfnamefont {A.}~\bibnamefont {Seytov}},\ and\ \bibinfo {author} {\bibfnamefont {J.}~\bibnamefont {Khasanov}},\ }\bibfield  {title} {\bibinfo {title} {{Spin effects on motion of charged particles around magnetized black holes in conformally coupled scalar fields}},\ }\href {https://doi.org/10.1016/j.nuclphysb.2026.117318} {\bibfield  {journal} {\bibinfo  {journal} {Nucl. Phys. B}\ }\textbf {\bibinfo {volume} {1023}},\ \bibinfo {pages} {117318} (\bibinfo {year} {2026})}\BibitemShut {NoStop}%
\bibitem [{\citenamefont {Chen}\ \emph {et~al.}(2025{\natexlab{b}})\citenamefont {Chen}, \citenamefont {Fan},\ and\ \citenamefont {Chew}}]{Chen:2024luw}%
  \BibitemOpen
  \bibfield  {author} {\bibinfo {author} {\bibfnamefont {H.}~\bibnamefont {Chen}}, \bibinfo {author} {\bibfnamefont {W.}~\bibnamefont {Fan}},\ and\ \bibinfo {author} {\bibfnamefont {X.~Y.}\ \bibnamefont {Chew}},\ }\bibfield  {title} {\bibinfo {title} {{Geodesic motion of test particles around the scalar hairy black holes with asymmetric vacua}},\ }\href {https://doi.org/10.1140/epjc/s10052-025-13948-x} {\bibfield  {journal} {\bibinfo  {journal} {Eur. Phys. J. C}\ }\textbf {\bibinfo {volume} {85}},\ \bibinfo {pages} {338} (\bibinfo {year} {2025}{\natexlab{b}})},\ \Eprint {https://arxiv.org/abs/2411.00565} {arXiv:2411.00565 [gr-qc]} \BibitemShut {NoStop}%
\bibitem [{\citenamefont {Turakhonov}\ \emph {et~al.}(2026)\citenamefont {Turakhonov}, \citenamefont {Ibrokhimov}, \citenamefont {Atamurotov}, \citenamefont {Abdujabbarov}, \citenamefont {Althukair},\ and\ \citenamefont {Zotos}}]{Turakhonov:2026lia}%
  \BibitemOpen
  \bibfield  {author} {\bibinfo {author} {\bibfnamefont {Z.}~\bibnamefont {Turakhonov}}, \bibinfo {author} {\bibfnamefont {T.}~\bibnamefont {Ibrokhimov}}, \bibinfo {author} {\bibfnamefont {F.}~\bibnamefont {Atamurotov}}, \bibinfo {author} {\bibfnamefont {A.}~\bibnamefont {Abdujabbarov}}, \bibinfo {author} {\bibfnamefont {A.~K.}\ \bibnamefont {Althukair}},\ and\ \bibinfo {author} {\bibfnamefont {E.~E.}\ \bibnamefont {Zotos}},\ }\bibfield  {title} {\bibinfo {title} {{Spinning particles as probes of quartic square-root Horndeski gravity: spin{\textendash}curvature coupling and orbital dynamics}},\ }\href {https://doi.org/10.1140/epjp/s13360-026-07875-3} {\bibfield  {journal} {\bibinfo  {journal} {Eur. Phys. J. Plus}\ }\textbf {\bibinfo {volume} {141}},\ \bibinfo {pages} {629} (\bibinfo {year} {2026})}\BibitemShut {NoStop}%
\bibitem [{\citenamefont {Umarov}\ \emph {et~al.}(2025)\citenamefont {Umarov}, \citenamefont {Atamurotov}, \citenamefont {Abdujabbarov},\ and\ \citenamefont {{\"O}vg{\"u}n}}]{Umarov:2025ihy}%
  \BibitemOpen
  \bibfield  {author} {\bibinfo {author} {\bibfnamefont {D.}~\bibnamefont {Umarov}}, \bibinfo {author} {\bibfnamefont {F.}~\bibnamefont {Atamurotov}}, \bibinfo {author} {\bibfnamefont {A.}~\bibnamefont {Abdujabbarov}},\ and\ \bibinfo {author} {\bibfnamefont {A.}~\bibnamefont {{\"O}vg{\"u}n}},\ }\bibfield  {title} {\bibinfo {title} {{Spinning particle dynamics around a black hole in Lorentz Gauge theory}},\ }\href {https://doi.org/10.1016/j.dark.2025.101945} {\bibfield  {journal} {\bibinfo  {journal} {Phys. Dark Univ.}\ }\textbf {\bibinfo {volume} {48}},\ \bibinfo {pages} {101945} (\bibinfo {year} {2025})}\BibitemShut {NoStop}%
\bibitem [{\citenamefont {Uktamov}\ \emph {et~al.}(2025)\citenamefont {Uktamov}, \citenamefont {Narzilloev}, \citenamefont {Hussain}, \citenamefont {Abdujabbarov},\ and\ \citenamefont {Ahmedov}}]{Uktamov:2025bth}%
  \BibitemOpen
  \bibfield  {author} {\bibinfo {author} {\bibfnamefont {U.}~\bibnamefont {Uktamov}}, \bibinfo {author} {\bibfnamefont {B.}~\bibnamefont {Narzilloev}}, \bibinfo {author} {\bibfnamefont {I.}~\bibnamefont {Hussain}}, \bibinfo {author} {\bibfnamefont {A.}~\bibnamefont {Abdujabbarov}},\ and\ \bibinfo {author} {\bibfnamefont {B.}~\bibnamefont {Ahmedov}},\ }\bibfield  {title} {\bibinfo {title} {{Spinning particle motion and MCMC analysis of S2 star orbiting Sgr A$^\ast$}},\ }\href {https://doi.org/10.1016/j.dark.2025.102022} {\bibfield  {journal} {\bibinfo  {journal} {Phys. Dark Univ.}\ }\textbf {\bibinfo {volume} {49}},\ \bibinfo {pages} {102022} (\bibinfo {year} {2025})}\BibitemShut {NoStop}%
\bibitem [{\citenamefont {Uktamov}\ \emph {et~al.}(2026)\citenamefont {Uktamov}, \citenamefont {{\"O}vg{\"u}n}, \citenamefont {Pantig},\ and\ \citenamefont {Ahmedov}}]{Uktamov:2026gtm}%
  \BibitemOpen
  \bibfield  {author} {\bibinfo {author} {\bibfnamefont {U.}~\bibnamefont {Uktamov}}, \bibinfo {author} {\bibfnamefont {A.}~\bibnamefont {{\"O}vg{\"u}n}}, \bibinfo {author} {\bibfnamefont {R.~C.}\ \bibnamefont {Pantig}},\ and\ \bibinfo {author} {\bibfnamefont {B.}~\bibnamefont {Ahmedov}},\ }\bibfield  {title} {\bibinfo {title} {{Spinning particle dynamics, epicyclic frequencies, and transient QPO signatures in Schwarzschild spacetime}},\ }\href@noop {} {\bibfield  {journal} {\bibinfo  {journal} {arXiv:2607.11993}\ } (\bibinfo {year} {2026})},\ \Eprint {https://arxiv.org/abs/2607.11993} {arXiv:2607.11993 [gr-qc]} \BibitemShut {NoStop}%
\bibitem [{\citenamefont {Yanchyshen}\ \emph {et~al.}(2026)\citenamefont {Yanchyshen}, \citenamefont {Hackmann},\ and\ \citenamefont {L{\"a}mmerzahl}}]{Yanchyshen:2026bmy}%
  \BibitemOpen
  \bibfield  {author} {\bibinfo {author} {\bibfnamefont {O.}~\bibnamefont {Yanchyshen}}, \bibinfo {author} {\bibfnamefont {E.}~\bibnamefont {Hackmann}},\ and\ \bibinfo {author} {\bibfnamefont {C.}~\bibnamefont {L{\"a}mmerzahl}},\ }\bibfield  {title} {\bibinfo {title} {{General orbital perturbation theory in Schwarzschild space-time}},\ }\href@noop {} {\bibfield  {journal} {\bibinfo  {journal} {arXiv:2601.16887}\ } (\bibinfo {year} {2026})},\ \Eprint {https://arxiv.org/abs/2601.16887} {arXiv:2601.16887 [gr-qc]} \BibitemShut {NoStop}%
\bibitem [{\citenamefont {Witzany}\ and\ \citenamefont {Piovano}(2024)}]{Witzany:2023bmq}%
  \BibitemOpen
  \bibfield  {author} {\bibinfo {author} {\bibfnamefont {V.}~\bibnamefont {Witzany}}\ and\ \bibinfo {author} {\bibfnamefont {G.~A.}\ \bibnamefont {Piovano}},\ }\bibfield  {title} {\bibinfo {title} {{Analytic Solutions for the Motion of Spinning Particles near Spherically Symmetric Black Holes and Exotic Compact Objects}},\ }\href {https://doi.org/10.1103/PhysRevLett.132.171401} {\bibfield  {journal} {\bibinfo  {journal} {Phys. Rev. Lett.}\ }\textbf {\bibinfo {volume} {132}},\ \bibinfo {pages} {171401} (\bibinfo {year} {2024})},\ \Eprint {https://arxiv.org/abs/2308.00021} {arXiv:2308.00021 [gr-qc]} \BibitemShut {NoStop}%
\bibitem [{\citenamefont {Armaza}\ \emph {et~al.}(2016)\citenamefont {Armaza}, \citenamefont {Ba{\~n}ados},\ and\ \citenamefont {Koch}}]{Armaza:2015eha}%
  \BibitemOpen
  \bibfield  {author} {\bibinfo {author} {\bibfnamefont {C.}~\bibnamefont {Armaza}}, \bibinfo {author} {\bibfnamefont {M.}~\bibnamefont {Ba{\~n}ados}},\ and\ \bibinfo {author} {\bibfnamefont {B.}~\bibnamefont {Koch}},\ }\bibfield  {title} {\bibinfo {title} {{Collisions of spinning massive particles in a Schwarzschild background}},\ }\href {https://doi.org/10.1088/0264-9381/33/10/105014} {\bibfield  {journal} {\bibinfo  {journal} {Class. Quant. Grav.}\ }\textbf {\bibinfo {volume} {33}},\ \bibinfo {pages} {105014} (\bibinfo {year} {2016})},\ \Eprint {https://arxiv.org/abs/1510.01223} {arXiv:1510.01223 [gr-qc]} \BibitemShut {NoStop}%
\bibitem [{\citenamefont {Skoup{\'y}}\ and\ \citenamefont {Witzany}(2025)}]{Skoupy:2024uan}%
  \BibitemOpen
  \bibfield  {author} {\bibinfo {author} {\bibfnamefont {V.}~\bibnamefont {Skoup{\'y}}}\ and\ \bibinfo {author} {\bibfnamefont {V.}~\bibnamefont {Witzany}},\ }\bibfield  {title} {\bibinfo {title} {{Analytic Solution for the Motion of Spinning Particles in Kerr Spacetime}},\ }\href {https://doi.org/10.1103/PhysRevLett.134.171401} {\bibfield  {journal} {\bibinfo  {journal} {Phys. Rev. Lett.}\ }\textbf {\bibinfo {volume} {134}},\ \bibinfo {pages} {171401} (\bibinfo {year} {2025})},\ \Eprint {https://arxiv.org/abs/2411.16855} {arXiv:2411.16855 [gr-qc]} \BibitemShut {NoStop}%
\bibitem [{\citenamefont {Piovano}(2026{\natexlab{a}})}]{Piovano:2025aro}%
  \BibitemOpen
  \bibfield  {author} {\bibinfo {author} {\bibfnamefont {G.~A.}\ \bibnamefont {Piovano}},\ }\bibfield  {title} {\bibinfo {title} {{Particles with precessing spin in Kerr spacetime: Analytic solutions for eccentric orbits and homoclinic motion near the equatorial plane}},\ }\href {https://doi.org/10.1103/jzbw-m1cp} {\bibfield  {journal} {\bibinfo  {journal} {Phys. Rev. D}\ }\textbf {\bibinfo {volume} {113}},\ \bibinfo {pages} {064024} (\bibinfo {year} {2026}{\natexlab{a}})},\ \Eprint {https://arxiv.org/abs/2510.09597} {arXiv:2510.09597 [gr-qc]} \BibitemShut {NoStop}%
\bibitem [{\citenamefont {Piovano}(2026{\natexlab{b}})}]{Piovano:2026wpz}%
  \BibitemOpen
  \bibfield  {author} {\bibinfo {author} {\bibfnamefont {G.~A.}\ \bibnamefont {Piovano}},\ }\bibfield  {title} {\bibinfo {title} {{Going into a tailspin near the abyss: analytic solutions for spinning particles on near equatorial, plunging orbits in Kerr spacetime}},\ }\href@noop {} {\bibfield  {journal} {\bibinfo  {journal} {arXiv:2603.04682}\ } (\bibinfo {year} {2026}{\natexlab{b}})},\ \Eprint {https://arxiv.org/abs/2603.04682} {arXiv:2603.04682 [gr-qc]} \BibitemShut {NoStop}%
\bibitem [{\citenamefont {Drummond}\ and\ \citenamefont {Hughes}(2022)}]{Drummond:2022xej}%
  \BibitemOpen
  \bibfield  {author} {\bibinfo {author} {\bibfnamefont {L.~V.}\ \bibnamefont {Drummond}}\ and\ \bibinfo {author} {\bibfnamefont {S.~A.}\ \bibnamefont {Hughes}},\ }\bibfield  {title} {\bibinfo {title} {{Precisely computing bound orbits of spinning bodies around black holes. I. General framework and results for nearly equatorial orbits}},\ }\href {https://doi.org/10.1103/PhysRevD.105.124040} {\bibfield  {journal} {\bibinfo  {journal} {Phys. Rev. D}\ }\textbf {\bibinfo {volume} {105}},\ \bibinfo {pages} {124040} (\bibinfo {year} {2022})},\ \Eprint {https://arxiv.org/abs/2201.13334} {arXiv:2201.13334 [gr-qc]} \BibitemShut {NoStop}%
\bibitem [{\citenamefont {van~de Meent}(2018)}]{vandeMeent:2017bcc}%
  \BibitemOpen
  \bibfield  {author} {\bibinfo {author} {\bibfnamefont {M.}~\bibnamefont {van~de Meent}},\ }\bibfield  {title} {\bibinfo {title} {{Gravitational self-force on generic bound geodesics in Kerr spacetime}},\ }\href {https://doi.org/10.1103/PhysRevD.97.104033} {\bibfield  {journal} {\bibinfo  {journal} {Phys. Rev. D}\ }\textbf {\bibinfo {volume} {97}},\ \bibinfo {pages} {104033} (\bibinfo {year} {2018})},\ \Eprint {https://arxiv.org/abs/1711.09607} {arXiv:1711.09607 [gr-qc]} \BibitemShut {NoStop}%
\bibitem [{\citenamefont {Wang}\ and\ \citenamefont {Feng}(2023)}]{Wang:2023eqj}%
  \BibitemOpen
  \bibfield  {author} {\bibinfo {author} {\bibfnamefont {K.}~\bibnamefont {Wang}}\ and\ \bibinfo {author} {\bibfnamefont {C.-J.}\ \bibnamefont {Feng}},\ }\bibfield  {title} {\bibinfo {title} {{Spin vector deviation and the gravitational wave memory effect between two free-falling gyroscopes in the plane wave spacetimes}},\ }\href {https://doi.org/10.1103/PhysRevD.107.084044} {\bibfield  {journal} {\bibinfo  {journal} {Phys. Rev. D}\ }\textbf {\bibinfo {volume} {107}},\ \bibinfo {pages} {084044} (\bibinfo {year} {2023})},\ \Eprint {https://arxiv.org/abs/2301.12341} {arXiv:2301.12341 [gr-qc]} \BibitemShut {NoStop}%
\bibitem [{\citenamefont {Wang}\ and\ \citenamefont {Feng}(2024)}]{Wang:2024dmn}%
  \BibitemOpen
  \bibfield  {author} {\bibinfo {author} {\bibfnamefont {K.}~\bibnamefont {Wang}}\ and\ \bibinfo {author} {\bibfnamefont {C.-J.}\ \bibnamefont {Feng}},\ }\bibfield  {title} {\bibinfo {title} {{Geometric deformation and redshift structure caused by plane gravitational waves}},\ }\href {https://doi.org/10.1016/j.physletb.2024.138875} {\bibfield  {journal} {\bibinfo  {journal} {Phys. Lett. B}\ }\textbf {\bibinfo {volume} {855}},\ \bibinfo {pages} {138875} (\bibinfo {year} {2024})},\ \Eprint {https://arxiv.org/abs/2404.07430} {arXiv:2404.07430 [gr-qc]} \BibitemShut {NoStop}%
\bibitem [{\citenamefont {Chen}\ \emph {et~al.}(2025{\natexlab{c}})\citenamefont {Chen}, \citenamefont {Wang},\ and\ \citenamefont {Feng}}]{Chen:2025tok}%
  \BibitemOpen
  \bibfield  {author} {\bibinfo {author} {\bibfnamefont {Y.}~\bibnamefont {Chen}}, \bibinfo {author} {\bibfnamefont {K.}~\bibnamefont {Wang}},\ and\ \bibinfo {author} {\bibfnamefont {C.-J.}\ \bibnamefont {Feng}},\ }\bibfield  {title} {\bibinfo {title} {{Higher order analysis of the gravitational wave velocity memory effect between two free-falling gyroscopes in the plane wave spacetime}},\ }\href {https://doi.org/10.1103/38k7-pwlr} {\bibfield  {journal} {\bibinfo  {journal} {Phys. Rev. D}\ }\textbf {\bibinfo {volume} {111}},\ \bibinfo {pages} {104085} (\bibinfo {year} {2025}{\natexlab{c}})},\ \Eprint {https://arxiv.org/abs/2501.15745} {arXiv:2501.15745 [gr-qc]} \BibitemShut {NoStop}%
\bibitem [{\citenamefont {Andrzejewski}(2026)}]{Andrzejewski:2026wmm}%
  \BibitemOpen
  \bibfield  {author} {\bibinfo {author} {\bibfnamefont {K.}~\bibnamefont {Andrzejewski}},\ }\bibfield  {title} {\bibinfo {title} {{Dynamics of spinning particles in pp-wave spacetimes}},\ }\href {https://doi.org/10.1140/epjc/s10052-026-16222-w} {\bibfield  {journal} {\bibinfo  {journal} {Eur. Phys. J. C}\ }\textbf {\bibinfo {volume} {86}},\ \bibinfo {pages} {967} (\bibinfo {year} {2026})},\ \Eprint {https://arxiv.org/abs/2602.24267} {arXiv:2602.24267 [gr-qc]} \BibitemShut {NoStop}%
\bibitem [{\citenamefont {Wang}(2026)}]{Wang:2026grz}%
  \BibitemOpen
  \bibfield  {author} {\bibinfo {author} {\bibfnamefont {K.}~\bibnamefont {Wang}},\ }\bibfield  {title} {\bibinfo {title} {{Analytic Solution for the Motion of Spinning Particles in Plane Gravitational Wave Spacetime}},\ }\href@noop {} {\bibfield  {journal} {\bibinfo  {journal} {arXiv:2601.21438}\ } (\bibinfo {year} {2026})},\ \Eprint {https://arxiv.org/abs/2601.21438} {arXiv:2601.21438 [gr-qc]} \BibitemShut {NoStop}%
\end{thebibliography}%

\end{document}